\documentclass[a4paper,fleqn]{cas-sc}

\usepackage[round]{natbib}
\usepackage{subfig}
\usepackage{algorithm}
\usepackage{algpseudocode}
\usepackage{mathtools}
\usepackage{bm}
\usepackage{tikz}
\usetikzlibrary{automata, positioning, shapes, shadows, arrows}
\usepackage{amsmath,amsfonts,amssymb,amsthm}

\def\tsc#1{\csdef{#1}{\textsc{\lowercase{#1}}\xspace}}
\tsc{WGM}
\tsc{QE}
\tsc{EP}
\tsc{PMS}
\tsc{BEC}
\tsc{DE}
\makeatletter
\ExplSyntaxOn
\RenewDocumentCommand \author { O{} m O{} }
   {
     \ResetMarks
     \tl_if_blank:nTF { #3 } { }
       { \keys_set:nn { stm / author } { #3 } }
     \int_gincr:N \g_stm_au_int  
     \tex_gdef:D \theau@ { \int_use:N \g_stm_au_int }
     \seq_gput_right:Nn \g_stm_prelimsau_seq { #2 }
     \bool_if:NTF \l_stm_au_collab_bool
       { \seq_gput_right:cn { g_stm_clau\int_use:N \g_stm_augr_int _seq } }
       { \seq_gput_right:cn { g_stm_au\int_use:N \g_stm_augr_int _seq } }
       {          
         \int_gincr:N \g_stm_aau_int  
         \tex_gdef:D \theauthor {\int_use:N \g_stm_aau_int }
         \keys_set:nn { stm /author } { #3 }
         \tl_if_head_eq_catcode:nNTF { #1 } a
            { \processAffRef { #1 } }
            { \processAffNum { #1 } }
         \tl_if_empty:NF \l_stm_au_prefix
            { \l_stm_au_prefix_tl \c_space_token } 
         \str_if_eq:VnTF \l_stm_au_style_tl  { chinese }
            { 
              \invparsename { #2 } 
              \textcolor{\l_stm_augroup_color_tl}{\surname}
              \no_break_space:
              \textcolor{\l_stm_augroup_color_tl}{\firstname}  
            }
            { 
              \parsename { #2 } 
              \textcolor{\l_stm_augroup_color_tl}{\firstname} 
              \no_break_space:
              \textcolor{\l_stm_augroup_color_tl}{\surname}
            }
        \tl_if_empty:NF \l_stm_au_suffix_tl
        { \c_space_token \l_stm_au_suffix_tl }
         \unskip
         \textsuperscript
         {
           \tl_if_blank:nTF { #1 }
           { \tex_def:D \sep{} }
           { {\itshape\listAff} \tex_def:D \sep{\unskip,} }
           \process@marks 
           \bool_if:NT \l_stm_au_deceased_bool 
           { \sep \maltese
             \tex_def:D \sep { \unksip, }
           }
         }
        \tl_if_empty:NF \l_stm_au_degree_tl
        { ,\c_space_token \l_stm_au_degree_tl }
        \tl_if_empty:NF \l_stm_au_role_tl
        { \c_space_token (\l_stm_au_role_tl) }
       \ResetMarks
      }
      \bool_if:NT \l_stm_au_deceased_bool 
       { 
        \seq_gput_right:Nn \g_stm_maltese_seq
         {
          \tex_def:D \thefootnote { \maltese }
          \footnotetext{Deceased~author.} 
         } 
       }
      \tl_if_empty:NTF \l_stm_au_facebook_tl { }
        { 
          \wrAux { \token_to_str:N \facebookauthor
            { \l_stm_au_facebook_tl } { \exp_not:n {#2} } }
        }
      \tl_if_empty:NF \l_stm_au_twitter_tl
        { 
          \wrAux { \token_to_str:N \twitterauthor
            { \l_stm_au_twitter_tl } { \exp_not:n {#2} } }
        }
      \tl_if_empty:NF \l_stm_au_gplus_tl
        { 
          \wrAux { \token_to_str:N \gplusauthor
            { \l_stm_au_gplus_tl } { \exp_not:n {#2} } }
        }
      \tl_if_empty:NF \l_stm_au_linkedin_tl
        { 
          \wrAux { \token_to_str:N \linkedinauthor
            { \l_stm_au_linkedin_tl } { \exp_not:n {#2} } }
        }
      \tl_if_empty:NF \l_stm_au_orcid_tl
        { 
          \wrAux { \token_to_str:N \orcidauthor
            { \l_stm_au_orcid_tl } { \exp_not:n {#2} } }
        }
        \@eadauthor={#2}
        \pdfstringdef\__info_au: { #2 }
        \int_compare:nNnTF { \theau@ } < { 4 }
        { \xappto \infoauthors { \__info_au: , ~ } }
        {
          \int_compare:nNnTF { \theau@ } = { 4 }
          { \xappto \infoauthors { et~al. } }
          { }
        }        
    }
\ExplSyntaxOff
\makeatother

\begin{document}
\let\WriteBookmarks\relax
\let\printorcid\relax 
\def\floatpagepagefraction{1}
\def\textpagefraction{.001}

\shorttitle{Foreign exchange rate forecasting with regression network}

\title [mode = title]{Forecasting foreign exchange rates with  regression networks tuned by Bayesian optimization}                      

%

\author[1, 2]{Linwei Li}[type=editor]
\fnmark[1]
\cormark[2]
\ead{Linwei.Li@campus.lmu.de}
\address[1]{Department of Statistics, University of Munich (LMU), Ludwigstr. 33, 80539, Munich, Gemany}
%
\author[2]{Paul-Amaury Matt}[style=normal]
\fnmark[2]
\ead{paul-amaury.matt@daimler.com}
%
\address[2]{Daimler AG, Breitwiesenstr. 5, Stuttgart, Germany}
\author[1]{Christian Heumann}
\fnmark[3]
\ead{chris@stat.uni-muenchen.de}




\begin{abstract}
The article is concerned with the problem of multi-step financial time series forecasting of Foreign Exchange (FX) rates. To address this problem, we introduce a regression network termed RegPred Net. The exchange rate to forecast is treated as a stochastic process. It is assumed to follow a generalization of Brownian motion and the mean-reverting process referred to as generalized \textit{Ornstein-Uhlenbeck} (OU) process, with time-dependent coefficients. Using past observed values of the input time series, these coefficients can be regressed online by the cells of the first half of the network (Reg). The regressed coefficients depend only on - but are very sensitive to - a small number of hyperparameters required to be set by a global optimization
 procedure for which, Bayesian optimization is an adequate heuristic. Thanks to its multi-layered architecture, the second half of the regression network (Pred) can project time-dependent values for the OU process coefficients and generate realistic trajectories of the time series. Predictions can be easily derived in the form of expected values estimated by averaging values obtained by Monte Carlo simulation. The forecasting accuracy on a $100$ days horizon is evaluated for several of the most important FX rates such as EUR/USD, EUR/CNY and EUR/GBP. Our experimental results show that the RegPred Net significantly outperforms ARMA, ARIMA, LSTMs, and Autoencoder-LSTM models in terms of metrics measuring the absolute error (RMSE) and correlation between predicted and actual values (\emph{Pearson's} R, R-squared, MDA). Compared to black-box deep learning models such as LSTM, RegPred Net has better interpretability, simpler structure, and fewer parameters. In addition, it can predict dynamic parameters that reflect trends in exchange rates over time, which provides decision-makers with important information when dealing with sequential decision-making tasks.
\end{abstract}

%

\begin{keywords}
Time series forecasting \sep Foreign Exchange rate \sep Regression network \sep Bayesian optimization \sep  Ornstein-Uhlenbeck process
\end{keywords}

\maketitle

\section{Introduction}

Foreign Exchange (FX) rate time series are considered in Academia and the Finance Industry to be some of the most challenging time series to forecast, due to their fast-changing trends, high volatilities and complex dependencies on a large number of macro-economic factors. Nevertheless, several important applications in Market Finance (FX trading \citep{donnelly2019art}, pricing and hedging of FX derivatives \citep{hull2018options}) and International Corporate Finance (Currency Risk Management) \citep{jacque2014international} rely on accurate long-term forecasts or realistic simulations of FX rates.

With the rise of Deep Learning \citep{dl}, \citep{LSTM} and its successes in Computer vision and Natural Language Processing in the last years, deep neural networks have started to be introduced in the area of Financial Time Series Forecasting, as with the works of \citep{ANN_forecasting}, \citep{Wavelet_LSTM} and \citep{financial_forecasting_dl}. Nevertheless, we observe experimentally that when applied to FX rates famous Deep Learning models for time series such as LSTMs \citep{LSTM} do not offer particularly good performance for multi-step forecasting. Additionally, these types of neural networks operate like “black boxes” and do not offer any insight regarding the dynamics of the time series considered.

To address the issues of performance and explainability, we propose a novel regression network called {RegPred Net}. The explainability of the model is achieved by design of the network's architecture, whereby it is assumed that the time series follows a stochastic (random) process whose parameters can be directly interpreted in terms of drift, mean-reversion level, mean-reversion rate and volatility. The network’s forecasting performance is essentially the result of i) using a sufficiently general stochastic process, referred to as generalized \textit{Ornstein-Uhlenbeck} (OU) process which encompasses Brownian motion with drift and the mean-reverting process as special cases, ii) the online regression of the parameters of the stochastic model by Regression Cells, iii) carefully choosing the network's hyperparameters by Bayesian optimization  \citep{BayOpt_ori}, and iv) using a multi-layer network architecture to also capture the time-dependency of the parameters of the stochastic process.

The RegPred Net is described by a number of hyperparameters playing the role of learning rates or initial regression cell states (RegCells). The choice of the hyperparameters' values has a huge impact on the network's predictions and accuracy, so instead of setting hyperparameters arbitrarily, we adopt Bayesian optimization as an efficient procedure to set their value optimally. {Bayesian optimization} is a global optimization heuristic that is gaining popularity in Machine Learning for hyperparameter-tuning.

The rest of this article is organized as follows. Section \ref{sec:related_work} is dedicated to a discussion of related work. In Section \ref{sec:OU}, we introduce the stochastic process used for modelling FX rates, viz. the generalized OU process, explain how its parameters can be regressed in an offline-fashion and then derive an online regression procedure. Section \ref{sec:RegPred_Net} provides a detailed description of the regression cells and architecture of the RegPred Net, as well as algorithms for simulation, forecasting and calculation of the network’s loss function. Section \ref{sec:intro_BayOpt} offers some background on Bayesian optimization and presents the method and algorithms used for training RegPred Net. Section \ref{sec:intro_experiments} offers a detailed experimental validation and we finally conclude in Section \ref{sec:conclusion}.

%
%

\section{{Related work}} \label{sec:related_work}

Time series forecasting has been since decades an important area of research and development in Academia and Industry. A particular difficulty with time series is multi-step predictions of time series that do not exhibit a clear and stable trend or seasonality, which is typically the case with financial time series such as stock prices or currency rates. 

In \citep{compare_ARIMA_LSTM}, the authors compared the performance of the {ARIMA} model with {LSTM} and concluded that {LSTM} outperforms {ARIMA}, with more than $80\%$ reduction of error on one-step-ahead time series forecasting. The authors of \citep{Auto-LSTM} compared {Auto-LSTM} with the standard {LSTM} on solar power data forecasting tasks and showed that in that case,  {Auto-LSTM} performed better than {LSTM} when predicting two-steps-ahead, although the performance of both models was similar. \citep{Wavelet_LSTM} proposed a novel {LSTM}-based structure that stacks wavelet transformation, autoencoders, and {LSTM} together and this new model outperformed {LSTM} on a one-step-ahead financial time series forecasting problem.

All the research works mentioned in the previous paragraph are limited to only one or two-step ahead predictions. Unfortunately, such approaches do not benefit a whole range of real-world activities that typically relate to risk analysis or sequential decision making. Regarding multi-steps forecasting models, \citep{multistep_LSTM_1} used the {LSTM} to predict different types of periodic time series and achieved better performance than the {ARIMA} model. \citep{multistep_LSTM_2} also proposed a {LSTM}-based model that combines variational mode decomposition, singular spectrum analysis, and extreme learning machine to make one to five-steps ahead wind speed prediction. However, five-steps ahead is still considered a short period, and time series like wind speed have intrinsic cycles, unlike currency rates.

Instead of predicting the values of a times series, a simpler approach often followed consists in assessing either the probability of an increase or decrease of the time series at some future point in time compared to the present, or the probability of being higher or lower than a reference value. In \citep{trend_forecast}, the authors use an {LSTM}-based model to predict whether a $S\&P500$ stock price will increase or decrease in the next time step and conclude that the {LSTM} performs better than other machine learning models such as random forest and logistic regression. In \citep{possibility_forecast_1}, the authors seek to predict the probability that a stock outperforms its cross-sectional median at the next time step. Their results also indicate that {LSTM}s outperforms other traditional machine learning models. \citep{possibility_forecast_2} forecasts the posterior distribution of future trajectories of time series given the past. The experiments were made on periodic electricity and traffic time series and showed that the proposed method performed well (especially on limited data) by modelling the seasonal structure of the dataset. 

The above-proposed forecasting approaches all are either limited to short term forecasts, to categorical forecasting, or applied to periodic data. To the best of our knowledge, frameworks suitable for numerically forecasting complicated non-stationary time series in the medium to long term (100 steps or more) are quite rare. The {RegPred Net} is a novel type of Recurrent Network that was developed to meet these requirements and that unlike some other RNNs extracts interpretable features of the predicted time series in the form of the parameters of an OU process, thereby providing accurate information about the trend, mean-reversion level or rate and volatility of the process. Unlike neural networks in Deep Learning that often have millions of weights to learn and store in memory, the {RegPred Net} is a completely weight-free network.

Finally, \citep{tutorial_BayOpt} provided a detailed tutorial on {Bayesian optimization} and discussed the pros and cons of this method in practice. \citep{phd_bayesian_optimization} explained in his Ph.D. thesis that Bayes-optimal acquisition criteria although being rarely studied can improve the efficiency of {Bayesian optimization} and indicates that using $\xi=0.01$ as exploration parameter of acquisition function performs well in most cases. Bayesian optimization is also widely used in Machine Learning. The authors in \citep{BayOpt_ML} show that Bayesian optimization outperforms human expert-level on parameter tuning of machine learning algorithms like SVMs, Convolutional Neural Networks, and Latent Dirichlet Allocations. \citep{BayOpt_ML_2} proposed a framework based on Bayesian optimization to automatically select the architecture for deep neural networks and their results show that their framework outperforms other baseline methods on several data sets.

\section{{Stochastic process for FX rates}} 
\label{sec:OU}

\subsection{Foreign exchange rates}

\begin{figure}[pos=htbp, width=12cm, align=\centering]
        \centering
        \begin{minipage}{0.32\linewidth}
        \centering
        \subfloat[EUR/CNY]{\includegraphics
        [scale=0.33]{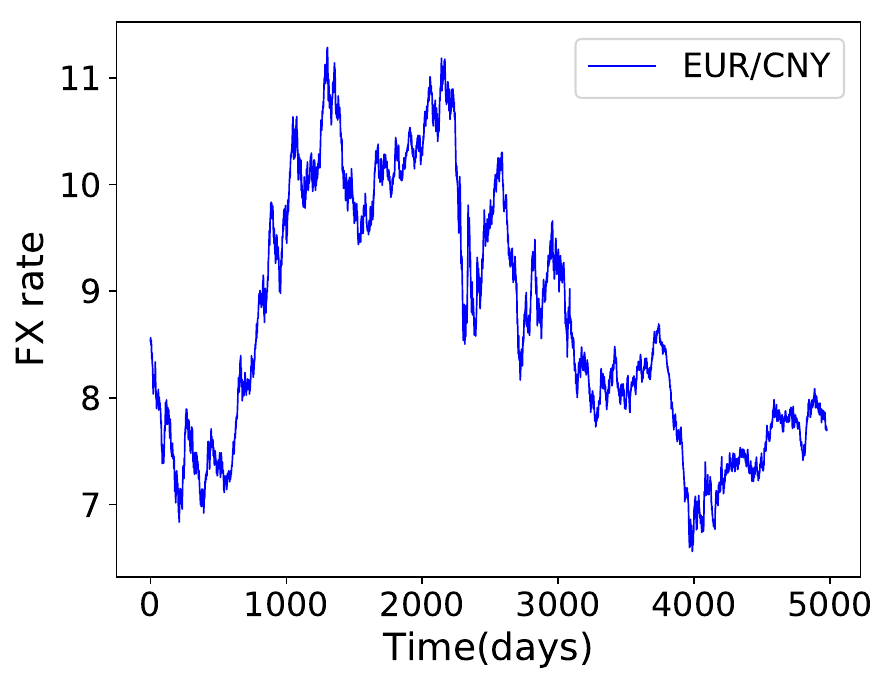}}
        \end{minipage}%
        \begin{minipage}{0.32\linewidth}
        \centering
        \subfloat[EUR/USD]{\includegraphics
        [scale=0.33]{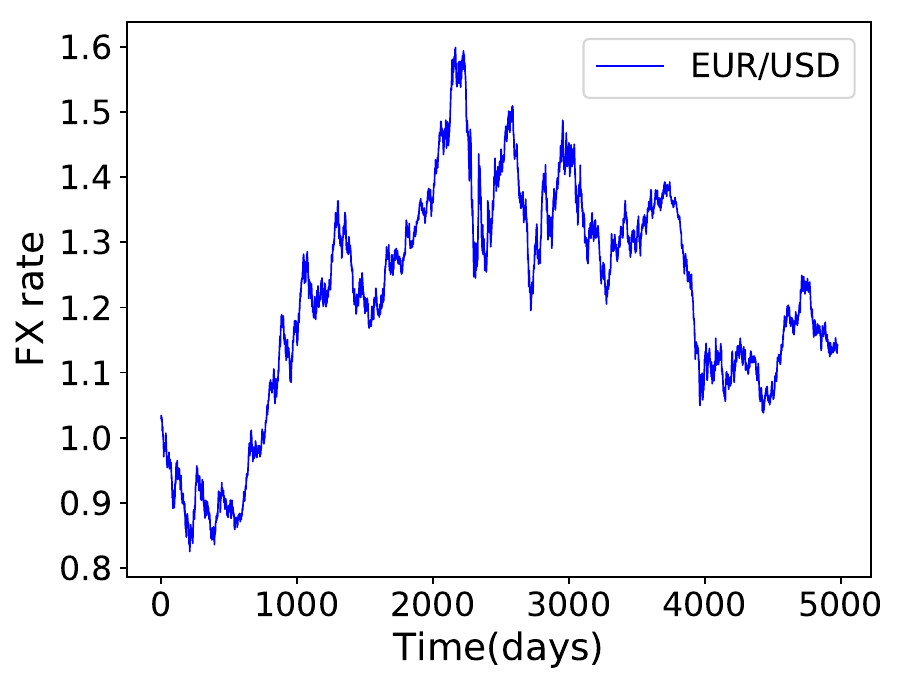}}
        \end{minipage}%
        \begin{minipage}{0.32\linewidth}
        \centering
        \subfloat[EUR/GBP]{\includegraphics
        [scale=0.33]{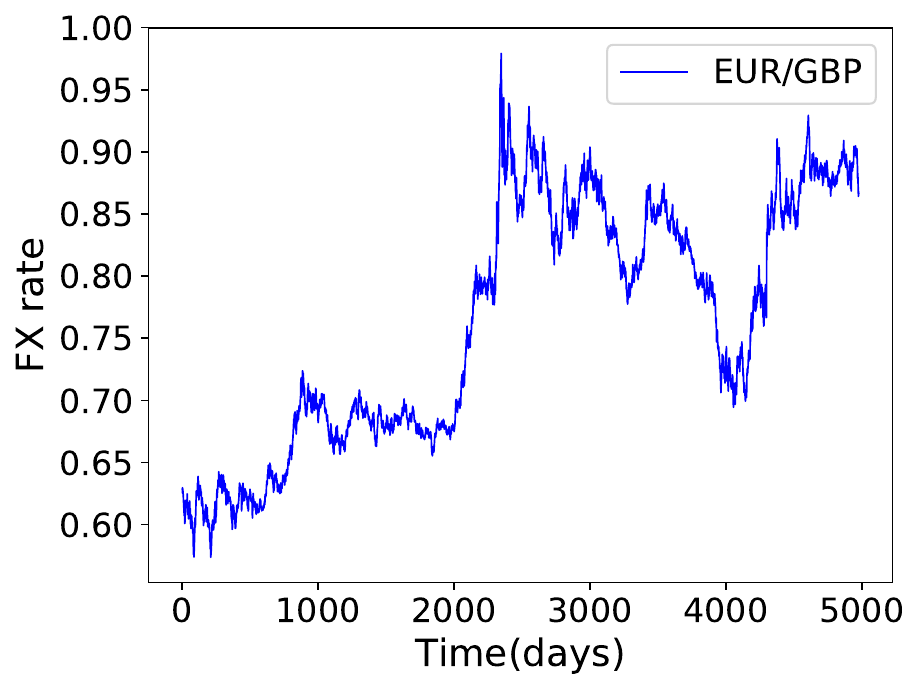}}
        \end{minipage}%
        \caption{The daily FX rates of EUR/CNY, EUR/USD, and EUR/GBP over 5000 days.} \label{fig:FX_example}
\end{figure}

A currency is a system of money in general use in a particular country. We refer to a given country’s currency as its domestic currency and refer to the currencies of other countries as foreign currencies. In Finance, a foreign exchange (FX) rate is the rate at which one currency is exchanged for another. It is also regarded as the value of one country's currency in relation to another currency. For example, the daily FX rate of EUR/CNY (Euro/Chinese Yuan) on Mar. 19, 2020 was 7.68, which means that 1 Euro was worth 7.68 Yuan. Fig. \ref{fig:FX_example} shows the 5000 days' daily FX rates of EUR/CNY, EUR/USD (US Dollar) and EUR/GBP (British Pound). The horizontal and vertical axes are time and FX rates, respectively. Since standardized currencies around the world float in value with demand, supply and consumer confidence, their relative values change over time, as illustrated in Fig. \ref{fig:FX_example}.

\subsection{Stochastic processes} 
\label{stochastic}

Daily FX rates can be modelled by discrete stochastic processes. The word stochastic is synonym of random. A discrete stochastic process is a system which evolves in time while undergoing random fluctuations over time. We describe such a system by defining a family of random variables $\{X_t \}_{t \in \mathbb{N}}$, where $X_t$  measures at time $t$ the aspect of the system which is of interest.

The discrete Wiener process \citep{wiener_process} $W_{t}$ is a discrete stochastic process defined for time steps $t$ where $t$ is a positive or null integer. The process is defined by the following properties: i) $W_0=0$, and ii) for every $t \geq 1$, the process increment given by the difference $\Delta W_t = W_{t} - W_{t-1}$ is independently and normally distributed:
\begin{equation}
\Delta W_t \sim \mathcal{N}(0, 1)
\end{equation}
Thus, the increment of the Wiener process is independent of its past values (Markov property).

Discrete Brownian motion with drift is another discrete stochastic process $X_t$ based on the Wiener process, defined by i) $X_0 \in \mathbb{R}$ and ii) for every $t \geq 1$, $\Delta X_t = X_{t} - X_{t-1} = \mu + \sigma \Delta W_t$, where the $\mu \in \mathbb{R}$ is a parameter called drift and $\sigma > 0$ is a second parameter called volatility. The Wiener process is a special case of Brownian motion where the drift is null ($\mu=0$) and the volatility is one ($\sigma=1$). Here, the drift parameter is interpreted as the deterministic trend of the process and the volatility as the amplitude of the noise or non-deterministic component in the process.

\begin{figure}[pos=htbp, width=14cm, align=\centering]
\centering
\begin{minipage}{0.32\linewidth}
\centering
\subfloat[Wiener process]{\includegraphics
[scale=0.33]{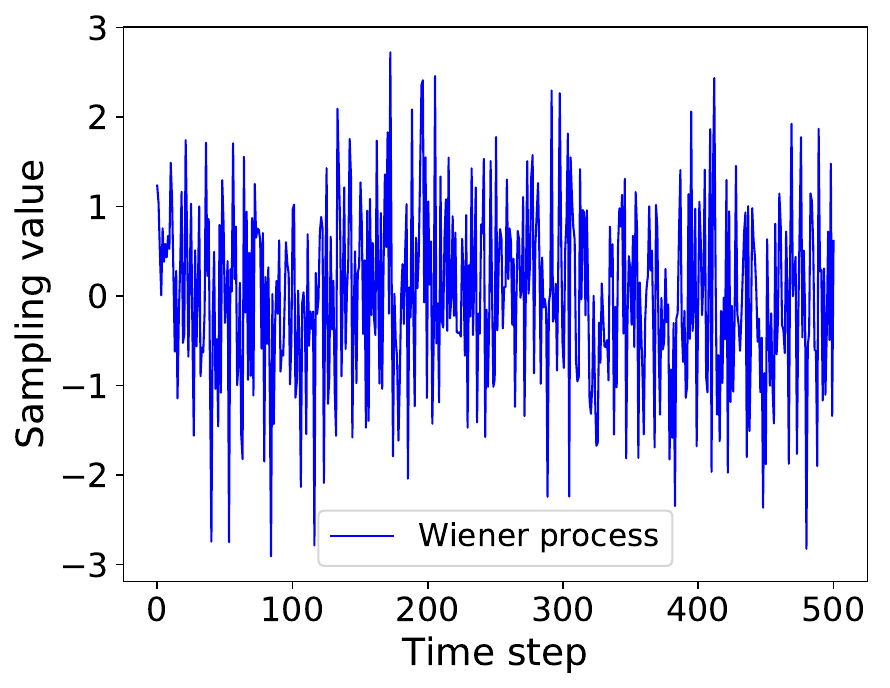}}
\end{minipage}%
\begin{minipage}{0.32\linewidth}
\centering
\subfloat[Brownian motion]{\includegraphics
[scale=0.33]{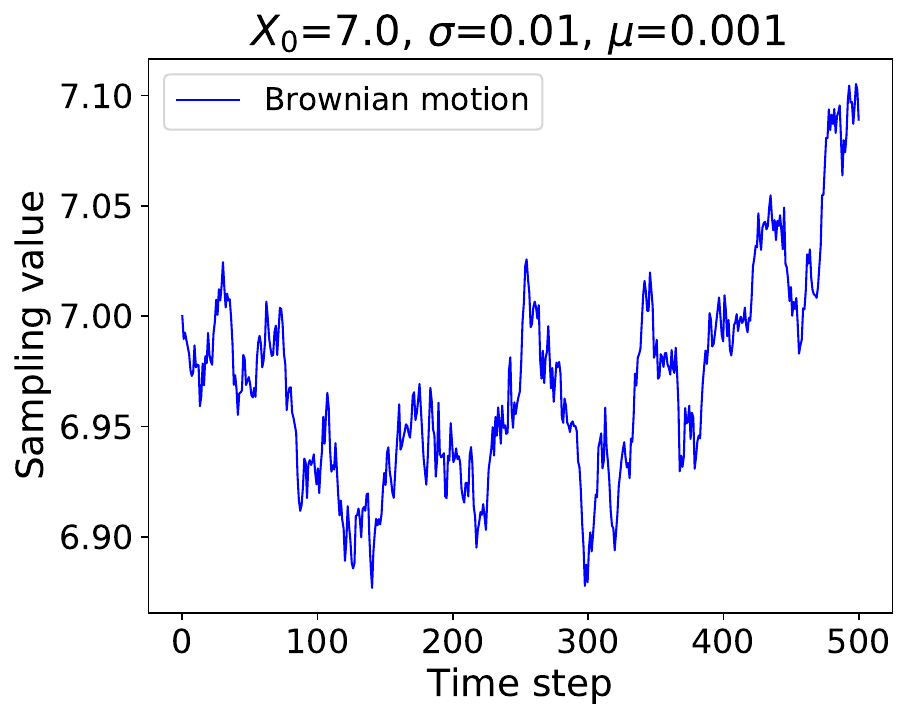}}
\end{minipage}%
\begin{minipage}{0.32\linewidth}
\centering
\subfloat[Mean-reverting process]{\includegraphics
[scale=0.33]{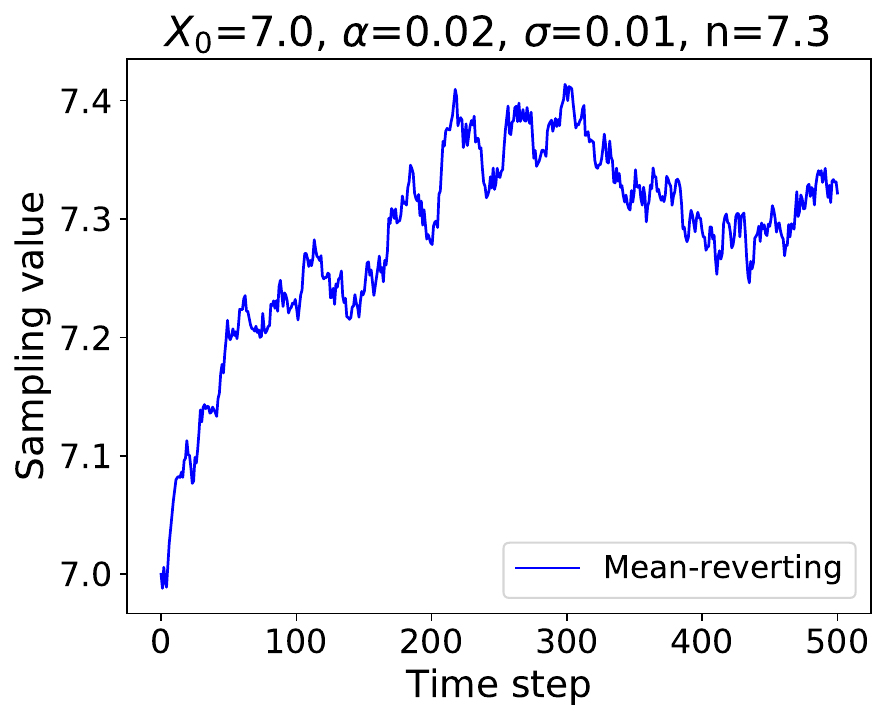}}
\end{minipage}%
\caption{Trajectories of a Wiener process, Brownian motion and mean-reverting process. }
\label{fig:Brownian}
\end{figure}

The mean reverting process, also called \textit{Ornstein-Uhlenbeck} (OU) process, is defined by i) $X_0 \in \mathbb{R}$ and ii) $\Delta X_t = X_{t} - X_{t-1} = \alpha (n - X_{t-1}) + \sigma \Delta W_t $ where $\alpha \in \mathbb{R}$ is a parameter called mean-reversion rate, $n \in \mathbb{R}$ is called mean-reversion level and $\sigma > 0$ is called volatility. This equation describes the dynamics of a variable that randomly fluctuates around some mean level $n$, follows random increments with an amplitude controlled by $\sigma$ and tends to revert back to $n$ at a speed controlled by $\alpha$ (assuming this parameter has a strictly positive value). FX rates are often modelled as mean reverting processes over long periods and as Brownian motion with drift over short periods. Fig. \ref{fig:Brownian} illustrates some trajectories of a Wiener process, a Brownian motion and a mean-reverting process over a period of 500 time steps.

\subsection{Generalized {Ornstein-Uhlenbeck} process} 
\label{sec:GOU}

We now generalize the OU process $X_t$ of dimension $1$ (univariate process) to a multivariate process $\mathbf{Y}_t$ of dimension $d$. Thus, $\mathbf{Y}_t$ denotes here a $d$-dimensional vector. The process is defined by $\mathbf{Y}_0 \in \mathbb{R}^d$ and 
\begin{equation} \label{generalized_OU}
\Delta \mathbf{Y}_t = \mathbf{Y}_t - \mathbf{Y}_{t-1} = \mathbf{A} \cdot \mathbf{Y}_{t-1} + \mathbf{N} + \bm\Sigma \cdot \Delta \mathbf{W}_t
\end{equation}
 where $\mathbf A=(\alpha_{i,j})$ is a real-valued square matrix of dimension $d \times d$, $\mathbf N$ is a real-valued vector of dimension $d$, $\bm\Sigma=(\sigma_{i,j})$ is a real-valued square matrix of dimension $d \times d$, and $\Delta \mathbf{W}_t = (\Delta W_{i,t})$ is a $d$-dimensional vector where the components $\Delta W_{i,t} \sim \mathcal{N}(0,1)$ are identically and independently normally distributed.

Observe that when $\mathbf A=0$, $\mathbf N=0$, and $\bm \Sigma=I_d$, the generalized OU process becomes a $d$-dimensional Wiener process. When $\mathbf A=0$, the generalized OU process becomes a $d$-dimensional Brownian motion with drift $\mathbf N$ and volatility $\mathbf \Sigma$. Finally when $\mathbf A=-\alpha$, $\mathbf N=\alpha \cdot n$ and $\bm\Sigma=\sigma$, the generalized OU process is a d-dimensional OU process with parameters $\alpha, n$ and $\sigma$.

\subsection{Offline regression of a generalized Ornstein-Uhlenbeck process}

Assume that we have observations for the values of the vector $Y_t$ for $t=1, …, T$, and wish to estimate the parameters $\mathbf A, \mathbf N, \bm\Sigma$ of the generalized OU process. A so called “offline” method such as Ordinary Least Squares can be used, where offline \citep{online_offline} means that the regression algorithm uses the whole data set of observations $Y_{1:d, 1:T}$ at once.

To calibrate the parameters in Eq. (\ref{generalized_OU}) using data $\mathbf{Y}_{{1:d, \ 1:T}{}}$, we first rewrite Eq. (\ref{generalized_OU}) as a linear equation of the form $\mathbf{y}_{{t}{}} = \bm{\beta} \mathbf{x}_{{t-1}{}} + \bm{\epsilon}_{{t}{}}$, where  
\begin{equation} \label{ols_OU}
\begin{aligned}
& \mathbf{y}_{{t}{}}  = \mathbf{Y}_{{t}{}} - \mathbf{Y}_{{t-1}{}} \\
&\bm{\beta} = 
\begin{bmatrix}
  n_{{1}{}} & a_{{1,1}{}}  & \dots & a_{{1,d}{}} \\
  \\
   n_{{2}{}} & a_{{2,1}{}}  & \dots & a_{{2,d}{}}  \\
   \\
   \vdots & \vdots  & \ddots & \vdots \\
   \\
   n_{{d}{}} & a_{{d,1}{}}  & \dots & a_{{d,d}{}}  \\
\end{bmatrix}
 \\
& \mathbf{x}_{{t-1}{}}  = \big[ 1, \ \mathbf{Y}_{{1, \ t-1}{}}, \ldots, \mathbf{Y}_{{d, \ t-1}} \big]^{{T}} 
\\
& \bm{\epsilon}_{{t}{}} = \bm\Sigma \cdot \Delta \mathbf{W}_{{t}{}}
 = 
\begin{bmatrix} \epsilon_{{1, t}{}} \\ \vdots \\ \epsilon_{{d, t}{}}
\end{bmatrix}
= 
\begin{bmatrix}
    \sum^{{d}{}}_{{k=1}{}} \sigma_{{1,k}{}} \cdot \Delta W_{{k,t}{}} \\ \vdots \\ \sum^{{d}{}}_{{k=1}{}} \sigma_{{d,k}{}} \cdot \Delta W_{{k,t}{}}
\end{bmatrix}
\end{aligned} 
\end{equation}

According to the ordinary least square method, the loss $L^{{OLS}{}}$ between observation $\mathbf{y}$ and function value of the linear model $\bm{\beta} \mathbf{x}$ is:
\begin{equation} \label{OLS_loss}
\begin{aligned}
L^{{OLS}} & = \frac{1}{2} \sum_{{t=1}{}}^{{T}{}} \big|\big| \mathbf{y}_{t} - \bm{\beta} \mathbf{x}_{{t-1}{}} \big|\big|^{2}_{2} 
\end{aligned}
\end{equation}
where $||\cdot||^{2}_{2}$ is the square of matrix norm 2 distance. Since $L^{{OLS}{}}$ is a positive quadratic function of $\bm{\beta}$ which admits a minimum, the optimal value of $\bm{\beta}$ can be computed by solving the quadratic problem:
\begin{equation} \label{solve_quadratic}
\cfrac{\partial L^{{OLS}{}}}{\partial \bm{\beta}} = 0
\end{equation}

The optimal value of $\bm{\beta}$ is:
\begin{equation}
\bm{\beta} = \bigg( \sum_{{t=1}{}}^{{T}{}} {y}_{{t}{}} \mathbf{x}_{{t-1}{}}^{{T}{}} \bigg) \bigg(\sum_{{t=1}{}}^{{T}{}} \mathbf{x}_{{t-1}{}} \mathbf{x}_{{t-1}{}}^{{T}{}} \bigg)^{{-1}{}}
\end{equation}

The parameters $\mathbf{A}$ and $\mathbf{N}$ can be retrieved from $\bm{\beta}$ by identification with Eq. (\ref{ols_OU}):
\begin{equation} \label{solve_ab}
\big[ \mathbf{N} \ ; \ \mathbf{A} \big] = \bm{\beta}
\end{equation}

To estimate $\bm{\Sigma}$, we first compute the covariance matrix $\bm{K}_{{\bm{\epsilon} \bm{\epsilon}, t}{}}$of $\bm{\epsilon}_{{t}{}} = \big[ \epsilon_{{1, t}{}}, \ldots, \epsilon_{{d, t}{}} \big]^{{T}{}}$:
\[
\bm{K}_{{\bm{\epsilon} \bm{\epsilon}, t}{}} =
\begin{bmatrix}
    cov(\epsilon_{{1, t}{}}, \epsilon_{{1, t}{}})   &  cov(\epsilon_{{1, t}{}}, \epsilon_{{2, t}{}}) & \dots & cov(\epsilon_{{1, t}{}}, \epsilon_{{d, t}{}}) \\
    \\
    cov(\epsilon_{{2, t}{}}, \epsilon_{{1, t}{}})       & cov(\epsilon_{{2, t}}, \epsilon_{{2, t}{}}) & \dots & cov(\epsilon_{{2, t}{}}, \epsilon_{{d, t}{}}) \\\\
   \vdots & \vdots  & \ddots & \vdots \\\\
    cov(\epsilon_{{d, t}{}}, \epsilon_{{1, t}{}})       & cov(\epsilon_{{d, t}{}}, \epsilon_{{2, t}{}}) & \dots & cov(\epsilon_{{d, t}{}}, \epsilon_{{d, t}{}})
\end{bmatrix}
\]
where
\begin{equation} \label{eq:cov_ep}
\begin{split}
cov ( \epsilon_{{i, t}{}}, \epsilon_{{j, t}{}} ) &= \mathbb{E} \Big[\big(\epsilon_{{i, t}{}} - \mathbb{E}(\epsilon_{{i, t}{}})\big)\big(\epsilon_{{j, t}{}} - \mathbb{E}(\epsilon_{{j, t}{}})\big) \Big] = \sum_{{k=1}{}}^{{d}{}} \sigma_{{i, k}{}} \cdot \sigma_{{j, k}{}} \\
\end{split}
\end{equation}

From Eq. (\ref{eq:cov_ep}) (Appx. \ref{cov_mat}) we can see that $\bm{K}_{{\bm{\epsilon} \bm{\epsilon}, t}{}}$ is not related to $t$, so $\bm{K}_{{\bm{\epsilon} \bm{\epsilon}}{}}$ can be further written as:
\begin{equation} \label{K_epsilon}
\bm{K}_{{\bm{\epsilon} \bm{\epsilon}}{}} = 
\begin{bmatrix}
\sum_{{k=1}{}}^{{d}{}} (\sigma_{{1, k}{}})^{2}        & \sum_{{k=1}{}}^{{d}{}} \sigma_{{1, k}{}}  \sigma_{{2, k}{}}   & \dots & \sum_{{k=1}{}}^{{d}{}} \sigma_{{1, k}{}}  \sigma_{{d, k}{}}   \\
    \\
    \sum_{{k=1}{}}^{{d}{}} \sigma_{{2, k}{}}  \sigma_{{1, k}{}}          & \sum_{{k=1}{}}^{{d}{}} \sigma_{{2, k}{}}  \sigma_{{2, k}{}}    & \dots & \sum_{{k=1}{}}^{{d}{}} \sigma_{{2, k}{}}  \sigma_{{d, k}{}}    \\\\
   \vdots & \vdots  & \ddots & \vdots \\\\
    \sum_{{k=1}{}}^{{d}{}} \sigma_{{d, k}{}}  \sigma_{{1, k}{}} & \sum_{{k=1}{}}^{{d}{}} \sigma_{{d, k}{}}  \sigma_{{2, k}{}}    & \dots & \sum_{{k=1}{}}^{{d}{}} (\sigma_{{d, k}{}})^{2}  
\end{bmatrix} 
\end{equation}

Since $\bm{K}_{\bm\epsilon \bm\epsilon}=\bm\Sigma \bm\Sigma^T$, it is a positive definite matrix and thus can be decomposed using the Cholesky decomposition into a product of the form $\mathbf M \mathbf M^T$, where $\mathbf M$ is a lower triangular matrix. So, we can estimate $\bm\Sigma$ by choosing $\bm\Sigma=\mathbf M$.

\subsection{Online regression of a generalized Ornstein-Uhlenbeck process} 
\label{online_reg}

The parameters $\mathbf{A}$, $\mathbf{N}$, $\bm{\Sigma}$ of the generalized OU process described in Eq. (\ref{generalized_OU}) can also be calibrated in an "online" fashion. The word online \citep{online_offline} refers to any method that estimates the result of an algorithm without having all input data at once, but step-by-step processes the input. In this way, at every time step $t$, the online algorithm updates the parameters from the previous time step $\mathbf{A}_{{t-1}}$, $\mathbf{N}_{{t-1}}$, $\bm{\Sigma}_{{t-1}}$ to $\mathbf{A}_{{t}}$, $\mathbf{N}_{{t}}$, $\bm{\Sigma}_{{t}}$. We replace in Eq. (\ref{generalized_OU}) the static parameters $\mathbf{A}$, $\mathbf{N}$, $\bm{\Sigma}$ by the time-dependent parameters as follows:
\begin{equation}
\Delta \mathbf{Y}_{{t}} = \mathbf{A}_{{t-1}} \cdot \mathbf{Y}_{{t-1}} + \mathbf{N}_{{t-1}} + \bm\Sigma_{{t-1}} \cdot \Delta \mathbf{W}_{{t}}
\end{equation}

The error term $\bm{\epsilon}_{{t}}$ is defined as:
\begin{equation} \label{update_epsilon}
\bm{\epsilon}_{{t}} = \Delta \mathbf{Y}_{{t}} - \big( \mathbf{A}_{{t-1}} \mathbf{Y}_{{t-1}} + \mathbf{N}_{{t-1}} \big)
=
\begin{bmatrix}
     \epsilon_{{1,t}}\\ 
      \vdots \\
      \epsilon_{{d,t}} 
\end{bmatrix} 
\end{equation}

To infer the update rule for $\mathbf{A}_{{t}}$ and $\mathbf{N}_{{t}}$, we define the quadratic loss $L_{{t}}(\mathbf{A}_{{t-1}}; \mathbf{N}_{{t-1}})$ at time $t$ as:
\begin{equation} \label{eq:L_AN}
\begin{aligned}
L_{{t}} \big( \mathbf{A}_{{t-1}}; \mathbf{N}_{{t-1}} \big)  = \epsilon_t^T \epsilon_t
\end{aligned}
\end{equation}

The update rules of $\mathbf{A}_{{t}}$ and $\mathbf{N}_{{t}}$ can be expressed as gradient descent steps:
\begin{equation} \label{update_A_ori}
\mathbf{A}_{{t}} \leftarrow \mathbf{A}_{{t-1}} - \eta_{{A}} \frac{ \partial L_{{t}} \big( \mathbf{A}_{{t-1}}; \mathbf{N}_{{t-1}} \big) } { \partial \mathbf{A}_{{t-1}} }
\end{equation}
\begin{equation} \label{update_N_ori}
\mathbf{N}_{{t}} \leftarrow \mathbf{N}_{{t-1}} - \eta_{{N}} \frac{ \partial L_{{t}} \big( \mathbf{A}_{{t-1}}; \mathbf{N}_{{t-1}} \big) } { \partial \mathbf{N}_{{t-1}} }
\end{equation}
where $\eta_{{A}}$, $\eta_{{N}}$ and $\eta_{{\Sigma}}$ are learning rates for $\mathbf{A}_{{t}}$, $\mathbf{N}_{{t}}$ and $\bm{\Sigma}_{{t}}$, respectively. 

The partial derivative of $L_{{t}}(\mathbf{A}_{{t-1}}; \mathbf{N}_{{t-1}})$ with respect to $\mathbf{A}_{{t-1}}$ is computed as:
\begin{equation} \label{partial_A}
\begin{split}
\frac{ \partial L_{{t}}(\mathbf{A}_{{t-1}}; \mathbf{N}_{{t-1}}) } { \partial \mathbf{A}_{{t-1}} } &= - 2 \cdot \bm{\epsilon}_{{t}} \mathbf{Y}_{{t-1}}^{{T}}
\end{split}
\end{equation}

Similarly, the partial derivative of $L_{{t}}(\mathbf{A}_{{t-1}}; \mathbf{N}_{{t-1}})$ with respect to $\mathbf{N}_{{t-1}}$ is:
\begin{equation} \label{partial_N}
\begin{split}
\frac{ \partial L_{{t}}(\mathbf{A}_{{t-1}}; \mathbf{N}_{{t-1}}) } { \partial \mathbf{N}_{{t-1}} } & = -2 \cdot \bm{\epsilon}_{{t}} 
\end{split}
\end{equation}
The calculation details are given in Appx. \ref{appx_gradients}. 

Substituting Eq. (\ref{partial_A}) and (\ref{partial_N}) into Eq. (\ref{update_A_ori}) and (\ref{update_N_ori}), we get the following update rules for $\mathbf{A}_{{t}}$ and $\mathbf{N}_{{t}}$:
\begin{equation}\label{update_A}
\mathbf{A}_{{t}} \leftarrow \mathbf{A}_{{t-1}}  + 2\eta_{{A}}  \bm{\epsilon}_{{t}}  \mathbf{Y}_{{t-1}}^{{T}} 
\end{equation}
\begin{equation}\label{update_N}
\mathbf{N}_{{t}}  \leftarrow \mathbf{N}_{{t-1}}  + 2\eta_{{N}} \bm{\epsilon}_{{t}} 
\end{equation}

Similarly, we use gradient descent to get the update rule for $\bm{\Sigma}_{{t}}$. We first define the loss $L_{{t}} \big( \bm{\Sigma}_{{t-1}} \big)$ at time $t$ as:
\begin{equation} \label{eq:L_sigma}
\begin{split}
L_{{t}} \big( \bm{\Sigma}_{{t-1}} \big) = \big|\big|\enskip \bm{\Sigma}_{{t-1}} \bm{\Sigma}^{{T}}_{{t-1}} - \hat{cov}( \bm{\epsilon}_{{t}} ) \enskip\big|\big|^{{2}}_{{2}}
\end{split}
\end{equation}
where $\hat{cov}( \bm{\epsilon}_{{t}} ) = \big(\big( \hat{cov}( \bm{\epsilon}_{{i, t}}, \bm{\epsilon}_{{j, t}} ) \big)\big)$, and $\hat{cov}( \bm{\epsilon}_{{i, t}}, \bm{\epsilon}_{{j, t}})$ is an online estimate at $t$ of the covariance of $\bm{\epsilon}_{{i, t}}$ and $\bm{\epsilon}_{{j, t}}$. 

The derivative of $L_{{t} }(\bm{\Sigma}_{{t-1} })$ over the whole matrix $\bm{\Sigma}_{{t-1} }$ is computed as (see Appx. \ref{appx_gradients}):
\begin{equation}
\frac{ \partial L_{{t} }(\bm{\Sigma}_{{t-1} }) }{ \partial \bm{\Sigma}_{{t-1} } } =
4 \cdot \big(\bm{\Sigma}_{{t-1} } \bm{\Sigma}_{{t-1} }^{{T} } - \hat{ cov}(\bm{\epsilon}_{{t} })\big) \cdot \bm{\Sigma}_{{t-1} }
\end{equation}

Thus, $\bm{\Sigma}_{t}$ can be updated by:
\begin{equation} \label{update_sigma}
\begin{aligned}
\bm{\Sigma}_{{t} } &\leftarrow \bm{\Sigma}_{{t-1} } - 4 \eta_{{\Sigma}} \cdot\big(\bm{\Sigma}_{{t-1} } \bm{\Sigma}_{{t-1} }^{{T} } - \hat{ cov}(\bm{\epsilon}_{{t} })\big) \cdot \bm{\Sigma}_{{t-1} }  
\end{aligned}
\end{equation}

To estimate the covariance matrix $\hat{cov}(\bm{\epsilon}_{{t} })$ in Eq. (\ref{update_sigma}), we first estimate the expectation of $\bm{\epsilon}_{{t} }$ by using an Exponential Moving Average (EMA) with weight $\varphi$:
\begin{equation} \label{update_E_epsilon}
\begin{aligned}
\hat{\mathbb{E}}(\bm{\epsilon}_{{t} }) & = EMA_{\varphi}(\bm{\epsilon}_{{t} }) = \varphi \cdot \bm{\epsilon}_{{t} } + (1-\varphi) \cdot EMA_{\varphi}(\bm{\epsilon}_{{t-1} })
\end{aligned}
\end{equation}

Since the covariance matrix $cov(\bm{\epsilon}_{{t} })$ is mathematically defined as 
\begin{equation} \label{covariance_epsilon_definition}
\begin{aligned}
cov(\bm{\epsilon}_t) &= \mathbb{E} \big[ \big( \bm{\epsilon}_t - \mathbb{E}[\bm{\epsilon}_t] \big) \big( \bm{\epsilon}_t - \mathbb{E}[\bm{\epsilon}_t] \big)^T \big]
\end{aligned}
\end{equation}
we can again introduce another EMA with weight $\rho$ for estimating also the outer expectation operator and estimate the covariance matrix as:
\begin{equation} \label{update_cov}
\begin{aligned}
\hat{cov}(\bm{\epsilon}_{t})  & = EMA_{\rho} \big[ \big(\bm{\epsilon}_{t} - \hat{\mathbb{E}}(\bm{\epsilon}_{{t} })  \big) \big(\bm{\epsilon}_{t} - \hat{\mathbb{E}}(\bm{\epsilon}_{{t} })  \big)^{T} \big]
\end{aligned}
\end{equation}

The online regression procedure allows to estimate the parameters $\mathbf{A}_{t}$, $\mathbf{N}_{t}$, $\bm{\Sigma}_{t}$
of the generalized OU process given only the following hyperparameters:
\begin{itemize}
\item initial values of $\mathbf{A}_0$, $\mathbf{N}_0$, $\bm{\Sigma}_0$, $\hat{\mathbb{E}}(\epsilon_0)$ and $\hat{cov}(\epsilon_0)$ 
\item a 5-dimensional vector of learning rates $\mathbf{H} = [\eta_{{A}}, \eta_{{N}} , \eta_{{\Sigma}}, {{\varphi}}, {{\rho}}]$ composed of three learning rates $\eta_{{A}}, \eta_{{N}} , \eta_{{\Sigma}}$ used in gradient descent update rules and the weights $\varphi$ and $\rho$ of two exponential moving averages for $\hat{\mathbb{E}}(\epsilon_t)$ and $\hat{cov}(\epsilon_t)$. 
\end{itemize}
Different values for the hyperparameters lead to different estimates for the parameters $\mathbf{A}_t$, $\mathbf{N}_t$ and $\bm{\Sigma}_t$, unlike the offline regression which always lead to the same result. For instance, with high values of $\mathbf{H}$, the parameters adapt very fast to the time series but contain more noise and tend to degrade the accuracy of long term forecasts. Conversely, small values of $\mathbf{H}$ lead to slowly changing stochastic parameter estimates, which is good for long term but tends to degrade the accuracy of short term forecasts. 

\section{RegPred Network} \label{sec:RegPred_Net}

This section introduces the two networks RegNet and PredNet that compose RegPred Net. We present their respective recurrent cells and network architecture. RegNet and PredNet are then simply juxtaposed to form the overall RegPred Net.

\subsection{Regression Cell (RegCell) and Regression Network (RegNet)} \label{sec:RegCell}

\tikzstyle{block} = [draw, rectangle, fill=black!15, text=black, text width=28 em,                           text centered, rounded corners, minimum                                 height=50mm, font=\fontsize{8}{0}\selectfont,                           node distance=5em]
\tikzstyle{blank} = [ fill=white, text=black, text width=1em, text centered, minimum height=1mm, font=\fontsize{8}{0}\selectfont, node distance=6em]

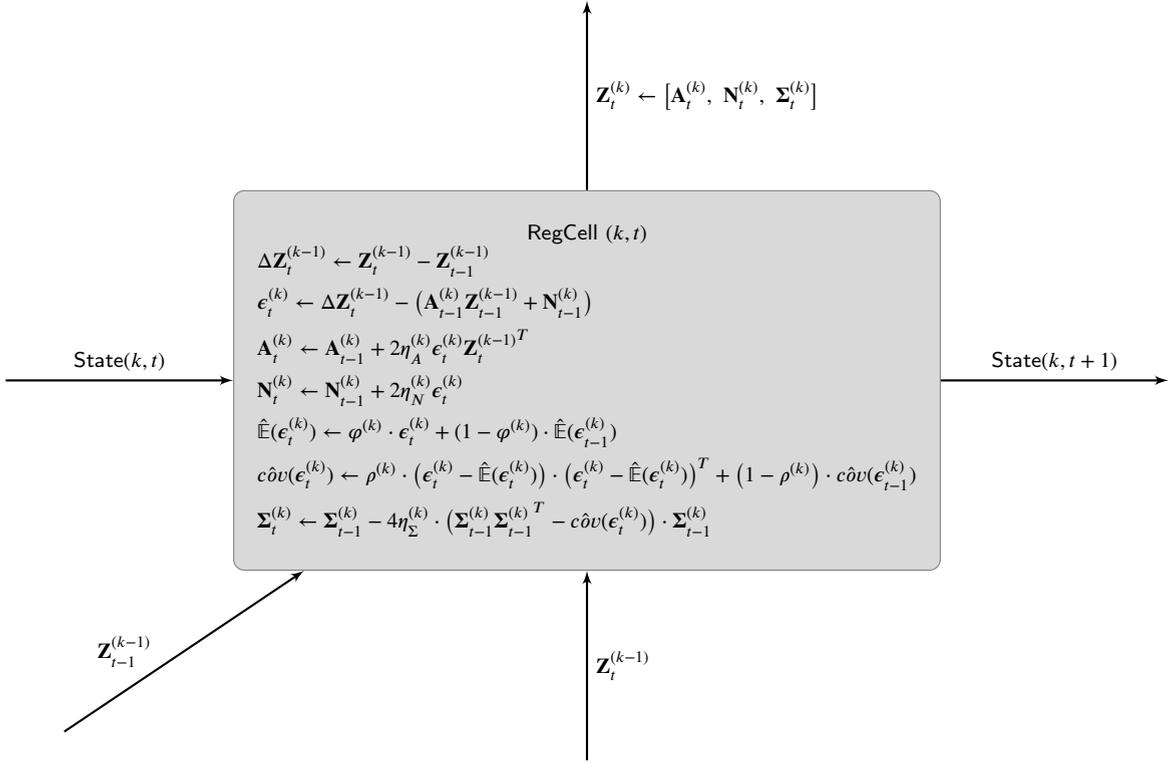
\begin{figure}[pos=htbp]
\centering
\begin{tikzpicture}[node distance=1cm, auto]

\node [block] (Z_11) 
{
RegCell $(k,t)$ \\
 $\begin{aligned}
  & \Delta \mathbf{Z}_{t}^{{(k-1)}} \leftarrow \mathbf{Z}_{t}^{{(k-1)}} - \mathbf{Z}_{{t-1}}^{{(k-1)}} \\
 & \bm{\epsilon}^{{(k)}}_{t} \leftarrow \Delta \mathbf{Z}_{t}^{{(k-1)}} - \big( \mathbf{A}_{{t-1}}^{{(k)}} \mathbf{Z}_{{t-1}}^{{(k-1)}} + \mathbf{N}_{{t-1}}^{{(k)}} \big) \\
 & \mathbf{A}_{t}^{{(k)}} \leftarrow \mathbf{A}_{{t-1}}^{{(k)}} + 2\eta_{{A}}^{{(k)}} \bm{\epsilon}_{t}^{{(k)}} {\mathbf{Z}_{t}^{{(k-1)}}}^{{T}} \\
 & \mathbf{N}_{t}^{{(k)}} \leftarrow \mathbf{N}_{{t-1}}^{{(k)}} + 2\eta_{{N}}^{{(k)}} \bm{\epsilon}_{t}^{{(k)}} \\
 & \hat{\mathbb{E}}(\bm{\epsilon}_{t}^{{(k)}})  \leftarrow  \varphi^{{(k)}} \cdot \bm{\epsilon}_{t}^{{(k)}} + (1-\varphi^{{(k)}}) \cdot \hat{\mathbb{E}}(\bm{\epsilon}_{{t-1}}^{{(k)}}) \\
 & \hat{cov}(\bm{\epsilon}_{t}^{{(k)}}) \leftarrow 
 \rho^{{(k)}} \cdot \big(\bm{\epsilon}_{t}^{{(k)}} - \hat{\mathbb{E}}(\bm{\epsilon}_{t}^{{(k)}}) \big) \cdot \big(\bm{\epsilon}_{t}^{{(k)}} - \hat{\mathbb{E}}(\bm{\epsilon}_{t}^{{(k)}} )\big)^{T} + \big( 1-\rho^{{(k)}} \big) \cdot \hat{cov} ( \bm{\epsilon}_{{t-1}}^{{(k)}} ) \\ 
 & \bm{\Sigma}_{t}^{{(k)}} \leftarrow \bm{\Sigma}_{{t-1}}^{{(k)}} - 4 \eta_{{\Sigma}}^{{(k)}} \cdot \big(\bm{\Sigma}_{{t-1}}^{{(k)}} {\bm{\Sigma}_{{t-1}}^{{(k)}}}^{{T}} - \hat{cov}(\bm{\epsilon}_{t}^{{(k)}})\big) \cdot \bm{\Sigma}_{{t-1}}^{{(k)}} \\
\end{aligned}$
};

\node[blank, left=3cm of Z_11] (blank_1){};
\draw [black, ->, thick, -latex'] (blank_1) -- node [above, midway, font=\fontsize{8}{0}\selectfont, text=black]
 {
        $\begin{aligned}
        \text{State} (k, t)
        \end{aligned}$
 }
(Z_11);

\node[blank, below left=3cm of Z_11] (blank_1_1){};
\draw [black, ->, thick, -latex'] (blank_1_1) -- node [above=0.2cm, near start, font=\fontsize{8}{0}\selectfont, text=black]
 {
 $\begin{aligned}
 \mathbf{Z}_{{t-1}}^{{(k-1)}}
\end{aligned}$
 }
(Z_11);

 \node[blank, right=3cm of Z_11] (blank_2){};
 \draw [black, ->, thick, -latex'] (Z_11) -- node [above, midway, font=\fontsize{8}{0}\selectfont, text=black]
 {
  $\begin{aligned}
  \text{State} (k, t+1)
\end{aligned}$
 }
 (blank_2);

\node[blank, above=2.5cm of Z_11] (blank_3){};
\draw [black, ->, thick, -latex'] (Z_11) -- node [right, midway, font=\fontsize{8}{0}\selectfont, text=black]
  {
 $\begin{aligned}
   & \mathbf{Z}_{t}^{(k)} \leftarrow \big[ \mathbf{A}_{t}^{{(k)}}, \ \mathbf{N}_{t}^{{(k)}}, \ \bm{\Sigma}_{t}^{{(k)}} \big] \\
 \end{aligned}$
 }
 (blank_3);

 \node[blank, below=2.5cm of Z_11] (blank_4){};
\draw [black, ->, thick, -latex'] (blank_4) -- node [right, midway, font=\fontsize{8}{0}\selectfont, text=black]
 {
 $\begin{aligned}
   & \mathbf{Z}_{t}^{{(k-1)}} \\
 \end{aligned}$
 }
 (Z_11);

\end{tikzpicture}
\caption{The RegCell $(k, t)$ updates the parameters $\mathbf A, \mathbf N, \bm\Sigma$ of the online regression model: $\Delta \mathbf{Z}_{t}^{{(k-1)}} = \mathbf{A}_{{t-1}}^{{(k)}} \mathbf{Z}_{{t-1}}^{{(k-1)}} + \mathbf{N}_{{t-1}}^{{(k)}} + \bm{\Sigma}_{{t-1}}^{{(k)}} \Delta \mathbf{W}_{{t}}^{{(k)}}$. $\text{State}(k,t)\coloneqq \mathbf{Z}_{{t-1}}^{{(k)}},  \hat{\mathbb{E}}(\bm{\epsilon}_{{t-1}}^{{(k)}}), \hat{cov}(\bm{\epsilon}_{{t-1}}^{{(k)}}) $.}
\label{fig:RegCell_kt}
\end{figure}

In this subsection, we introduce a recurrent network termed RegNet for the online estimation of parameters of a generalized OU process $\mathbf{Y}_t$. The basic RegNet is a single layer network using a recurrent cell called RegCell. The RegCell at time $t$ and layer $k$, which is defined in Algorithm \ref{alg:Algorithm_RegCell} and illustrated in Fig. \ref{fig:RegCell_kt}, simply encapsulates all the update rules needed for the online regression of the parameters $\mathbf A$, $\mathbf N$, $\bm\Sigma$ as described in the previous section. Several layers of the basic RegNet can be stacked on top of each other to form a multi-layered RegNet (see Fig. \ref{fig:RegNet}). In this case, the $k$-th layer performs online regression of the parameters $\mathbf{A}^{{(k)}}$, $\mathbf{N}^{{(k)}}$, $\bm{\Sigma}^{{(k)}}$ of an generalized OU process for the multivariate input series $\mathbf{Z}^{{(k-1)}}_{{t}}$, defined as the flattened vector of regressed parameters from the OU process in $k-1$-th layer:
\begin{equation}
  \forall k=1,..., K
    \begin{cases}
      \Delta \mathbf{Z}_{{t}}^{{(k-1)}} = \mathbf{A}_{{t-1}}^{{(k)}} \mathbf{Z}_{{t-1}}^{{(k-1)}} + \mathbf{N}_{{t-1}}^{{(k)}} + \bm{\Sigma}_{{t-1}}^{{(k)}} \Delta \mathbf{W}_{{t}}^{{(k)}} \\
      \\
     \mathbf{Z}_{{t}}^{{(k-1)}} = \big[ \mathbf{A}_{{t}}^{{(k-1)}}, \ \mathbf{N}_{{t}}^{{(k-1)}}, \ \bm{\Sigma}_{{t}}^{{(k-1)}} \big] \\
    \end{cases}       
\end{equation}
with $\mathbf{Z}_{{t}}^{{(0)}} = \mathbf{Y}_{{t}}$.

\begin{algorithm}[!htbp]
\caption{RegCell in layer $k$ at time $t$}
\begin{algorithmic}[1]  
\renewcommand{\algorithmicrequire}{\textbf{Input:}}
\renewcommand{\algorithmicensure}{\textbf{Output:}}

\Require {State$(k, t) \coloneqq  \Big\{ \ \mathbf{Z}^{{(k)}} _{{t-1}} \coloneqq \big[ \mathbf{A}^{{(k)}}_{{t-1}}, \ \mathbf{N}^{{(k)}}_{{t-1}}, \ \mathbf{\Sigma}^{{(k)}}_{{t-1}} \big], \ \hat{\mathbb{E}}(\bm{\epsilon}^{{(k)}}_{{t-1}}), \ \hat{cov}(\bm{\epsilon}^{{(k)}}_{{t-1}}) \ \Big\}$, 

$\mathbf{Z}^{{(k-1)}} _{{t-1  :  t}} \coloneqq \begin{cases} \big[ \mathbf{A}^{{(k-1)}} _{{t-1  :  t}} \ , \ \mathbf{N}^{{(k-1)}} _{{t-1  :  t}} \ , \ \mathbf{\Sigma}^{{(k-1)}} _{{t-1  :  t}} \big],  &\text{if} \quad k > 1, \\
 \ \mathbf{Y}_{{t-1  :  t}} \ , &\text{elif} \quad k = 1.
\end{cases}$ }

\Ensure State$(k, t+1) \coloneqq  \Big\{ \ \mathbf{Z}^{{(k)}} _{{t}} \coloneqq \big[ \mathbf{A}^{{(k)}} _{{t}}, \ \mathbf{N}^{{(k)}} _{{t}}, \ \mathbf{\Sigma}^{{(k)}} _{{t}} \big], \ \hat{\mathbb{E}}(\bm{\epsilon}^{{(k)}}_{t}), \ \hat{cov}(\bm{\epsilon}^{{(k)}}_{t}) \ \Big\}$

\State $\Delta \mathbf{Z}^{{(k-1)}}_{{t}} \leftarrow \mathbf{Z}^{{(k-1)}}_{{t}} - \mathbf{Z}^{{(k-1)}}_{{t-1}}$

 \State $ \bm{\epsilon}^{{(k)}}_{{t}} \leftarrow \Delta \mathbf{Z}^{{(k-1)}}_{{t}} - \big( \mathbf{A}^{{(k)}}_{{t-1}} \mathbf{Z}^{{(k-1)}}_{{t-1}} + \mathbf{N}^{{(k)}}_{{t-1}} \big)$
  \State $ \mathbf{A}^{{(k)}}_{t} \leftarrow \mathbf{A}^{{(k)}}_{t-1} + 2\eta^{{(k)}}_{{A}} \bm{\epsilon}^{{(k)}}_{t} \mathbf{Z}_{t}^{{{(k-1)}}^{T}} $
  \State $ \mathbf{N}^{{(k)}}_{t} \leftarrow \mathbf{N}^{{(k)}}_{t-1} + 2\eta^{{(k)}}_{{N}} \bm{\epsilon}^{{(k)}}_{t} $
  \State $ \hat{\mathbb{E}}(\bm{\epsilon}^{{(k)}}_{t-1}) \coloneqq EMA_{\varphi}(\bm{\epsilon}^{{(k)}}_{t-1}) $
  \State $ \hat{\mathbb{E}}(\bm{\epsilon}^{{(k)}}_{t}) \leftarrow \varphi^{{(k)}} \cdot \bm{\epsilon}^{{(k)}}_{t} + (1-\varphi^{{(k)}}) \cdot \hat{\mathbb{E}}(\bm{\epsilon}^{{(k)}}_{t-1}) $
  \State $ \hat{cov}(\bm{\epsilon}^{{(k)}}_{t-1}) \coloneqq EMA_{\rho}\big(\hat{cov}(\bm{\epsilon}^{{(k)}}_{t-1})\big) $
 \State $ \hat{cov}(\bm{\epsilon}^{{(k)}}_{t}) \leftarrow \rho^{{(k)}} \cdot \big(\bm{\epsilon}^{{(k)}}_{t} - \hat{\mathbb{E}}(\bm{\epsilon}^{{(k)}}_{t})\big) \cdot \big(\bm{\epsilon}^{{(k)}}_{t} - \hat{\mathbb{E}}(\bm{\epsilon}^{{(k)}}_{t})\big) + (1-\rho^{{(k)}})\cdot \hat{cov}(\bm{\epsilon}^{{(k)}}_{t-1}) $
  \State $ \bm{\Sigma}^{{(k)}}_{t} \leftarrow \bm{\Sigma}^{{(k)}}_{t-1} - 4 \eta^{{(k)}}_{{\Sigma}} \cdot\big(\bm{\Sigma}^{{(k)}}_{t-1} \bm{\Sigma}_{t-1}^{{{(k)}}^{T}} - \hat{cov}(\bm{\epsilon}^{{(k)}}_{t})\big) \cdot \bm{\Sigma}^{{(k)}}_{t-1} $
  \State $ \mathbf{Z}^{{(k)}}_{{t}} \leftarrow \big[ \mathbf{A}^{{(k)}}_{t}, \mathbf{N}^{{(k)}}_{t}, \bm{\Sigma}^{{(k)}}_{t} \big]$

\end{algorithmic} \label{alg:Algorithm_RegCell}
\end{algorithm}

\tikzstyle{block_reg} = [draw, 
                             rectangle, 
                             fill=black!15, 
                             text width=1.5em, 
                             text centered, 
                             text=black,
                             minimum height={width("RegCell")+3pt},
                             font=\fontsize{7}{0}\selectfont, 
                             node distance=5em]
\tikzstyle{blank} = [fill=white, text width=1em,  text=black, text centered, minimum height=1mm, font=\fontsize{7}{0}\selectfont, node distance=6em]

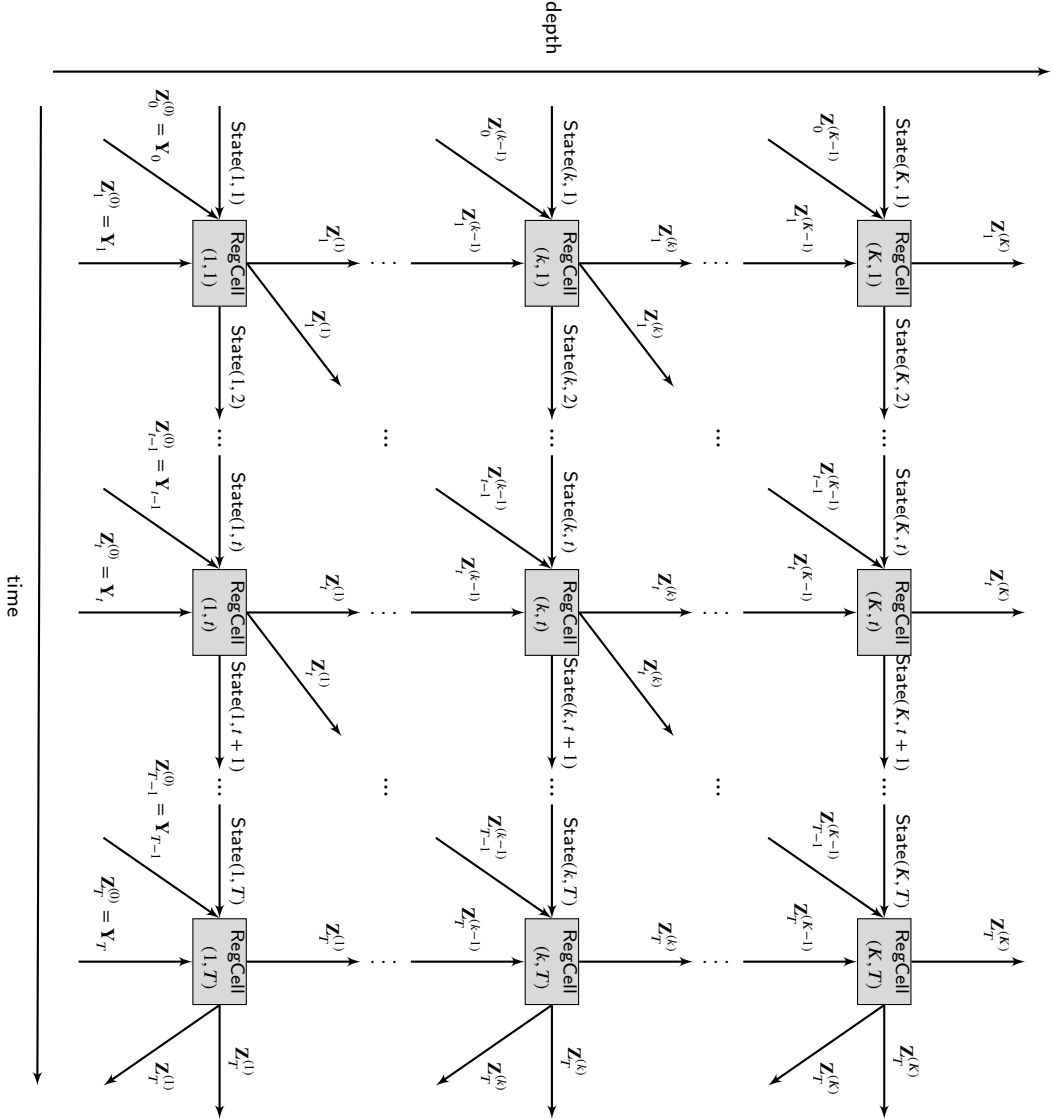
\begin{figure}[pos=htbp, width=11cm, align=\centering] 
\centering
\begin{tikzpicture} 

\node [black, block_reg] (Z_1) 
{
\rotatebox{-90}{$\begin{aligned}
\text{Reg} &\text{Cell} \\
(1&, 1)
\end{aligned}$}
 };

 \node [black, below=1.5cm of Z_1] (vdots_1) {\vdots};

  \node[black, blank, above left=1.5cm of Z_1] (blank_s1){};
  \draw [black, ->, thick, -latex'] (blank_s1) -- node [black, above=0.1cm, midway, font=\fontsize{7}{0}\selectfont] 
  {\rotatebox{-90}{$\begin{aligned}
  \mathbf{Z}_{{0}}^{{(0)}} = \mathbf{Y}_{{0}}
\end{aligned}$}}
(Z_1.north);

  \node[black, blank, below right=1.5cm of Z_1] (blank_s2){};
  \draw [black, ->, thick, -latex'] (Z_1.east) -- node [black, above=0.1cm, near end, font=\fontsize{7}{0}\selectfont]
  {\rotatebox{-90}{$\begin{aligned}
  \mathbf{Z}_{{1}}^{{(1)}} 
  \end{aligned}$}}
(blank_s2);

\node [black, right=1.8cm of vdots_1] (vdots_s1) {\vdots};

  \node[black, blank, above=1.5cm of Z_1] (blank_1){};
  \draw [black, ->, thick, -latex'] (blank_1) -- node [black,right, midway, font=\fontsize{7}{0}\selectfont] 
  {\rotatebox{-90}{$\begin{aligned}
  & \text{State}(1, 1) \\ 
\end{aligned}$}}
(Z_1);

\draw [black,->, thick, -latex'] (Z_1) -- node [black, right, midway, font=\fontsize{7}{0}\selectfont]
{\rotatebox{-90}{
$\begin{aligned}
 & \text{State}(1, 2) \\ 
\end{aligned}$}}
(vdots_1);

 \node[black, blank, left=1.5cm of Z_1] (blank_2){};
 \draw [black, ->, thick, -latex'] (blank_2) -- node [black,above, near start, font=\fontsize{7}{0}\selectfont] 
 {\rotatebox{-90}{$\begin{aligned}
  \mathbf{Z}_{1}^{(0)} = \mathbf{Y}_{1}
\end{aligned}$}}
(Z_1);
\node [black, right=1.5cm of Z_1] (ldots_1) {\ldots};
\draw [black, ->, thick, -latex'] (Z_1) -- node [black, above, near end, font=\fontsize{7}{0}\selectfont] {\rotatebox{-90}{$\begin{aligned}
\mathbf{Z}_{1}^{(1)}
\end{aligned}$}}
(ldots_1);

\node [black, block_reg, below=1.5cm of vdots_1] (Z_1t) 
{
\rotatebox{-90}{$\begin{aligned}
\text{Reg} &\text{Cell} \\
(1&, t)
\end{aligned}$}
  };

 \node [black, below=1.5cm of Z_1t] (vdots_1t) {\vdots};

\draw [black, ->, thick, -latex'] (Z_1t) -- node [black, right, midway, font=\fontsize{7}{0}\selectfont] {\rotatebox{-90}{
$\begin{aligned}
& \text{State}(1, t+1) \\ 
\end{aligned}$}}
(vdots_1t);

  \draw [black, ->, thick, -latex'] (vdots_1) -- node [black, right, midway, font=\fontsize{7}{0}\selectfont] {\rotatebox{-90}
  {$\begin{aligned}
  & \text{State}(1, t) \\ 
  \end{aligned}$}}
(Z_1t);

  \node[black, blank, above left=1.5cm of Z_1t] (blank_s1t){};
  \draw [black, ->, thick, -latex'] (blank_s1t) -- node [black, above=0.1cm, midway, font=\fontsize{7}{0}\selectfont] 
  {\rotatebox{-90}{$\begin{aligned}
  \mathbf{Z}_{{t-1}}^{{(0)}} = \mathbf{Y}_{{t-1}}
\end{aligned}$}}
(Z_1t.north);

  \node[black, blank, below right=1.5cm of Z_1t] (blank_s2t){};
  \draw [black, ->, thick, -latex'] (Z_1t.east) -- node [black, above=0.1cm, near end, font=\fontsize{7}{0}\selectfont] 
  {\rotatebox{-90}{$\begin{aligned}
  \mathbf{Z}_{{t}}^{{(1)}} 
  \end{aligned}$}}
(blank_s2t);

\node [black,right=1.8cm of vdots_1t] (vdots_s1) {\vdots};

\node[black,blank, left=1.5cm of Z_1t] (blank_1tl){};
 \draw [black, ->, thick, -latex'] (blank_1tl) -- node [black, above, near start, font=\fontsize{7}{0}\selectfont] {\rotatebox{-90}
 {$\begin{aligned}
  \mathbf{Z}_{t}^{(0)} = \mathbf{Y}_{t}
\end{aligned}$}}
(Z_1t);

  \node [black, right=1.5cm of Z_1t] (ldots_1t) {\ldots};
\draw [black, ->, thick, -latex'] (Z_1t) -- node [black, above, near end, font=\fontsize{7}{0}\selectfont] {\rotatebox{-90}{$\begin{aligned}
 \mathbf{Z}_{t}^{(1)}
\end{aligned}$}}
(ldots_1t);

\node [black, block_reg, below=1.5cm of vdots_1t] (Z_1T) 
{
\rotatebox{-90}{$\begin{aligned}
\text{Reg} &\text{Cell} \\
(1&, T)
\end{aligned}$}
  };

  \draw [black, ->, thick, -latex'] (vdots_1t) -- node [black, right, midway, font=\fontsize{7}{0}\selectfont] {\rotatebox{-90}
  {$\begin{aligned}
  & \text{State}(1, T) \\ 
\end{aligned}$}}
(Z_1T);

  \node[black,blank, below=1.5cm of Z_1T] (blank_d1){};
  \draw [black,->, thick, -latex'] (Z_1T) -- node [black,right, midway, font=\fontsize{7}{0}\selectfont] {\rotatebox{-90}
  {$\mathbf{Z}_{{T}}^{{(1)}}$}}
(blank_d1);

\node[black, blank, left=1.5cm of Z_1T] (blank_1Tl){};
 \draw [black, ->, thick, -latex'] (blank_1Tl) -- node [black, above, near start, font=\fontsize{7}{0}\selectfont] {\rotatebox{-90}
 {$\begin{aligned}
  \mathbf{Z}_{{T}}^{{(0)}}
 = \mathbf{Y}_{{T}}
\end{aligned}$}}
(Z_1T);

  \node[black, blank, above left=1.5cm of Z_1T] (blank_s1T){};
  \draw [black, ->, thick, -latex'] (blank_s1T) -- node [black, above=0.1cm, midway, font=\fontsize{7}{0}\selectfont] 
  {\rotatebox{-90}{$\begin{aligned}
  \mathbf{Z}_{{T-1}}^{{(0)}} = \mathbf{Y}_{{T-1}}
\end{aligned}$}}
(Z_1T.north);

  \node[black, blank, below left=1.5cm of Z_1T] (blank_s1T_2){};
  \draw [black, ->, thick, -latex'] (Z_1T.south) -- node [black, below=0.1cm, midway, font=\fontsize{7}{0}\selectfont] 
  {\rotatebox{-90}{$\begin{aligned}
  \mathbf{Z}_{{T}}^{{(1)}} \\ 
\end{aligned}$}}
(blank_s1T_2);

  \node [black, right=1.5cm of Z_1T] (ldots_1T) {\ldots};
\draw [black, ->, thick, -latex'] (Z_1T) -- node [black, above, near end, font=\fontsize{7}{0}\selectfont] {\rotatebox{-90}
{$\begin{aligned}
 \mathbf{Z}_{{T}}^{{(1)}}
\end{aligned}$}}
(ldots_1T);

  \node[black, blank,  left=1.8cm of blank_1] (blank_tl1){};
  \node[black, blank,  left=0.15cm of blank_s1T_2] (blank_tl2){};
  \draw [black, ->, thick, -latex'] (blank_tl1) -- node [black, left=0.1cm, midway, font=\fontsize{8}{0}\selectfont] 
  {\rotatebox{-90}{$\begin{aligned}
  \text{time}
  \end{aligned}$}}
(blank_tl2);

\node [black, block_reg, right=1.5cm of ldots_1] (Z_k0) 
{
\rotatebox{-90}{$\begin{aligned}
\text{Reg} &\text{Cell} \\
(k&, 1)
\end{aligned}$}
  };

 \node [black, below=1.5cm of Z_k0] (vdots_k0) {\vdots};

  \node[black, blank, above=1.5cm of Z_k0] (blank_k01){};
  \draw [black, ->, thick, -latex'] (blank_k01) -- node [black, right, midway, font=\fontsize{7}{0}\selectfont] {\rotatebox{-90}
  {$\begin{aligned}
  & \text{State}(k, 1) \\ 
  \end{aligned}$}}
(Z_k0);

\draw [black, ->, thick, -latex'] (Z_k0) -- node [black, right, midway, font=\fontsize{7}{0}\selectfont] {\rotatebox{-90}{
$\begin{aligned}
& \text{State}(k, 2) \\ 
\end{aligned}$}}
(vdots_k0);

  \node[black, blank, above left=1.5cm of Z_k0] (blank_s1t){};
  \draw [black, ->, thick, -latex'] (blank_s1t) -- node [black, above=0.1cm, midway, font=\fontsize{7}{0}\selectfont] 
  {\rotatebox{-90}{$\begin{aligned}
  \mathbf{Z}_{{0}}^{{(k-1)}}
\end{aligned}$}}
(Z_k0.north);

  \node[black, blank, below right=1.5cm of Z_k0] (blank_s2t){};
  \draw [black, ->, thick, -latex'] (Z_k0.east) -- node [black, above=0.1cm, near end, font=\fontsize{7}{0}\selectfont] 
  {\rotatebox{-90}{$\begin{aligned}
  \mathbf{Z}_{{1}}^{{(k)}} 
  \end{aligned}$}}
(blank_s2t);

\node [black, right=1.8cm of vdots_k0] (vdots_sk1) {\vdots};

 \draw [black, ->, thick, -latex'] (ldots_1) -- node [black, above, midway, font=\fontsize{7}{0}\selectfont]  {\rotatebox{-90}
 {$\begin{aligned}
  \mathbf{Z}_{{1}}^{{(k-1)}}
\end{aligned}$}}
(Z_k0);

\node [black, right=1.5cm of Z_k0] (ldots_k0) {\ldots};
\draw [black, ->, thick, -latex'] (Z_k0) -- node [black, above, near end, font=\fontsize{7}{0}\selectfont]  {\rotatebox{-90}
{$\begin{aligned}
\mathbf{Z}_{{1}}^{{(k)}}
\end{aligned}$}}
(ldots_k0);

\node [black, block_reg, below=1.5cm of vdots_k0] (Z_kt) 
{
\rotatebox{-90}{$\begin{aligned}
\text{Reg} &\text{Cell} \\
(k&, t)
\end{aligned}$}
  };

 \node [black, below=1.5cm of Z_kt] (vdots_kt) {\vdots};

  \draw [black, ->, thick, -latex'] (vdots_k0) -- node [black, right, midway, font=\fontsize{7}{0}\selectfont] {\rotatebox{-90}
  {$\begin{aligned}
  & \text{State}(k, t) \\ 
 \end{aligned}$}}
(Z_kt);

\draw [black, ->, thick, -latex'] (Z_kt) -- node [black, right, midway, font=\fontsize{7}{0}\selectfont] {\rotatebox{-90}
{$\begin{aligned}
& \text{State}(k, t+1) \\ 
\end{aligned}$}}
(vdots_kt);

  \node[black, blank, above left=1.5cm of Z_kt] (blank_s1t){};
  \draw [black, ->, thick, -latex'] (blank_s1t) -- node [black, above=0.1cm, midway, font=\fontsize{7}{0}\selectfont] 
  {\rotatebox{-90}{$\begin{aligned}
  \mathbf{Z}_{{t-1}}^{{(k-1)}}
\end{aligned}$}}
(Z_kt.north);

  \node[black, blank, below right=1.5cm of Z_kt] (blank_s2t){};
  \draw [black, ->, thick, -latex'] (Z_kt.east) -- node [black, above=0.1cm, near end, font=\fontsize{7}{0}\selectfont] 
  {\rotatebox{-90}{$\begin{aligned}
  \mathbf{Z}_{{t}}^{{(k)}} 
  \end{aligned}$}}
(blank_s2t);

\node [black, right=1.8cm of vdots_kt] (vdots_sk2) {\vdots};

 \draw [black, ->, thick, -latex'] (ldots_1t) -- node [black, above, midway, font=\fontsize{7}{0}\selectfont] {\rotatebox{-90}
 {$\begin{aligned}
  \mathbf{Z}_{{t}}^{{(k-1)}}
\end{aligned}$}}
(Z_kt);
  \node [black, right=1.5cm of Z_kt] (ldots_kt) {\ldots};
\draw [black, ->, thick, -latex'] (Z_kt) -- node [black, above, near end, font=\fontsize{7}{0}\selectfont] {\rotatebox{-90}
{$\begin{aligned}
 \mathbf{Z}_{{t}}^{{(k)}}
\end{aligned}$}}
(ldots_kt);

\node [black, block_reg, below=1.5cm of vdots_kt] (Z_kT) 
{
\rotatebox{-90}{$\begin{aligned}
\text{Reg} &\text{Cell} \\
(k&, T)
\end{aligned}$}
  };

  \draw [black, ->, thick, -latex'] (vdots_kt) -- node [black, right, midway, font=\fontsize{7}{0}\selectfont] {\rotatebox{-90}
  {$\begin{aligned} 
  & \text{State}(k, T) \\ 
  \end{aligned}$}}
(Z_kT);

\node[black, blank, below=1.5cm of Z_kT] (blank_Z_kT_below){};
\draw [black, ->, thick, -latex'] (Z_kT) -- node [black, right, midway, font=\fontsize{7}{0}\selectfont] {\rotatebox{-90}
  {$\begin{aligned}
     \mathbf{Z}_{{T}}^{{(k)}} \\ 
     \end{aligned}$}}
(blank_Z_kT_below);

  \node[black, blank, above left=1.5cm of Z_kT] (blank_s1T){};
  \draw [black, ->, thick, -latex'] (blank_s1T) -- node [black, above=0.1cm, midway, font=\fontsize{7}{0}\selectfont] 
  {\rotatebox{-90}{$\begin{aligned}
  \mathbf{Z}_{{T-1}}^{{(k-1)}}
\end{aligned}$}}
(Z_kT.north);

  \node[black, blank, below left=1.5cm of Z_kT] (blank_skT_2){};
  \draw [black, ->, thick, -latex'] (Z_kT.south) -- node [black, below=0.1cm, midway, font=\fontsize{7}{0}\selectfont] 
  {\rotatebox{-90}{$\begin{aligned}
  \mathbf{Z}_{{T}}^{{(k)}} \\ 
\end{aligned}$}}
(blank_skT_2);

 \draw [black, ->, thick, -latex'] (ldots_1T) -- node [black, above, midway, font=\fontsize{7}{0}\selectfont] {\rotatebox{-90}
 {$\begin{aligned}
  \mathbf{Z}_{{T}}^{{(k-1)}}
\end{aligned}$}}
(Z_kT);

  \node [black, right=1.5cm of Z_kT] (ldots_kT) {\ldots};
\draw [black, ->, thick, -latex'] (Z_kT) -- node [black, above, near end, font=\fontsize{7}{0}\selectfont] {\rotatebox{-90}
{$\begin{aligned}
 \mathbf{Z}_{{T}}^{{(k)}}
\end{aligned}$}}
(ldots_kT);

\node [black, block_reg, right=1.5cm of ldots_k0] (Z_K0) 
{
\rotatebox{-90}{$\begin{aligned}
\text{Reg} &\text{Cell} \\
(K&, 1)
\end{aligned}$}
  };

\node [black, below=1.5cm of Z_K0] (vdots_K0) {\vdots};

  \node[black, blank, above left=2.6cm of Z_1] (blank_dl1){};
  \node[black, blank, above right=2.6cm of Z_K0] (blank_dl2){};
  \draw [black, ->, thick, -latex'] (blank_dl1) -- node [black, above=0.1cm, midway, font=\fontsize{8}{0}\selectfont] 
  {\rotatebox{-90}{$\begin{aligned}
  \text{depth}
  \end{aligned}$}}
(blank_dl2);

    \node[black, blank, above=1.5cm of Z_K0] (blank_K01){};
  \draw [black, ->, thick, -latex'] (blank_K01) -- node [black, right, midway, font=\fontsize{7}{0}\selectfont] {\rotatebox{-90}
  {$\begin{aligned}
  & \text{State}(K, 1) \\ 
   \end{aligned}$}}
(Z_K0);

 \draw [black, ->, thick, -latex'] (Z_K0) -- node [black, right, midway, font=\fontsize{7}{0}\selectfont] {\rotatebox{-90}
 {$\begin{aligned}
& \text{State}(K, 2) \\ 
\end{aligned}$}}
(vdots_K0);

  \node[black, blank, above left=1.5cm of Z_K0] (blank_s1t){};
  \draw [black, ->, thick, -latex'] (blank_s1t) -- node [black, above=0.1cm, midway, font=\fontsize{7}{0}\selectfont] 
  {\rotatebox{-90}{$\begin{aligned}
  \mathbf{Z}_{{0}}^{{(K-1)}}
\end{aligned}$}}
(Z_K0.north);

 \draw [black, ->, thick, -latex'] (ldots_k0) -- node [black, above, midway, font=\fontsize{7}{0}\selectfont] {\rotatebox{-90}
 {$\begin{aligned}
 \mathbf{Z}_{{1}}^{{(K-1)}}
\end{aligned}$}}
(Z_K0);
\node[black, blank, right=1.5cm of Z_K0] (blank_K0r){};
\draw [black, ->, thick, -latex'] (Z_K0) -- node [black, above, near end, font=\fontsize{7}{0}\selectfont] {\rotatebox{-90}
{$\begin{aligned}
\mathbf{Z}_{{1}}^{{(K)}}
\end{aligned}$}}
(blank_K0r);

\node [black, block_reg, below=1.5cm of vdots_K0] (Z_Kt) 
{
\rotatebox{-90}{$\begin{aligned}
\text{Reg} &\text{Cell} \\
(K&, t)
\end{aligned}$}
  };

 \node [black, below=1.5cm of Z_Kt] (vdots_Kt) {\vdots};

\draw [black, ->, thick, -latex'] (Z_Kt) -- node [black, right, midway, font=\fontsize{7}{0}\selectfont] {\rotatebox{-90}
{$\begin{aligned}
& \text{State}(K, t+1) \\ 
\end{aligned}$}}
(vdots_Kt);

\draw [black, ->, thick, -latex'] (vdots_K0) -- node [black, right, midway, font=\fontsize{7}{0}\selectfont] {\rotatebox{-90}
  {$\begin{aligned}
   & \text{State}(K, t) \\ 
   \end{aligned}$}}
(Z_Kt);

  \node[black, blank, above left=1.5cm of Z_Kt] (blank_sKT){};
  \draw [black, ->, thick, -latex'] (blank_sKT) -- node [black, above=0.1cm, midway, font=\fontsize{7}{0}\selectfont] 
  {\rotatebox{-90}{$\begin{aligned}
  \mathbf{Z}_{{t-1}}^{{(K-1)}}
\end{aligned}$}}
(Z_Kt.north);

 \draw [black, ->, thick, -latex'] (ldots_kt) -- node [black, above, midway, font=\fontsize{7}{0}\selectfont] {\rotatebox{-90}
 {$\begin{aligned}
  \mathbf{Z}_{{t}}^{{(K-1)}}
\end{aligned}$}}
(Z_Kt);
  \node[black, blank, right=1.5cm of Z_Kt] (blank_Kt1){};
\draw [black, ->, thick, -latex'] (Z_Kt) -- node [black, above, near end, font=\fontsize{7}{0}\selectfont] {\rotatebox{-90}
{$\begin{aligned}
 \mathbf{Z}_{{t}}^{{(K)}}
\end{aligned}$}}
(blank_Kt1);

\node [black, block_reg, below=1.5cm of vdots_Kt] (Z_KT) 
{
\rotatebox{-90}{$\begin{aligned}
\text{Reg} &\text{Cell} \\
(K&, T)
\end{aligned}$}
  };

  \draw [black, ->, thick, -latex'] (vdots_Kt) -- node [black, right, midway, font=\fontsize{7}{0}\selectfont] {\rotatebox{-90}
  {$\begin{aligned}
     & \text{State}(K, T) \\ 
     \end{aligned}$}}
(Z_KT);

\node[black, blank, below=1.5cm of Z_KT] (blank_Z_KT_below){};
\draw [black, ->, thick, -latex'] (Z_KT) -- node [black, right, midway, font=\fontsize{7}{0}\selectfont] {\rotatebox{-90}
  {$\begin{aligned}
     \mathbf{Z}_{{T}}^{{(K)}} \\ 
     \end{aligned}$}}
(blank_Z_KT_below);

  \node[black, blank, above left=1.5cm of Z_KT] (blank_sKT){};
  \draw [black, ->, thick, -latex'] (blank_sKT) -- node [black, above=0.1cm, midway, font=\fontsize{7}{0}\selectfont] 
  {\rotatebox{-90}{$\begin{aligned}
  \mathbf{Z}_{{T-1}}^{{(K-1)}}
\end{aligned}$}}
(Z_KT.north);

  \node[black, blank, below left=1.5cm of Z_KT] (blank_sKT_2){};
  \draw [black, ->, thick, -latex'] (Z_KT.south) -- node [black,below=0.1cm, midway, font=\fontsize{7}{0}\selectfont] 
  {\rotatebox{-90}{$\begin{aligned}
  \mathbf{Z}_{{T}}^{{(K)}}
\end{aligned}$}}
(blank_sKT_2);

 \draw [black, ->, thick, -latex'] (ldots_kT) -- node [black,above, midway, font=\fontsize{7}{0}\selectfont] {\rotatebox{-90}
 {$\begin{aligned}
  \mathbf{Z}_{{T}}^{{(K-1)}}
  \end{aligned}$}}
(Z_KT);

\node[black,blank, right=1.5cm of Z_KT] (blank_Z_KT){};
\draw [black,->, thick, -latex'] (Z_KT) -- node [black,above, near end, font=\fontsize{7}{0}\selectfont] {\rotatebox{-90}
{$\begin{aligned}
 \mathbf{Z}_{{T}}^{{(K)}}
\end{aligned}$}}
(blank_Z_KT);

\end{tikzpicture}
\caption{The structure of RegNet. $\text{State}(k,t)= \mathbf{Z}_{{t-1}}^{{(k)}}, \hat{\mathbb{E}}(\bm{\epsilon}_{{t-1}}^{{(k)}}),  \hat{cov}(\bm{\epsilon}_{{t-1}}^{{(k)}}) $.}
\label{fig:RegNet}
\end{figure}

In a RegNet (Fig. \ref{fig:RegNet}), the RegCell is replicated $T$ times along the time axis at each time step from $t = 1$ to $T$, which allows for an iterative regression of the coefficients of an OU process modeling the univariate input time series $Y_{1}, \cdots , Y_{T}$.
These $T$ RegCells form the first layer $k = 1$ of the RegNet. The outputs of layer $k = 1$ are the regressed coefficients $\mathbf A_t$, $\mathbf N_{t}$, $\bm\Sigma_t$ ($1$-dimensional for each) for the time steps $t = 1,\cdots,T$. We append and flatten these coefficients into a $3$-dimensional vector of coefficients denoted $\mathbf{Z}^{(1)}_{t} = \big[\mathbf A_t^{(1)}, \mathbf N^{(1)}_{t}, \bm\Sigma^{(1)}_t \big]$. The output of layer $k = 1$ at $t$ is now a multivariate time series $\mathbf{Z}^{(1)}_{t}$ of dimension $3$.

The dynamics of $\mathbf{Z}^{(1)}_{t}$ for the time steps $t = 1,\cdots,T$ can be analyzed in the same way as $Y_t$ . Thus, we treat $\mathbf{Z}^{(1)}_{t}$ as a multivariate OU process and regress its coefficients in an online fashion using a second layer composed of $T$ RegCells. Several layers can then be stacked on top of each other as shown in Fig. \ref{fig:RegNet}, allowing each layer $k$ to regress the vector of parameters $\Delta \mathbf{Z}^{(k)}_{t} = \big[\mathbf A^{(k)}_t, \mathbf N^{(k)}_{t}, \bm\Sigma^{(k)}_t\big]$ of an OU process $\Delta \mathbf{Z}^{(k-1)}_{t} = \mathbf{A}_{t-1}^{(k)} \mathbf{Z}^{(k-1)}_{t-1} + \mathbf{N}^{(k)}_{t-1} + \bm{\Sigma}^{(k)}_{t-1} \Delta \mathbf{W}^{(k)}_{t}$, modeling the dynamics $\Delta \mathbf{Z}^{(k-1)}_{t} = \mathbf{Z}^{(k-1)}_{t} - \mathbf{Z}^{(k-1)}_{t-1}$ of the time series output $\mathbf{Z}^{(k-1)}_{t}$ by the previous layer $k-1$. Since we denote generally $\mathbf{Z}^{(k)}_{t}$ the output of 
a RegCell at time step $t$ in layer $k$ for any integer $k \geq 0$, we adopt the convention ${Z}^{(0)}_{t} = Y_{t}$. Observe that each layer k increases the dimensionality of the vector of coefficients from $d_{k-1}$ to $d_k = 2{d_{k - 1}}^{2} + d_{k - 1}$. The sequence of dimensions $(d_{k}) = 1, 3, 21, 903, 1631721, \cdots$ quickly diverges towards infinity, thus limiting in practice RegNet to a maximum of 2 or 3 layers.

The motivation for using multiple layers in the RegNet is that it allows us to extract more information from the time series $Y_{t}$. Since this information is passed on to the PredNet, the resulting RegPred Net can potentially yield better forecasts in the long run. This can be best understood perhaps by analogy with a function $f$ defined on a time interval $[0,T]$ that we would like to extrapolate to $[T, +\infty]$. If the function is $2$ times continuously differentiable on $[0,T]$, we could extrapolate it for $t > T$ with the Taylor series expansion of order $2$.
The higher the order of the Taylor series, the more accurate the extrapolation becomes. Similarly, the higher the number of layers used the RegPred Net, the better the forecasts may get, as primarily, each layer $k$ is modeling the $k$-th discrete derivative of $Y$ with respect to $t$. 

\subsection{Prediction Cell (PredCell) in Prediction Network (PredNet)} \label{sec:PredCell}

For prediction, we need another type of cell that uses the information extracted by the RegCell, which we call PredCell. Assume $K$ is the number of total layers of the RegNet, $T$ is the last time step of an input time series $\mathbf{Y}$, and at time $T$ RegNet outputs $\mathbf{Z}_T^{1:K}$. PredNet starts making predictions of the multivariate process $\mathbf{Z}^{(k)}_{t}$ at the last layer $k=K$ and ends making predictions for the process at the first layer $k=0$ where the process $\mathbf{Z}^{(0)}_{t}=\mathbf{Y}_t$. In the last layer $k=K$, since the process to forecast is not stochastically modeled, we assume that the outputs of the PredCell $\mathbf{Z}_ {{T}}^{{(K)}}$ is constant for any time step $T+i$, $i>0$:  
\begin{equation} \label{identity_layer}
\mathbf{Z}_ {{T+i}}^{{(K)}} \leftarrow \mathbf{Z}_ {{T}}^{{(K)}}
\end{equation}
The PredCells in all other layers where $k<K$ follow the update rule:
\begin{equation} \label{decode_update}
\mathbf{Z}_{{t}}^{{(k)}} \leftarrow \mathbf{Z}_{{t-1}}^{{(k)}} + \mathbf{A}_{{t-1}}^{{(k+1)}} \mathbf{Z}_{{t-1}}^{{(k)}} + \mathbf{N}_{{t-1}}^{{(k+1)}} + \bm{\Sigma}^{{(k+1)}}_{{t-1}} \Delta \mathbf{W}^{{(k)}}_{{t}}
\end{equation} 
which simply randomly generates a new value for the process using its previous value and the equation for the increment of an OU process.

\tikzstyle{block} = [draw, rectangle, fill=cyan!15, text width=18 em, 
                             text centered, rounded corners, minimum height=16mm,
                             font=\fontsize{8}{0}\selectfont, node distance=5em]
\tikzstyle{blank} = [ fill=white, text width=1em, text centered, minimum height=1mm, font=\fontsize{8}{0}\selectfont, node distance=6em]

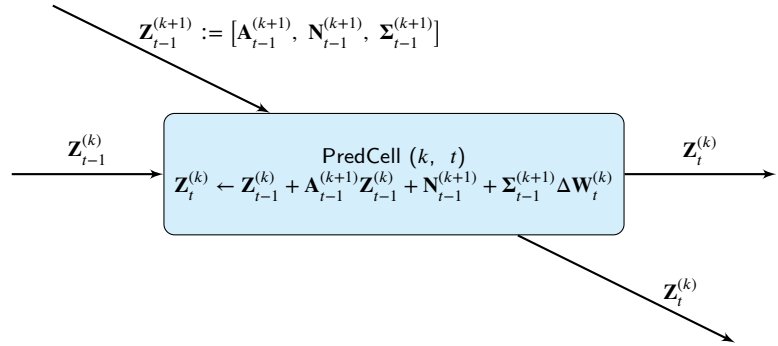
\begin{figure}[pos=htbp, width=6.cm, align=\centering]
\centering
\begin{tikzpicture}

\node [black, block] (D_1) 
{
PredCell ($k$, \ $t$) \\
 $\begin{aligned}
 \mathbf{Z}_{t}^{(k)} \leftarrow \mathbf{Z}_{t-1}^{(k)} + \mathbf{A}_{t-1}^{(k+1)}\mathbf{Z}_{t-1}^{(k)} + \mathbf{N}_{t-1}^{(k+1)} + \bm{\Sigma}^{{(k+1)}}_{{t-1}} \Delta \mathbf{W}^{{(k)}}_{{t}}
\end{aligned}$ 
};

\node[black,blank, left=2cm of D_1] (blank_1){};
 \draw [black,->, thick, -latex'] (blank_1) -- node [black,above, font=\fontsize{8}{0}\selectfont]
 {
 $\begin{aligned}
\mathbf{Z}_{{t-1}}^{{(k)}}
\end{aligned}$
 }
 (D_1);

 \node[black,blank, right=2cm of D_1] (blank_2){};
 \draw [black,->, thick, -latex'] (D_1) -- node [black,above, font=\fontsize{8}{0}\selectfont] 
 {
 $\begin{aligned}
 \mathbf{Z}_{{t}}^{{(k)}}
\end{aligned}$
 }
 (blank_2);

\node[black,blank, below right=2cm of D_1] (blank_3){};
\node[black,blank,  below=0cm of D_1] (blank_3_1){};
 \draw [black,->, thick, -latex'] (D_1) -- node [black,above, near end, font=\fontsize{8}{0}\selectfont] 
  {
 $\begin{aligned}
 \mathbf{Z}_{{t}}^{{(k)}}
\end{aligned}$
 }
 (blank_3);

\node[black,blank, above left=2cm of D_1] (blank_4){};
 \draw [black,->, thick, -latex'] (blank_4) -- node [black,near start, right=0.3cm, font=\fontsize{8}{0}\selectfont] 
 {
 $\begin{aligned}
 \mathbf{Z}_{{t-1}}^{{(k+1)}} \coloneqq \big[ \mathbf{A}_{{t-1}}^{{(k+1)}}, \ \mathbf{N}_{{t-1}}^{{(k+1)}}, \ \bm{\Sigma}_{{t-1}}^{{(k+1)}} \big]
\end{aligned}$
 }
 (D_1);
\end{tikzpicture}
\caption{The PredCell ($k$, $t$).}
\label{fig:PredCell}
\end{figure}

Fig. \ref{fig:PredCell} illustrates the function of a PredCell in layer $k$ at time $t$ and {Algorithm} \ref{alg:PredCell} shows how to implement such a cell.
In the calculation of the output $\mathbf{Z}^{(k)}_{t}$ of PredCell$(k,t)$, the term $\Delta \mathbf{W}^{(k)}_{t}$ is a randomly generated vector of the same dimension as $\mathbf{Z}^{(k)}_{t}$, which we denote by $d_{k}$. Each component of this vector follows an independent standard normal distribution. It is called a \textit{random factor} of $\mathbf{Z}_{t}$ and can be seen as the source of randomness of the process. So remember that $\Delta \mathbf{W}^{(k)}_{t}$ has dimension $d_k$ and its components are i.i.d. with distribution $\mathcal{N}(0, 1)$. Thus, the PredCells generate random outputs. In that sense, the PredNet constitutes a generative network. This is why the PredNet can only be used for simulation of trajectories and not directly for prediction. The diagonal arrow above the cell represents the input $\mathbf{Z}_{{t-1}}^{{(k+1)}}$, which is from layer $k+1$ and time $t-1$. $\mathbf{Z}_{{t-1}}^{{(k+1)}}$ contains the parameters $\mathbf{A}_{{t-1}}^{{(k+1)}}$, $\mathbf{N}_{{t-1}}^{{(k+1)}}$, $\bm{\Sigma}_{{t-1}}^{{(k+1)}}$ that are needed in Eq. (\ref{decode_update}). The left horizontal arrow indicates the input $\mathbf{Z}_{{t-1}}^{{(k)}}$ from layer $k$, time $t-1$. Notice here at the initial prediction step $T+1$, $\mathbf{Z}_{{t-1}}^{{(k)}}$ is the outputs of RegCell in layer $k$ at step $T$. The diagonal arrow below the cell indicates the output $\mathbf{Z}_{{t}}^{{(k)}}$. This will be divided into $\mathbf{A}_{{t}}^{{(k)}}$, $\mathbf{N}_{{t}}^{{(k)}}$, $\bm{\Sigma}_{{t}}^{{(k)}}$, and be used for the PredCell of layer $k-1$, time $t+1$. $\mathbf{Z}_{{t}}^{{(k)}}$ is also transferred to the next time step for PredCell $(k, t+1)$ at the right horizontal arrow. The multi-layer PredNet is illustrated in Fig. \ref{fig:PredNet}. 

\begin{algorithm}[!htbp]
\caption{PredCell in layer $k$ at time $t$}
\label{alg:PredCell}
\begin{algorithmic}[1]
\renewcommand{\algorithmicrequire}{\textbf{Input:}}
\renewcommand{\algorithmicensure}{\textbf{Output:}}

\Require $\mathbf{Z}^{{(k+1)}} _{{t-1}}$, $\mathbf{Z}^{{(k)}} _{{t-1}}$, $\Delta \mathbf{W}^{{(k)}} _{{t}}$
\Ensure $\mathbf{Z}^{{(k)}}_{{t}}$

\If{layer $k$ is the last layer}

        \State $\mathbf{Z}^{{(k)}}_{{t}} \leftarrow \mathbf{Z}^{{(k)}}_{{t-1}}$

\Else
\State $\mathbf{Z}_{{t}}^{{(k)}} \leftarrow \mathbf{Z}_{{t-1}}^{{(k)}} + \mathbf{A}_{{t-1}}^{{(k+1)}} \mathbf{Z}_{{t-1}}^{{(k)}} + \mathbf{N}_{{t-1}}^{{(k+1)}} + \bm{\Sigma}^{{(k+1)}}_{{t-1}} \Delta \mathbf{W}^{{(k)}}_{{t}}$
\EndIf

\end{algorithmic}
\end{algorithm}

\tikzstyle{block_pred} = [draw, 
                             rectangle, 
                             fill=cyan!15, 
                             text width=1.5em, 
                             text centered, 
                             minimum height={width("(K-1, T+1)")+3pt},
                             font=\fontsize{6}{0}\selectfont, 
                             node distance=5em]
\tikzstyle{blank} = [fill=white, text width=1em, text centered, minimum height=1mm, font=\fontsize{6}{0}\selectfont, node distance=6em]
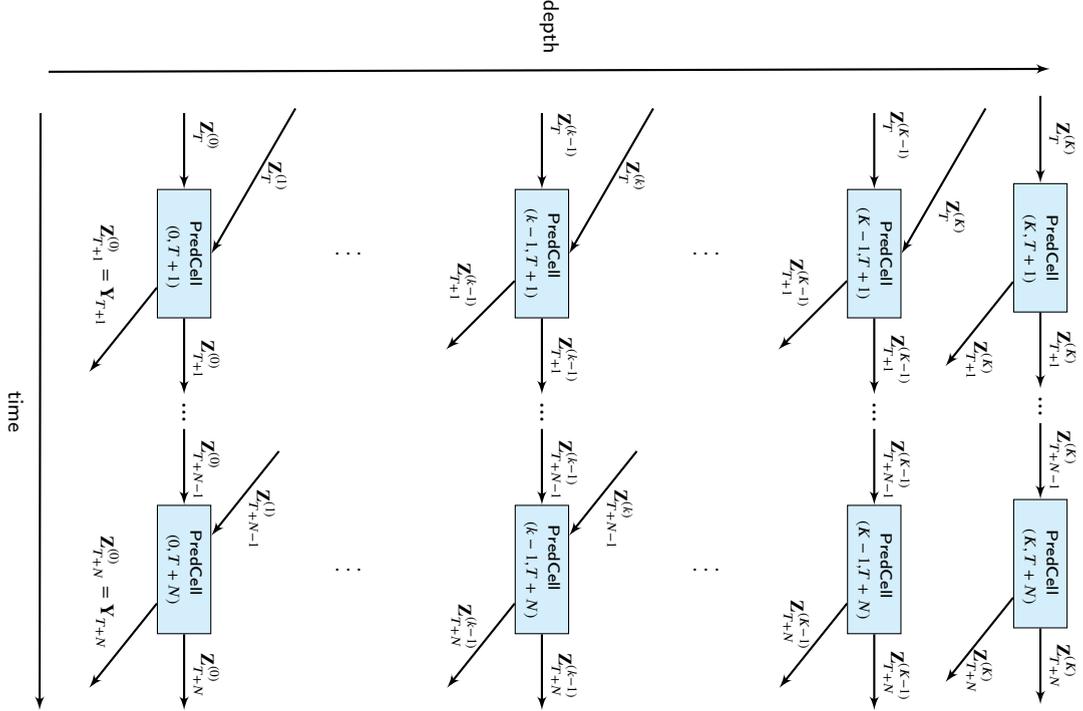
\begin{figure}[pos=htbp, width=5cm, align=\centering] 
\centering
\begin{tikzpicture} 

\node [black, block_pred, below=2.cm of ldots_1T] (Z_kT1) 
{
\rotatebox{-90}{$\begin{aligned}
\text{Pred} &\text{Cell} \\
(k-1&, T+1)
\end{aligned}$}
};

\node [black,below=1.cm of Z_kT1] (vdots_kT1) {\vdots};

\draw [black,->, thick, -latex'] (Z_kT1) -- node [black,right, midway, font=\fontsize{7}{0}\selectfont] {\rotatebox{-90}{
$\begin{aligned}
&\mathbf{Z}_{{T+1}}^{{(k-1)}} \\
\end{aligned}$}}
(vdots_kT1);

  \node [black,above=1.cm of Z_kT1] (blank_kT) {};
\draw [black,->, thick, -latex'] (blank_kT) -- node [black,right, near start, font=\fontsize{7}{0}\selectfont] {\rotatebox{-90}{
$\begin{aligned}
&\mathbf{Z}_{{T}}^{{(k-1)}} \\
\end{aligned}$}}
(Z_kT1);

 \node[black,blank, below left of=Z_kT1] (blank_10){};
 \draw [black,->, thick, -latex', rotate=45] (Z_kT1) -- node [black,above=0.2cm, near end, font=\fontsize{7}{0}\selectfont] {\rotatebox{-90}
 {$\begin{aligned}
  \mathbf{Z}_{{T+1}}^{{(k-1)}}
  \end{aligned}$}}
(blank_10);

  \node [black, right=1.5cm of Z_kT1] (ldots_kT1) {\ldots};
  \node [black,above right=1.5cm of Z_kT1] (blank_20){};
\draw [black,->, thick, -latex'] (blank_20) -- node [black,below=0.1cm, near start, font=\fontsize{7}{0}\selectfont] {\rotatebox{-90}{$\begin{aligned}
 \mathbf{Z}_{{T}}^{{(k)}}
\end{aligned}$}}
(Z_kT1.east);

\node [black,block_pred, below=1.cm of vdots_kT1] (Z_kTN) 
{
\rotatebox{-90}{$\begin{aligned}
\text{Pred} &\text{Cell} \\
(k-1&, T+N)
\end{aligned}$}
  };

 \node[black,blank, below=1.cm of Z_kTN] (blank_kTN1){};
\draw [black,->, thick, -latex'] (Z_kTN) -- node [black,right, midway, font=\fontsize{7}{0}\selectfont] {\rotatebox{-90}{
$\begin{aligned}
&\mathbf{Z}_{{T+N}}^{{(k-1)}} \\
\end{aligned}$}}
(blank_kTN1);

\draw [black,->, thick, -latex'] (vdots_kT1) -- node [black,right, midway, font=\fontsize{7}{0}\selectfont] {\rotatebox{-90}{
$\begin{aligned}
&\mathbf{Z}_{{T+N-1}}^{{(k-1)}} \\
\end{aligned}$}}
(Z_kTN);

 \node[black,blank, below left=1.cm of Z_kTN] (blank_kTN0){};
 \draw [black,->, thick, -latex'] (Z_kTN) -- node [black,above=0.1cm, near end, font=\fontsize{7}{0}\selectfont]  {\rotatebox{-90}
 {$\begin{aligned}
  \mathbf{Z}_{{T+N}}^{{(k-1)}}
\end{aligned}$}}
(blank_kTN0);

  \node[black,blank, above right=1.cm of Z_kTN] (blank_kTNr){};
  \node [black,right=1.5cm of Z_kTN] (ldots_kTN) {\ldots};
\draw [black,->, thick, -latex'] (blank_kTNr) -- node [black,below=0.1cm, near start, font=\fontsize{7}{0}\selectfont]  {\rotatebox{-90}{$\begin{aligned}
 \mathbf{Z}_{{T+N-1}}^{{(k)}}
\end{aligned}$}}
(Z_kTN);

\node [black,block_pred, left=4.cm of Z_kT1] (Z_1T1) 
{
\rotatebox{-90}{$\begin{aligned}
\text{Pred}&\text{Cell} \\
 \ \quad (0, T&+1) \quad
\end{aligned}$}
};

 \node [black,below=1.cm of Z_1T1] (vdots_1T1) {\vdots};

\draw [black,->, thick, -latex'] (Z_1T1) -- node [black,right, midway, font=\fontsize{7}{0}\selectfont] {\rotatebox{-90}{
$\begin{aligned}
\mathbf{Z}_{{T+1}}^{{(0)}}
\end{aligned}$}}
(vdots_1T1);

 \node [black,above=1.cm of Z_1T1] (blank_1T1) {};
\draw [black,->, thick, -latex'] (blank_1T1) -- node [black,right, near start, font=\fontsize{7}{0}\selectfont]{\rotatebox{-90}{
$\begin{aligned}
\mathbf{Z}_{{T}}^{{(0)}}
\end{aligned}$}}
(Z_1T1);

\node[black,blank, below left=1.cm of Z_1T1] (blank_1T1l){};
 \draw [black,->, thick, -latex'] (Z_1T1) -- node [black,above=0.2cm, near end, font=\fontsize{7}{0}\selectfont] {\rotatebox{-90}
 {$\begin{aligned}
  \mathbf{Z}_{{T+1}}^{{(0)}} = \mathbf{Y}_{{T+1}}
\end{aligned}$}}
(blank_1T1l);

  \node [black,right=1.5cm of Z_1T1] (ldots_1T1) {\ldots};
  \node [black,above right=1.5cm of Z_1T1] (blank_30){};
\draw [black,->, thick, -latex'] (blank_30) -- node [black,below=0.1cm, near start, font=\fontsize{7}{0}\selectfont] {\rotatebox{-90}{$\begin{aligned}
 \mathbf{Z}_{{T}}^{{(1)}}
\end{aligned}$}}
(Z_1T1.east);

\node [black,block_pred, below=1.cm of vdots_1T1] (Z_1TN) 
{
\rotatebox{-90}{$\begin{aligned}
\text{Pred}&\text{Cell} \\
 \quad (0, T&+N) \quad
\end{aligned}$}
  };

 \node[black,blank, below=1.cm of Z_1TN] (blank_1TN1){};
\draw [black,->, thick, -latex'] (Z_1TN) -- node [black,right, midway, font=\fontsize{7}{0}\selectfont]{\rotatebox{-90}{
$\begin{aligned}
\mathbf{Z}_{{T+N}}^{{(0)}}
\end{aligned}$}}
(blank_1TN1);

  \node[black,blank,  left=1.5cm of blank_1T1] (blank_tl1){};
  \node[black,blank,  left=1.35cm of blank_1TN1] (blank_tl2){};
  \draw [black,->, thick, -latex'] (blank_tl1) -- node [black,left=0.1cm, midway, font=\fontsize{8}{0}\selectfont] 
  {\rotatebox{-90}{$\begin{aligned}
  \text{time}
  \end{aligned}$}}
(blank_tl2);

\draw [black,->, thick, -latex'] (vdots_1T1) -- node [black,right, midway, font=\fontsize{7}{0}\selectfont]{\rotatebox{-90}{
$\begin{aligned}
\mathbf{Z}_{{T+N-1}}^{{(0)}}
\end{aligned}$}}
(Z_1TN);

\node[black,blank, below left=1.cm of Z_1TN] (blank_1TNl){};
 \draw [black,->, thick, -latex'] (Z_1TN) -- node [black,above=0.1cm, near end, font=\fontsize{7}{0}\selectfont] {\rotatebox{-90}
 {$\begin{aligned}
  \mathbf{Z}_{{T+N}}^{{(0)}} = \mathbf{Y}_{{T+N}}
\end{aligned}$}}
(blank_1TNl);

  \node [black,right=1.5cm of Z_1TN] (ldots_1TN) {\ldots};
  \node[black,blank, above right=1.cm of Z_1TN] (blank_1TN0){};
\draw [black,->, thick, -latex'] (blank_1TN0) -- node [black,below=0.1cm, near start, font=\fontsize{7}{0}\selectfont] {\rotatebox{-90}{$\begin{aligned}
 \mathbf{Z}_{{T+N-1}}^{{(1)}} 
\end{aligned}$}}
(Z_1TN);

\node [black,block_pred, below=1.48cm of Z_KT] (Z_IKT1) 
{
\rotatebox{-90}{$\begin{aligned}
\ \text{Pred}&\text{Cell} \ \\
 (K, T&+1)
\end{aligned}$}
  };

\node [black,above=1.15cm of Z_IKT1] (blank_40){};
\draw [black,->, thick, -latex'] (blank_40) -- node [black,right, midway, font=\fontsize{7}{0}\selectfont] {\rotatebox{-90}
{$\begin{aligned}
 \mathbf{Z}_{{T}}^{{(K)}}
\end{aligned}$}}
(Z_IKT1);

\node [black,below=1.cm of Z_IKT1] (vdots_IKT1) {\vdots};

\draw [black,->, thick, -latex'] (Z_IKT1) -- node [black,right, midway, font=\fontsize{7}{0}\selectfont] {\rotatebox{-90}
{$\begin{aligned}
 \mathbf{Z}_{{T+1}}^{{(K)}}
\end{aligned}$}}
(vdots_IKT1);

\node[black,blank, below left=1cm of Z_IKT1] (blank_Z_IKT){};
\draw [black,->, thick, -latex'] (Z_IKT1) -- node [black,below=0.1cm,  midway, font=\fontsize{7}{0}\selectfont] {\rotatebox{-90}
{$\begin{aligned}
 \mathbf{Z}_{{T+1}}^{{(K)}}
\end{aligned}$}}
(blank_Z_IKT);

\node [black,block_pred, below=1.0cm of vdots_IKT1] (Z_IKTN) 
{
\rotatebox{-90}{$\begin{aligned}
\ \text{Pred}&\text{Cell} \ \\
(K, T&+N) 
\end{aligned}$}
};

\draw [black,->, thick, -latex'] (vdots_IKT1.south) -- node [black,right,  midway, font=\fontsize{7}{0}\selectfont] {\rotatebox{-90}
{$\begin{aligned}
 \mathbf{Z}_{{T+N-1}}^{{(K)}}
\end{aligned}$}}
(Z_IKTN);

\node[black,blank, below=1.cm of Z_IKTN] (blank_Z_IKTN){};
\draw [black,->, thick, -latex'] (Z_IKTN.south) -- node [black,right,  midway, font=\fontsize{7}{0}\selectfont] {\rotatebox{-90}
{$\begin{aligned}
 \mathbf{Z}_{{T+N}}^{{(K)}}
\end{aligned}$}}
(blank_Z_IKTN);

\node[black,blank, below left =1.cm of Z_IKTN] (blank_Z_IKTNl){};
\draw [black,->, thick, -latex'] (Z_IKTN) -- node [black,below,  midway, font=\fontsize{7}{0}\selectfont] {\rotatebox{-90}
{$\begin{aligned}
 \mathbf{Z}_{{T+N}}^{{(K)}}
\end{aligned}$}}
(blank_Z_IKTNl);

\node [black,block_pred, below=2.cm of ldots_kT] (Z_KT1) 
{
\rotatebox{-90}{$\begin{aligned}
\text{Pred}&\text{Cell} \\
\ (K-1, &T+1) \
\end{aligned}$}
  };

 \node [black,below=1.cm of Z_KT1] (vdots_KT1) {\vdots};

\draw [black,->, thick, -latex'] (Z_KT1) -- node [black,right, midway, font=\fontsize{7}{0}\selectfont] {\rotatebox{-90}{
$\begin{aligned}
&\mathbf{Z}_{{T+1}}^{{(K-1)}} \\
\end{aligned}$}}
(vdots_KT1);

   \node[black,blank, above=1.cm of Z_KT1] (blank_KT1){};
\draw [black,->, thick, -latex'] (blank_KT1) -- node [black,right, near start, font=\fontsize{7}{0}\selectfont] {\rotatebox{-90}{
$\begin{aligned}
&\mathbf{Z}_{{T}}^{{(K-1)}} \\
\end{aligned}$}}
(Z_KT1);

 \node[black,blank, below left of=Z_KT1] (blank_vdots_kT1){};
 \draw [black,->, thick, -latex'] (Z_KT1) -- node [black,above=0.2cm, near end, font=\fontsize{7}{0}\selectfont] {\rotatebox{-90}
 {$\begin{aligned}
  \mathbf{Z}_{{T+1}}^{{(K-1)}}
\end{aligned}$}}
(blank_vdots_kT1);

\node [black,above right=1.5cm of Z_KT1] (blank_50){};
\draw [black,->, thick, -latex'] (blank_50) -- node [black,right=0.1cm, near end , font=\fontsize{7}{0}\selectfont] {\rotatebox{-90}
{$\begin{aligned}
 \mathbf{Z}_{T}^{(K)}
\end{aligned}$}}
(Z_KT1.east);

\node [black,block_pred, below=1.cm of vdots_KT1] (Z_KTN) 
{
\rotatebox{-90}{$\begin{aligned}
\text{Pred}&\text{Cell} \\
  (K-1, &T+N) 
\end{aligned}$}
  };

 \node[black,blank, below=1.cm of Z_KTN] (blank_KTN1){};
\draw [black,->, thick, -latex'] (Z_KTN) -- node [black,right, midway, font=\fontsize{7}{0}\selectfont] {\rotatebox{-90}{
$\begin{aligned}
&\mathbf{Z}_{{T+N}}^{{(K-1)}} \\
\end{aligned}$}}
(blank_KTN1);

\draw [black,->, thick, -latex'] (vdots_KT1) -- node [right, midway, font=\fontsize{7}{0}\selectfont] {\rotatebox{-90}{
$\begin{aligned}
&\mathbf{Z}_{{T+N-1}}^{{(K-1)}} \\
\end{aligned}$}}
(Z_KTN);

 \node[black,blank, below left=1.cm of Z_KTN] (blank_KTNl){};
 \draw [black,->, thick, -latex'] (Z_KTN) -- node [black,above=0.1cm, near end, font=\fontsize{7}{0}\selectfont] {\rotatebox{-90}
 {$\begin{aligned}
 &\mathbf{Z}_{{T+N}}^{{(K-1)}} \\
  \end{aligned}$}}
(blank_KTNl);

  \node[black,blank, above left=2.02cm of Z_1T1] (blank_dl1){};
  \node[black,blank, above right=0.015cm of blank_40] (blank_dl2){};
  \draw [black,->, thick, -latex'] (blank_dl1) -- node [black,above=0.1cm, midway, font=\fontsize{8}{0}\selectfont] 
  {\rotatebox{-90}{$\begin{aligned}
  \text{depth}
  \end{aligned}$}}
(blank_dl2);

\end{tikzpicture}
\caption{The structure of PredNet.}
\label{fig:PredNet}
\end{figure}

\subsection{Regression-Prediction Network (RegPred Net)} \label{sec:RegPred_Network}

We combine the RegNet in Fig. \ref{fig:RegNet} and PredNet in Fig. \ref{fig:PredNet} together to get the overall RegPred Network, as in Fig. \ref{RegPred_Network}. The regression part of the network starts from RegCell $(1,1)$ at the bottom left and ends at RegCell $(K,T)$.

\begin{algorithm}[!htbp]
\caption{$K$ layer(s) RegPred Net (prediction)}
\begin{algorithmic}[1]
\renewcommand{\algorithmicrequire}{\textbf{Input:}}
\renewcommand{\algorithmicensure}{\textbf{Output:}}

\Require Input series $\mathbf{Y}$ with length $T$, number of prediction steps $N$, $\mathbf{H}^{{(1: K)}}$, $\mathbf{Z}^{{(1 : K)}}_{{0}}$, i.i.d. noises $\Delta \mathbf W^{{(0 :   K-1)}} _{{T+1:T+N}}$ (size $[K,N]$) following the standard normal distribution

\Ensure Simulated trajectory $\mathbf{Z}^{{(0)}}_{{T+1:T+N}} = \big[ {Y}_{{T+1}}, ..., {Y}_{{T+N}} \big] $

\State {\textbf{Initialization:} State$({{1:K}}, {{0}{}}) = \big[ \mathbf{Z}^{{(1: K)}}_{{0}}, \ \hat{\mathbb{E}}(\bm{\epsilon}^{{(1:K)}}_{{0}}) = \mathbf{0}, \ \hat{cov}(\bm{\epsilon}^{{(1:K)}}_{{0}}) = \mathbf{0} \big]$, $\bm{\epsilon}^{{(1:K)}}_{{0}} = \mathbf0$, $\mathbf{Z}^{{(0)}}_{{0:T}} = \mathbf{Y}_{{0:T}}$}

\For{$t = 1, \ldots, T$}
        \For{$k = 1, \ldots, K$}

\State Run {Algorithm} \ref{alg:Algorithm_RegCell} with corresponding inputs to get State$({{1:k}{}}, {{t}{}})$

        \EndFor
\EndFor

\For{$t = T+1, \ldots, T+N$}

\For{$k = K, \ldots, 0$}

\State Run {Algorithm} \ref{alg:PredCell} with corresponding inputs to get $\mathbf{Z}^{{(k)}}_{{t}}$

        \EndFor
\EndFor
\end{algorithmic}\label{alg:RegPred_pred}
\end{algorithm}

\tikzstyle{block_reg} = [draw, 
                             rectangle, 
                             fill=black!15, 
                             text width=1.5em, 
                             text centered, 
                             minimum height={width("RegCell")+3pt},
                             font=\fontsize{6}{0}\selectfont, 
                             node distance=5em]
\tikzstyle{block_pred} = [draw, 
                             rectangle, 
                             fill=cyan!15, 
                             text width=1.5em, 
                             text centered, 
                             minimum height={width("(K-1, T+1)")+3pt},
                             font=\fontsize{6}{0}\selectfont, 
                             node distance=5em]
\tikzstyle{blank} = [fill=white, text width=1em, text centered, minimum height=1mm, font=\fontsize{6}{0}\selectfont, node distance=6em]
\begin{figure}[pos=htbp, width=10cm, align=\centering] 
\centering
\begin{tikzpicture} 

\node [black, block_reg] (Z_1) 
{
\rotatebox{-90}{$\begin{aligned}
\text{Reg} &\text{Cell} \\
(1&, 1)
\end{aligned}$}
 };

 \node [black,below=1.5cm of Z_1] (vdots_1) {\vdots};

  \node[black,blank, above left=1.5cm of Z_1] (blank_s1){};
  \draw [black,->, thick, -latex'] (blank_s1) -- node [black,above=0.1cm, midway, font=\fontsize{7}{0}\selectfont] 
  {\rotatebox{-90}{$\begin{aligned}
  \mathbf{Z}_{{0}}^{{(0)}} = \mathbf{Y}_{{0}}
\end{aligned}$}}
(Z_1.north);

  \node[black,blank, below right=1.5cm of Z_1] (blank_s2){};
  \draw [black,->, thick, -latex'] (Z_1.east) -- node [black,above=0.1cm, near end, font=\fontsize{7}{0}\selectfont]
  {\rotatebox{-90}{$\begin{aligned}
  \mathbf{Z}_{{1}}^{{(1)}} 
  \end{aligned}$}}
(blank_s2);

\node [black,right=1.8cm of vdots_1] (vdots_s1) {\vdots};

  \node[black,blank, above=1.5cm of Z_1] (blank_1){};
  \draw [black,->, thick, -latex'] (blank_1) -- node [black,right, midway, font=\fontsize{7}{0}\selectfont] 
  {\rotatebox{-90}{$\begin{aligned}
  & \text{State}(1, 1) \\ 
\end{aligned}$}}
(Z_1);

\draw [black,->, thick, -latex'] (Z_1) -- node [black,right, midway, font=\fontsize{7}{0}\selectfont]
{\rotatebox{-90}{
$\begin{aligned}
 & \text{State}(1, 2) \\ 
\end{aligned}$}}
(vdots_1);

 \node[black,blank, left=1.5cm of Z_1] (blank_2){};
 \draw [black,->, thick, -latex'] (blank_2) -- node [black,above, near start, font=\fontsize{7}{0}\selectfont] 
 {\rotatebox{-90}{$\begin{aligned}
  \mathbf{Z}_{1}^{(0)} = \mathbf{Y}_{1}
\end{aligned}$}}
(Z_1);
\node [black,right=1.5cm of Z_1] (ldots_1) {\ldots};
\draw [black,->, thick, -latex'] (Z_1) -- node [black,above, near end, font=\fontsize{7}{0}\selectfont] {\rotatebox{-90}{$\begin{aligned}
\mathbf{Z}_{1}^{(1)}
\end{aligned}$}}
(ldots_1);

\node [black,block_reg, below=1.5cm of vdots_1] (Z_1t) 
{
\rotatebox{-90}{$\begin{aligned}
\text{Reg} &\text{Cell} \\
(1&, t)
\end{aligned}$}
  };

 \node [black,below=1.5cm of Z_1t] (vdots_1t) {\vdots};

\draw [black,->, thick, -latex'] (Z_1t) -- node [black,right, midway, font=\fontsize{7}{0}\selectfont] {\rotatebox{-90}{
$\begin{aligned}
& \text{State}(1, t+1) \\ 
\end{aligned}$}}
(vdots_1t);

  \draw [black,->, thick, -latex'] (vdots_1) -- node [black,right, midway, font=\fontsize{7}{0}\selectfont] {\rotatebox{-90}
  {$\begin{aligned}
  & \text{State}(1, t) \\ 
  \end{aligned}$}}
(Z_1t);

  \node[black,blank, above left=1.5cm of Z_1t] (blank_s1t){};
  \draw [black,->, thick, -latex'] (blank_s1t) -- node [black,above=0.1cm, midway, font=\fontsize{7}{0}\selectfont] 
  {\rotatebox{-90}{$\begin{aligned}
  \mathbf{Z}_{{t-1}}^{{(0)}} = \mathbf{Y}_{{t-1}}
\end{aligned}$}}
(Z_1t.north);

  \node[black,blank, below right=1.5cm of Z_1t] (blank_s2t){};
  \draw [black,->, thick, -latex'] (Z_1t.east) -- node [black,above=0.1cm, near end, font=\fontsize{7}{0}\selectfont] 
  {\rotatebox{-90}{$\begin{aligned}
  \mathbf{Z}_{{t}}^{{(1)}} 
  \end{aligned}$}}
(blank_s2t);

\node [black,right=1.8cm of vdots_1t] (vdots_s1) {\vdots};

\node[black,blank, left=1.5cm of Z_1t] (blank_1tl){};
 \draw [black,->, thick, -latex'] (blank_1tl) -- node [black,above, near start, font=\fontsize{7}{0}\selectfont] {\rotatebox{-90}
 {$\begin{aligned}
  \mathbf{Z}_{t}^{(0)} = \mathbf{Y}_{t}
\end{aligned}$}}
(Z_1t);

  \node [black,right=1.5cm of Z_1t] (ldots_1t) {\ldots};
\draw [black,->, thick, -latex'] (Z_1t) -- node [black,above, near end, font=\fontsize{7}{0}\selectfont] {\rotatebox{-90}{$\begin{aligned}
 \mathbf{Z}_{t}^{(1)}
\end{aligned}$}}
(ldots_1t);

\node [black,block_reg, below=1.5cm of vdots_1t] (Z_1T) 
{
\rotatebox{-90}{$\begin{aligned}
\text{Reg} &\text{Cell} \\
(1&, T)
\end{aligned}$}
  };

  \draw [black,->, thick, -latex'] (vdots_1t) -- node [black,right, midway, font=\fontsize{7}{0}\selectfont] {\rotatebox{-90}
  {$\begin{aligned}
  & \text{State}(1, T) \\ 
\end{aligned}$}}
(Z_1T);

  \node[black,blank, below=1.5cm of Z_1T] (blank_d1){};
  \draw [black,->, thick, -latex'] (Z_1T) -- node [black,right, midway, font=\fontsize{7}{0}\selectfont] {\rotatebox{-90}
  {$\mathbf{Z}_{{T}}^{{(1)}}$}}
(blank_d1);

\node[black,blank, left=1.5cm of Z_1T] (blank_1Tl){};
 \draw [black,->, thick, -latex'] (blank_1Tl) -- node [black,above, near start, font=\fontsize{7}{0}\selectfont] {\rotatebox{-90}
 {$\begin{aligned}
  \mathbf{Z}_{{T}}^{{(0)}}
 = \mathbf{Y}_{{T}}
\end{aligned}$}}
(Z_1T);

  \node[black,blank, above left=1.5cm of Z_1T] (blank_s1T){};
  \draw [black,->, thick, -latex'] (blank_s1T) -- node [black,above=0.1cm, midway, font=\fontsize{7}{0}\selectfont] 
  {\rotatebox{-90}{$\begin{aligned}
  \mathbf{Z}_{{T-1}}^{{(0)}} = \mathbf{Y}_{{T-1}}
\end{aligned}$}}
(Z_1T.north);

  \node [black,right=1.5cm of Z_1T] (ldots_1T) {\ldots};
\draw [black,->, thick, -latex'] (Z_1T) -- node [black,above, near end, font=\fontsize{7}{0}\selectfont] {\rotatebox{-90}
{$\begin{aligned}
 \mathbf{Z}_{{T}}^{{(1)}}
\end{aligned}$}}
(ldots_1T);

\node [black,block_reg, right=1.5cm of ldots_1] (Z_k0) 
{
\rotatebox{-90}{$\begin{aligned}
\text{Reg} &\text{Cell} \\
(k&, 1)
\end{aligned}$}
  };

 \node [black,below=1.5cm of Z_k0] (vdots_k0) {\vdots};

  \node[black,blank, above=1.5cm of Z_k0] (blank_k01){};
  \draw [black,->, thick, -latex'] (blank_k01) -- node [black,right, midway, font=\fontsize{7}{0}\selectfont] {\rotatebox{-90}
  {$\begin{aligned}
  & \text{State}(k, 1) \\ 
  \end{aligned}$}}
(Z_k0);

\draw [black,->, thick, -latex'] (Z_k0) -- node [black,right, midway, font=\fontsize{7}{0}\selectfont] {\rotatebox{-90}{
$\begin{aligned}
& \text{State}(k, 2) \\ 
\end{aligned}$}}
(vdots_k0);

  \node[black,blank, above left=1.5cm of Z_k0] (blank_s1t){};
  \draw [black,->, thick, -latex'] (blank_s1t) -- node [black,above=0.1cm, midway, font=\fontsize{7}{0}\selectfont] 
  {\rotatebox{-90}{$\begin{aligned}
  \mathbf{Z}_{{0}}^{{(k-1)}}
\end{aligned}$}}
(Z_k0.north);

  \node[black,blank, below right=1.5cm of Z_k0] (blank_s2t){};
  \draw [black,->, thick, -latex'] (Z_k0.east) -- node [black,above=0.1cm, near end, font=\fontsize{7}{0}\selectfont] 
  {\rotatebox{-90}{$\begin{aligned}
  \mathbf{Z}_{{1}}^{{(k)}} 
  \end{aligned}$}}
(blank_s2t);

\node [black,right=1.8cm of vdots_k0] (vdots_sk1) {\vdots};

 \draw [black,->, thick, -latex'] (ldots_1) -- node [black,above, midway, font=\fontsize{7}{0}\selectfont]  {\rotatebox{-90}
 {$\begin{aligned}
  \mathbf{Z}_{{1}}^{{(k-1)}}
\end{aligned}$}}
(Z_k0);

\node [black,right=1.5cm of Z_k0] (ldots_k0) {\ldots};
\draw [black,->, thick, -latex'] (Z_k0) -- node [black,above, near end, font=\fontsize{7}{0}\selectfont]  {\rotatebox{-90}
{$\begin{aligned}
\mathbf{Z}_{{1}}^{{(k)}}
\end{aligned}$}}
(ldots_k0);

\node [black,block_reg, below=1.5cm of vdots_k0] (Z_kt) 
{
\rotatebox{-90}{$\begin{aligned}
\text{Reg} &\text{Cell} \\
(k&, t)
\end{aligned}$}
  };

 \node [black,below=1.5cm of Z_kt] (vdots_kt) {\vdots};

  \draw [black,->, thick, -latex'] (vdots_k0) -- node [black,right, midway, font=\fontsize{7}{0}\selectfont] {\rotatebox{-90}
  {$\begin{aligned}
  & \text{State}(k, t) \\ 
 \end{aligned}$}}
(Z_kt);

\draw [black,->, thick, -latex'] (Z_kt) -- node [black,right, midway, font=\fontsize{7}{0}\selectfont] {\rotatebox{-90}
{$\begin{aligned}
& \text{State}(k, t+1) \\ 
\end{aligned}$}}
(vdots_kt);

  \node[black,blank, above left=1.5cm of Z_kt] (blank_s1t){};
  \draw [black,->, thick, -latex'] (blank_s1t) -- node [black,above=0.1cm, midway, font=\fontsize{7}{0}\selectfont] 
  {\rotatebox{-90}{$\begin{aligned}
  \mathbf{Z}_{{t-1}}^{{(k-1)}}
\end{aligned}$}}
(Z_kt.north);

  \node[black,blank, below right=1.5cm of Z_kt] (blank_s2t){};
  \draw [black,->, thick, -latex'] (Z_kt.east) -- node [black,above=0.1cm, near end, font=\fontsize{7}{0}\selectfont] 
  {\rotatebox{-90}{$\begin{aligned}
  \mathbf{Z}_{{t}}^{{(k)}} 
  \end{aligned}$}}
(blank_s2t);

\node [black,right=1.8cm of vdots_kt] (vdots_sk2) {\vdots};

 \draw [black,->, thick, -latex'] (ldots_1t) -- node [black,above, midway, font=\fontsize{7}{0}\selectfont] {\rotatebox{-90}
 {$\begin{aligned}
  \mathbf{Z}_{{t}}^{{(k-1)}}
\end{aligned}$}}
(Z_kt);
  \node [black,right=1.5cm of Z_kt] (ldots_kt) {\ldots};
\draw [black,->, thick, -latex'] (Z_kt) -- node [black,above, near end, font=\fontsize{7}{0}\selectfont] {\rotatebox{-90}
{$\begin{aligned}
 \mathbf{Z}_{{t}}^{{(k)}}
\end{aligned}$}}
(ldots_kt);

\node [black,block_reg, below=1.5cm of vdots_kt] (Z_kT) 
{
\rotatebox{-90}{$\begin{aligned}
\text{Reg} &\text{Cell} \\
(k&, T)
\end{aligned}$}
  };

  \draw [black,->, thick, -latex'] (vdots_kt) -- node [black,right, midway, font=\fontsize{7}{0}\selectfont] {\rotatebox{-90}
  {$\begin{aligned} 
  & \text{State}(k, T) \\ 
  \end{aligned}$}}
(Z_kT);

  \node[black,blank, below=1.5cm of Z_kT] (blank_dk){};
  \draw [black,->, thick, -latex'] (Z_kT) -- node [black,right, midway, font=\fontsize{7}{0}\selectfont] {\rotatebox{-90}
  {$\mathbf{Z}_{{T}}^{{(k)}}$}}
(blank_dk);

  \node[black,blank, above left=1.5cm of Z_kT] (blank_s1T){};
  \draw [black,->, thick, -latex'] (blank_s1T) -- node [black,above=0.1cm, midway, font=\fontsize{7}{0}\selectfont] 
  {\rotatebox{-90}{$\begin{aligned}
  \mathbf{Z}_{{T-1}}^{{(k-1)}}
\end{aligned}$}}
(Z_kT.north);

 \draw [black,->, thick, -latex'] (ldots_1T) -- node [black,above, midway, font=\fontsize{7}{0}\selectfont] {\rotatebox{-90}
 {$\begin{aligned}
  \mathbf{Z}_{{T}}^{{(k-1)}}
\end{aligned}$}}
(Z_kT);

  \node [black,right=1.5cm of Z_kT] (ldots_kT) {\ldots};
\draw [black,->, thick, -latex'] (Z_kT) -- node [black,above, near end, font=\fontsize{7}{0}\selectfont] {\rotatebox{-90}
{$\begin{aligned}
 \mathbf{Z}_{{T}}^{{(k)}}
\end{aligned}$}}
(ldots_kT);

\node [black,block_pred, below=2.cm of ldots_1T] (Z_kT1) 
{
\rotatebox{-90}{$\begin{aligned}
\text{Pred} &\text{Cell} \\
(k-1&, T+1)
\end{aligned}$}
};

 \node [black,below=1.cm of Z_kT1] (vdots_kT1) {\vdots};

\draw [black,->, thick, -latex'] (Z_kT1) -- node [black,right, midway, font=\fontsize{7}{0}\selectfont] {\rotatebox{-90}{
$\begin{aligned}
&\mathbf{Z}_{{T+1}}^{{(k-1)}} \\
\end{aligned}$}}
(vdots_kT1);

  \node [black,above=1.cm of Z_kT1] (blank_kT) {};
\draw [black,->, thick, -latex'] (blank_kT) -- node [black,right, near start, font=\fontsize{7}{0}\selectfont] {\rotatebox{-90}{
$\begin{aligned}
&\mathbf{Z}_{{T}}^{{(k-1)}} \\
\end{aligned}$}}
(Z_kT1);

 \node[black,blank, below left of=Z_kT1] (blank_10){};
 \draw [black,->, thick, -latex', rotate=45] (Z_kT1) -- node [above=0.2cm, near end, font=\fontsize{7}{0}\selectfont] {\rotatebox{-90}
 {$\begin{aligned}
  \mathbf{Z}_{{T+1}}^{{(k-1)}}
  \end{aligned}$}}
(blank_10);
  \node [black,right=1.5cm of Z_kT1] (ldots_kT1) {\ldots};
\draw [black,->, thick, -latex'] (Z_kT.south) -- node [black,below=0.1cm, near start, font=\fontsize{7}{0}\selectfont] {\rotatebox{-90}{$\begin{aligned}
 \mathbf{Z}_{{T}}^{{(k)}}
\end{aligned}$}}
(Z_kT1.east);

\node [black,block_pred, below=1.cm of vdots_kT1] (Z_kTN) 
{
\rotatebox{-90}{$\begin{aligned}
\text{Pred} &\text{Cell} \\
(k-1&, T+N)
\end{aligned}$}
  };

 \node[black,blank, below=1.cm of Z_kTN] (blank_kTN1){};
\draw [black,->, thick, -latex'] (Z_kTN) -- node [black,right, midway, font=\fontsize{7}{0}\selectfont] {\rotatebox{-90}{
$\begin{aligned}
&\mathbf{Z}_{{T+N}}^{{(k-1)}} \\
\end{aligned}$}}
(blank_kTN1);

\draw [black,->, thick, -latex'] (vdots_kT1) -- node [black,right, midway, font=\fontsize{7}{0}\selectfont] {\rotatebox{-90}{
$\begin{aligned}
&\mathbf{Z}_{{T+N-1}}^{{(k-1)}} \\
\end{aligned}$}}
(Z_kTN);

 \node[black,blank, below left=1.cm of Z_kTN] (blank_kTN0){};
 \draw [black,->, thick, -latex'] (Z_kTN) -- node [black,above=0.1cm, near end, font=\fontsize{7}{0}\selectfont]  {\rotatebox{-90}
 {$\begin{aligned}
  \mathbf{Z}_{{T+N}}^{{(k-1)}}
\end{aligned}$}}
(blank_kTN0);

  \node[black,blank, above right=1.cm of Z_kTN] (blank_kTNr){};
  \node [black,right=1.5cm of Z_kTN] (ldots_kTN) {\ldots};
\draw [black,->, thick, -latex'] (blank_kTNr) -- node [black,below=0.1cm, near start, font=\fontsize{7}{0}\selectfont]  {\rotatebox{-90}{$\begin{aligned}
 \mathbf{Z}_{{T+N-1}}^{{(k)}}
\end{aligned}$}}
(Z_kTN);

\node [black,block_pred, left=4.cm of Z_kT1] (Z_1T1) 
{
\rotatebox{-90}{$\begin{aligned}
\text{Pred}&\text{Cell} \\
 \ \quad (0, T&+1) \quad
\end{aligned}$}
};

 \node [black,below=1.cm of Z_1T1] (vdots_1T1) {\vdots};

\draw [black,->, thick, -latex'] (Z_1T1) -- node [black,right, midway, font=\fontsize{7}{0}\selectfont] {\rotatebox{-90}{
$\begin{aligned}
\mathbf{Z}_{{T+1}}^{{(0)}}
\end{aligned}$}}
(vdots_1T1);

 \node [black,above=1.cm of Z_1T1] (blank_1T1) {};
\draw [black,->, thick, -latex'] (blank_1T1) -- node [black,right, near start, font=\fontsize{7}{0}\selectfont]{\rotatebox{-90}{
$\begin{aligned}
\mathbf{Z}_{{T}}^{{(0)}}
\end{aligned}$}}
(Z_1T1);

\node[black,blank, below left=1.cm of Z_1T1] (blank_1T1l){};
 \draw [black,->, thick, -latex'] (Z_1T1) -- node [black,above=0.2cm, near end, font=\fontsize{7}{0}\selectfont] {\rotatebox{-90}
 {$\begin{aligned}
  \mathbf{Z}_{{T+1}}^{{(0)}} = \mathbf{Y}_{{T+1}}
\end{aligned}$}}
(blank_1T1l);
  \node [black,right=1.5cm of Z_1T1] (ldots_1T1) {\ldots};
\draw [black,->, thick, -latex'] (Z_1T.south) -- node [black,below=0.1cm, near start, font=\fontsize{7}{0}\selectfont] {\rotatebox{-90}{$\begin{aligned}
 \mathbf{Z}_{{T}}^{{(1)}}
\end{aligned}$}}
(Z_1T1.east);

\node [black,block_pred, below=1.cm of vdots_1T1] (Z_1TN) 
{
\rotatebox{-90}{$\begin{aligned}
\text{Pred}&\text{Cell} \\
 \quad (0, T&+N) \quad
\end{aligned}$}
  };

  \node[black,blank,  left=3.4cm of Z_1] (blank_tl1){};
  \node[black,blank,  left=1.cm of Z_1TN] (blank_tl2){};
  \draw [black,->, thick, -latex'] (blank_tl1) -- node [black,left=0.1cm, midway, font=\fontsize{8}{0}\selectfont] 
  {\rotatebox{-90}{$\begin{aligned}
  \text{time}
  \end{aligned}$}}
(blank_tl2);

 \node[black,blank, below=1.cm of Z_1TN] (blank_1TN1){};
\draw [black,->, thick, -latex'] (Z_1TN) -- node [black,right, midway, font=\fontsize{7}{0}\selectfont]{\rotatebox{-90}{
$\begin{aligned}
\mathbf{Z}_{{T+N}}^{{(0)}}
\end{aligned}$}}
(blank_1TN1);

\draw [black,->, thick, -latex'] (vdots_1T1) -- node [black,right, midway, font=\fontsize{7}{0}\selectfont]{\rotatebox{-90}{
$\begin{aligned}
\mathbf{Z}_{{T+N-1}}^{{(0)}}
\end{aligned}$}}
(Z_1TN);

\node[black,blank, below left=1.cm of Z_1TN] (blank_1TNl){};
 \draw [black,->, thick, -latex'] (Z_1TN) -- node [black,above=0.1cm, near end, font=\fontsize{7}{0}\selectfont] {\rotatebox{-90}
 {$\begin{aligned}
  \mathbf{Z}_{{T+N}}^{{(0)}} = \mathbf{Y}_{{T+N}}
\end{aligned}$}}
(blank_1TNl);

  \node [black,right=1.5cm of Z_1TN] (ldots_1TN) {\ldots};
  \node[black,blank, above right=1.cm of Z_1TN] (blank_1TN0){};
\draw [black,->, thick, -latex'] (blank_1TN0) -- node [black,below=0.1cm, near start, font=\fontsize{7}{0}\selectfont] {\rotatebox{-90}{$\begin{aligned}
 \mathbf{Z}_{{T+N-1}}^{{(1)}} 
\end{aligned}$}}
(Z_1TN);

\node [black,block_reg, right=1.5cm of ldots_k0] (Z_K0) 
{
\rotatebox{-90}{$\begin{aligned}
\text{Reg} &\text{Cell} \\
(K&, 1)
\end{aligned}$}
  };

\node [black,below=1.5cm of Z_K0] (vdots_K0) {\vdots};

  \node[black,blank, above left=2.5cm of Z_1] (blank_dl1){};
  \node[black,blank, above right=2.5cm of Z_K0] (blank_dl2){};
  \draw [black,->, thick, -latex'] (blank_dl1) -- node [black,above=0.1cm, midway, font=\fontsize{8}{0}\selectfont] 
  {\rotatebox{-90}{$\begin{aligned}
  \text{depth}
  \end{aligned}$}}
(blank_dl2);

    \node[black,blank, above=1.5cm of Z_K0] (blank_K01){};
  \draw [black,->, thick, -latex'] (blank_K01) -- node [black,right, midway, font=\fontsize{7}{0}\selectfont] {\rotatebox{-90}
  {$\begin{aligned}
  & \text{State}(K, 1) \\ 
   \end{aligned}$}}
(Z_K0);

 \draw [black,->, thick, -latex'] (Z_K0) -- node [black,right, midway, font=\fontsize{7}{0}\selectfont] {\rotatebox{-90}
 {$\begin{aligned}
& \text{State}(K, 2) \\ 
\end{aligned}$}}
(vdots_K0);

  \node[black,blank, above left=1.5cm of Z_K0] (blank_s1t){};
  \draw [black,->, thick, -latex'] (blank_s1t) -- node [black,above=0.1cm, midway, font=\fontsize{7}{0}\selectfont] 
  {\rotatebox{-90}{$\begin{aligned}
  \mathbf{Z}_{{0}}^{{(K-1)}}
\end{aligned}$}}
(Z_K0.north);

 \draw [black,->, thick, -latex'] (ldots_k0) -- node [black,above, midway, font=\fontsize{7}{0}\selectfont] {\rotatebox{-90}
 {$\begin{aligned}
 \mathbf{Z}_{{1}}^{{(K-1)}}
\end{aligned}$}}
(Z_K0);
\node[black,blank, right=1.5cm of Z_K0] (blank_K0r){};
\draw [black,->, thick, -latex'] (Z_K0) -- node [black,above, near end, font=\fontsize{7}{0}\selectfont] {\rotatebox{-90}
{$\begin{aligned}
\mathbf{Z}_{{1}}^{{(K)}}
\end{aligned}$}}
(blank_K0r);

\node [black,block_reg, below=1.5cm of vdots_K0] (Z_Kt) 
{
\rotatebox{-90}{$\begin{aligned}
\text{Reg} &\text{Cell} \\
(K&, t)
\end{aligned}$}
  };

 \node [black,below=1.5cm of Z_Kt] (vdots_Kt) {\vdots};

\draw [black,->, thick, -latex'] (Z_Kt) -- node [black,right, midway, font=\fontsize{6.5}{0}\selectfont] {\rotatebox{-90}
{$\begin{aligned}
& \text{State}(K, t+1) \\ 
\end{aligned}$}}
(vdots_Kt);

\draw [black,->, thick, -latex'] (vdots_K0) -- node [black,right, midway, font=\fontsize{7}{0}\selectfont] {\rotatebox{-90}
  {$\begin{aligned}
   & \text{State}(K, t) \\ 
   \end{aligned}$}}
(Z_Kt);

  \node[black,blank, above left=1.5cm of Z_Kt] (blank_sKT){};
  \draw [black,->, thick, -latex'] (blank_sKT) -- node [black,above=0.1cm, midway, font=\fontsize{7}{0}\selectfont] 
  {\rotatebox{-90}{$\begin{aligned}
  \mathbf{Z}_{{t-1}}^{{(K-1)}}
\end{aligned}$}}
(Z_Kt.north);

 \draw [black,->, thick, -latex'] (ldots_kt) -- node [black,above, midway, font=\fontsize{7}{0}\selectfont] {\rotatebox{-90}
 {$\begin{aligned}
  \mathbf{Z}_{{t}}^{{(K-1)}}
\end{aligned}$}}
(Z_Kt);
  \node[black,blank, right=1.5cm of Z_Kt] (blank_Kt1){};
\draw [black,->, thick, -latex'] (Z_Kt) -- node [black,above, near end, font=\fontsize{7}{0}\selectfont] {\rotatebox{-90}
{$\begin{aligned}
 \mathbf{Z}_{{t}}^{{(K)}}
\end{aligned}$}}
(blank_Kt1);

\node [black,block_reg, below=1.5cm of vdots_Kt] (Z_KT) 
{
\rotatebox{-90}{$\begin{aligned}
\text{Reg} &\text{Cell} \\
(K&, T)
\end{aligned}$}
  };

  \draw [black,->, thick, -latex'] (vdots_Kt) -- node [black,right, midway, font=\fontsize{7}{0}\selectfont] {\rotatebox{-90}
  {$\begin{aligned}
     & \text{State}(K, T) \\ 
     \end{aligned}$}}
(Z_KT);

  \node[black,blank, above left=1.5cm of Z_KT] (blank_sKT){};
  \draw [black,->, thick, -latex'] (blank_sKT) -- node [black,above=0.1cm, midway, font=\fontsize{7}{0}\selectfont] 
  {\rotatebox{-90}{$\begin{aligned}
  \mathbf{Z}_{{T-1}}^{{(K-1)}}
\end{aligned}$}}
(Z_KT.north);

 \draw [black,->, thick, -latex'] (ldots_kT) -- node [black,above, midway, font=\fontsize{7}{0}\selectfont] {\rotatebox{-90}
 {$\begin{aligned}
  \mathbf{Z}_{{T}}^{{(K-1)}}
  \end{aligned}$}}
(Z_KT);

\node[black,blank, right=1.5cm of Z_KT] (blank_Z_KT){};
\draw [black,->, thick, -latex'] (Z_KT) -- node [black,above, near end, font=\fontsize{7}{0}\selectfont] {\rotatebox{-90}
{$\begin{aligned}
 \mathbf{Z}_{{T}}^{{(K)}}
\end{aligned}$}}
(blank_Z_KT);

\node [black,block_pred, below=1.48cm of Z_KT] (Z_IKT1) 
{
\rotatebox{-90}{$\begin{aligned}
\ \text{Pred}&\text{Cell} \ \\
 (K, T&+1)
\end{aligned}$}
  };

\draw [black,->, thick, -latex'] (Z_KT) -- node [black,right, midway, font=\fontsize{7}{0}\selectfont] {\rotatebox{-90}
{$\begin{aligned}
 \mathbf{Z}_{{T}}^{{(K)}}
\end{aligned}$}}
(Z_IKT1);

\node [black,below=1.cm of Z_IKT1] (vdots_IKT1) {\vdots};

\draw [black,->, thick, -latex'] (Z_IKT1) -- node [black,right, midway, font=\fontsize{7}{0}\selectfont] {\rotatebox{-90}
{$\begin{aligned}
 \mathbf{Z}_{{T+1}}^{{(K)}}
\end{aligned}$}}
(vdots_IKT1);

\node[black,blank, below left=1cm of Z_IKT1] (blank_Z_IKT){};
\draw [black,->, thick, -latex'] (Z_IKT1) -- node [black,below=0.1cm,  midway, font=\fontsize{7}{0}\selectfont] {\rotatebox{-90}
{$\begin{aligned}
 \mathbf{Z}_{{T+1}}^{{(K)}}
\end{aligned}$}}
(blank_Z_IKT);

\node [black,block_pred, below=1.0cm of vdots_IKT1] (Z_IKTN) 
{
\rotatebox{-90}{$\begin{aligned}
\ \text{Pred}&\text{Cell} \ \\
(K, T&+N) 
\end{aligned}$}
};

\draw [black,->, thick, -latex'] (vdots_IKT1.south) -- node [black,right,  midway, font=\fontsize{7}{0}\selectfont] {\rotatebox{-90}
{$\begin{aligned}
 \mathbf{Z}_{{T+N-1}}^{{(K)}}
\end{aligned}$}}
(Z_IKTN);

\node[black,blank, below=1.cm of Z_IKTN] (blank_Z_IKTN){};
\draw [black,->, thick, -latex'] (Z_IKTN.south) -- node [black,right,  midway, font=\fontsize{7}{0}\selectfont] {\rotatebox{-90}
{$\begin{aligned}
 \mathbf{Z}_{{T+N}}^{{(K)}}
\end{aligned}$}}
(blank_Z_IKTN);

\node[black,blank, below left =1.cm of Z_IKTN] (blank_Z_IKTNl){};
\draw [black,->, thick, -latex'] (Z_IKTN) -- node [black,below,  midway, font=\fontsize{7}{0}\selectfont] {\rotatebox{-90}
{$\begin{aligned}
 \mathbf{Z}_{{T+N}}^{{(K)}}
\end{aligned}$}}
(blank_Z_IKTNl);

\node [black,block_pred, below=2.cm of ldots_kT] (Z_KT1) 
{
\rotatebox{-90}{$\begin{aligned}
\text{Pred}&\text{Cell} \\
\ (K-1, &T+1) \
\end{aligned}$}
  };

 \node [black,below=1.cm of Z_KT1] (vdots_KT1) {\vdots};

\draw [black,->, thick, -latex'] (Z_KT1) -- node [black,right, midway, font=\fontsize{7}{0}\selectfont] {\rotatebox{-90}{
$\begin{aligned}
&\mathbf{Z}_{{T+1}}^{{(K-1)}} \\
\end{aligned}$}}
(vdots_KT1);

   \node[black,blank, above=1.cm of Z_KT1] (blank_KT1){};
\draw [black,->, thick, -latex'] (blank_KT1) -- node [black,right, near start, font=\fontsize{7}{0}\selectfont] {\rotatebox{-90}{
$\begin{aligned}
&\mathbf{Z}_{{T}}^{{(K-1)}} \\
\end{aligned}$}}
(Z_KT1);

 \node[black,blank, below left of=Z_KT1] (blank_vdots_kT1){};
 \draw [black,->, thick, -latex'] (Z_KT1) -- node [black,above=0.2cm, near end, font=\fontsize{7}{0}\selectfont] {\rotatebox{-90}
 {$\begin{aligned}
  \mathbf{Z}_{{T+1}}^{{(K-1)}}
\end{aligned}$}}
(blank_vdots_kT1);

\draw [black,->, thick, -latex'] (Z_KT.south) -- node [black,right=0.1cm, near end , font=\fontsize{7}{0}\selectfont] {\rotatebox{-90}
{$\begin{aligned}
 \mathbf{Z}_{T}^{(K)}
\end{aligned}$}}
(Z_KT1.east);

\node [black,block_pred, below=1.cm of vdots_KT1] (Z_KTN) 
{
\rotatebox{-90}{$\begin{aligned}
\text{Pred}&\text{Cell} \\
  (K-1, &T+N) 
\end{aligned}$}
  };

 \node[black,blank, below=1.cm of Z_KTN] (blank_KTN1){};
\draw [black,->, thick, -latex'] (Z_KTN) -- node [black,right, midway, font=\fontsize{7}{0}\selectfont] {\rotatebox{-90}{
$\begin{aligned}
&\mathbf{Z}_{{T+N}}^{{(K-1)}} \\
\end{aligned}$}}
(blank_KTN1);

\draw [black,->, thick, -latex'] (vdots_KT1) -- node [black,right, midway, font=\fontsize{7}{0}\selectfont] {\rotatebox{-90}{
$\begin{aligned}
&\mathbf{Z}_{{T+N-1}}^{{(K-1)}} \\
\end{aligned}$}}
(Z_KTN);

 \node[black,blank, below left=1.cm of Z_KTN] (blank_KTNl){};
 \draw [black,->, thick, -latex'] (Z_KTN) -- node [black,above=0.1cm, near end, font=\fontsize{7}{0}\selectfont] {\rotatebox{-90}
 {$\begin{aligned}
 &\mathbf{Z}_{{T+N}}^{{(K-1)}} \\
  \end{aligned}$}}
(blank_KTNl);

 \node[black,blank, above right=1.cm of Z_KTN] (blank_KTN1){};
\draw [black,->, thick, -latex'] (blank_KTN1) -- node [black,below=0.2cm, near start, font=\fontsize{7}{0}\selectfont] {\rotatebox{-90}
{$\begin{aligned}
 \mathbf{Z}_{{T+N-1}}^{{(K)}}
\end{aligned}$}}
(Z_KTN);

\end{tikzpicture}
\caption{RegPred Net. $\text{State}(k,t)= \hspace{0.1em} \mathbf{Z}_{{t-1}}^{{(k)}}, \hspace{0.1em} \hat{\mathbb{E}}(\bm{\epsilon}_{{t-1}}^{{(k)}}), \hspace{0.1em} \hat{cov}(\bm{\epsilon}_{{t-1}}^{{(k)}}) $}
\label{RegPred_Network}
\end{figure}
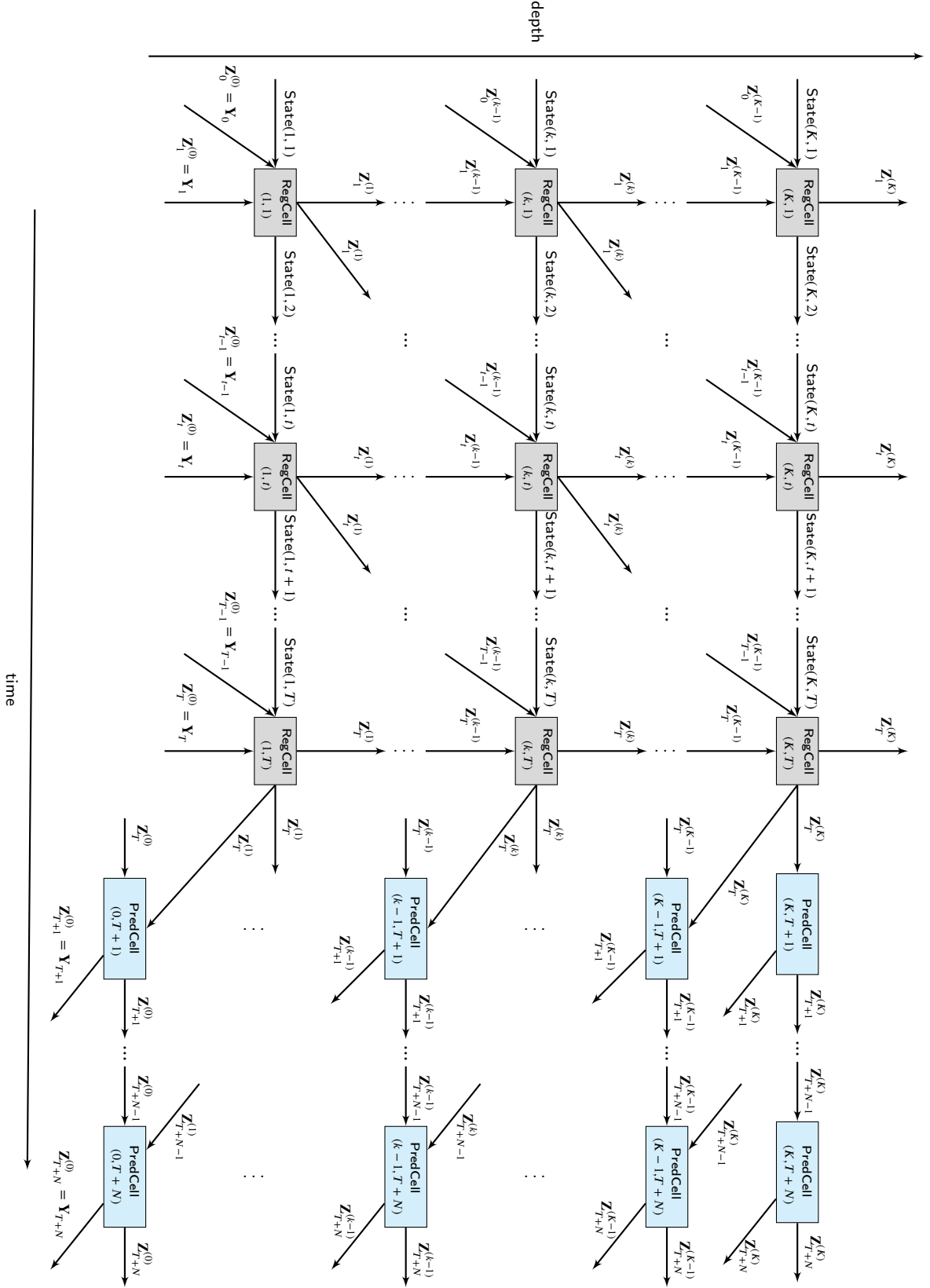

{Algorithm} \ref{alg:RegPred_pred} describes how a $K$ layer(s) RegPred Net in Fig. \ref{RegPred_Network} predicts the future $N$ steps given an input time series. The inputs of {Algorithm} \ref{alg:RegPred_pred} include: $\mathbf Y_{{1:T}}$ as input time series, $\mathbf{H}^{{(1: K)}}$ as the learning rates of layer $1$ to $K$, $\mathbf{Z}^{{(1 : K)}}_{{0}}$ are the initial  input vector of layer $1$ to $K$. We initialize State$({{1:K}}, {{0}})$ as the concatenation of $\mathbf{H}^{{(1: K)}}$, $\mathbf{Z}^{{(1 : K)}}_{{0}}$, $\hat{\mathbb{E}}(\bm{\epsilon}^{{(1:K)}}_{{0}})$, and $\hat{cov}(\bm{\epsilon}^{{(1:K)}}_{{0}})$. Among them $\hat{\mathbb{E}}(\bm{\epsilon}^{{(1:K)}}_{0})$, $\hat{cov}(\bm{\epsilon}^{{(1:K)}}_{0})$ are set as vector and matrix of $\mathbf{0}$. The errors  $\bm{\epsilon}^{{(1:K)}}_{{0}}$ are also initialized as $\mathbf{0}$. $\mathbf{Z}^{{(0)}}_{{0:T}}$ are always equal to $\mathbf{Y}_{{0:T}}$. At the start of {Algorithm} \ref{alg:RegPred_pred}, we generate $K \times N$ normally distributed random noises $\Delta \mathbf{W}^{{(0  :   K-1)}} _{{T+1:T+N}}$, then take $\mathbf{Y}$ as the input series of RegNet and run RegCell described in {Algorithm} \ref{alg:Algorithm_RegCell} from layer $1$ to layer $K$ and times $1$ to $T$. After the regression, we get State$({{1:K}}, {{T+1}})$. They are the inputs for the prediction. The predictions are calculated in the reverse order of the regression: run Algorithm \ref{alg:PredCell} downwards from layer $K$ to $0$, time $T+1$ to $T+N$. At layer $0$, we obtain a simulated trajectory $\mathbf{Z}^{{(0)}}_{{T+1:T+N}} = \mathbf{Y}_{{T+1:T+N}}$ as output. Like any Recurrent Neural Network, the RegPred Net can handle input series with different input lengths $T$ and  make predictions of arbitrary length $N$.

\begin{algorithm} [!htbp]
\caption{Calculating the loss of mean $L^\mathbb{E}$ and the loss of variance $L^\mathbb{V}$ using Monte Carlo simulation}
\begin{algorithmic}[1]
\renewcommand{\algorithmicrequire}{\textbf{Input:}}
\renewcommand{\algorithmicensure}{\textbf{Output:}}

\Require The number $n_W$ of simulated trajectories, input time series $\mathbf Y_1,\dotsc,\mathbf Y_T, \dotsc, \mathbf Y_{T+N}$, $\mathbf{H}^{{(1: K)}}$, $\mathbf{Z}^{{(1 : K)}}_{{0}}$

\Ensure $L^{{\mathbb{E}}}$,  $L^{{\mathbb{V}}}$, $\mathbb{E}_{{T+1:T+N}}$, $\mathbb{V}_{{T+1:T+N}}$

\State Compute $\mathbf Z_T^{(1:K)}$ by online regression with Algorithm \ref{alg:Algorithm_RegCell}

\State Generate $n_{{W}}$ random noises $\Delta \mathbf{W}^{{(0  :   K-1)}} _{{T+1:T+N}}$

\For{ i = 1, \ldots, $n_W$}

\State Use Algorithm \ref{alg:PredCell} to generate the $i$-th trajectory $\mathbf Z_{T+1,i}^{(0)}, \ldots, \mathbf Z_{T+N,i}^{(0)}$

\EndFor

\State Calculate mean $\mathbb{E}_t$ and variance $\mathbb{V}_t$ at every time step using the $n_W$ trajectories

\State Calculate the losses for the mean and the variance:
 $$L^{\mathbb{E}} = \sqrt{ \frac{1}{N} \sum_{{t=T+1}}^{{T+N}} (Y_{t} - \mathbb{E}_{t})^{2}}\ , \quad 
L^{\mathbb{V}} = \sqrt{ \frac{1}{N} \sum_{{t=T+1}}^{{T+N}} \big[(Y_{t} - \mathbb{E}_{t})^{2} - \mathbb{V}_{t}\big]^{2}}$$
\end{algorithmic} \label{alg:loss}
\end{algorithm}

Algorithm \ref{alg:loss} explains how to calculate losses for the mean and variance of the trajectories using Monte Carlo simulation. Assume we simulate $n_{{W}{}}$ trajectories, the loss of mean $L^{{\mathbb{E}}{}}$ between the mean of samples $\mathbb{E}$ and the target series $\mathbf{Y}_{T+1:T+N}$, and the loss of variance $L^{{\mathbb{V}}{}}$ between the variance of samples $\mathbb{V}$ and the target series $\mathbf{Y}_{T+1:T+N}$ can be calculated as:
\begin{equation} \label{loss_calculation}
\begin{aligned}
L^{{\mathbb{E}}} &= \sqrt{ \frac{{1}}{{N}} \sum_{{t=T+1}}^{{T+N}} \big(Y_{t} - \mathbb{E}_{t}\big)^{2}} \\
L^{{\mathbb{V}}} &= \sqrt{ \frac{{1}}{{N}} \sum_{{t=T+1}{}}^{{T+N}{}} \Big[ \big( Y_{t} - \mathbb{E}_{t} \big)^{2} - \mathbb{V}_{t} \Big]^{2}}
\end{aligned}
\end{equation}
where $\mathbf Y_{t}$ is the value of series $\mathbf{Y}$ at time $t$. The loss of mean $L^{{\mathbb{E}}}$ is evaluated by taking the average of the difference between values of the target series and their corresponding statistical mean of predicted samples. Similarly, the loss of variance $L^{{\mathbb{V}}}$ is computed by taking the mean of the difference between calculated variance $\big( Y_{t} - \mathbb{E}_{t} \big)^{2}$ and the statistical variance of samples $\mathbb{V}_{t}$. 

We improve the basic definition of the loss in {Algorithm} \ref{alg:loss} to a more statistically robust and meaningful loss by computing an average loss over several prediction horizons, as in {Algorithm} \ref{alg:avg_loss}. In {Algorithm} \ref{alg:avg_loss}, $\mathbf{Y}_{1:t}$ are the first $t$ values in series $\mathbf{Y}$, where $t=2, \cdots, T$. Use each sub-series $\mathbf{Y}_{1:t+N}$ as input, RegPred Net can predict the mean $\mathbb{E}_{t+1:t+N}$ and the variance $\mathbb{V}_{t+1:t+N}$ of the next $N$ time steps, the label for $\mathbb{E}_{t+1:t+N}$ is actually the steps ${t+1:t+N}$ of series $\mathbf{Y}_{1:t+N}$. The average loss of mean $L^{{\mathbb{E}}{}}_{avg}$ and the average loss of variance $L^{{\mathbb{V}}{}}_{avg}$ are then calculated by averaging $T-1$ losses computed by running {Algorithm} \ref{alg:loss} with $Y_{1:2+N}, \cdots, Y_{1:T+N}$ as inputs.

\begin{algorithm}[!htbp]
\caption{Calculating the average loss of mean $L^{{\mathbb{E}}{}}_{avg}$ and the average loss of variance $L^{{\mathbb{V}}{}}_{avg}$}
\begin{algorithmic}[1]
\renewcommand{\algorithmicrequire}{\textbf{Input:}}
\renewcommand{\algorithmicensure}{\textbf{Output:}}

\Require number $n_W$ of trajectories per Monte Carlo simulation, input time series $\mathbf Y_1,\dotsc, \mathbf Y_T,\dotsc, \mathbf Y_{T+N}$, $\mathbf{H}^{{(1: K)}}$, $\mathbf{Z}^{{(1 : K)}}_{{0}}$

\Ensure $L^{{\mathbb{E}}}_{avg}$, $L^{{\mathbb{V}}}_{avg}$

\For{ t = 2, \ldots, $T$}

\State Run Algorithm \ref{alg:loss} with $\mathbf Y_{1:t+N}$ as input series and get loss of mean $L_{t+1:t+N}^{\mathbb{E}}$ and loss of variance $L_{t+1:t+N}^{\mathbb{V}}$ for a regression window $[1, t]$, and prediction window $[t, t+N]$ estimated over $n_W$ trajectories

\EndFor

\State Calculate the average loss of mean and the average loss of variance:

 $L^{\mathbb{E}}_{avg} =  \frac{1}{T-1} \sum_{{t=2}}^{{T}} L^{{\mathbb{E}}}_{{t+1:t+N}} \ , \quad 
L^{\mathbb{V}}_{avg}= \frac{1}{T-1} \sum_{{t=2}}^{{T}} L^{{\mathbb{V}}}_{{t+1:t+N}}$

\end{algorithmic}\label{alg:avg_loss}
\end{algorithm}

\section{Optimization of hyperparameters in RegPred Net}
\label{sec:intro_BayOpt}

As explained in Sec. \ref{sec:RegPred_Net}, {RegPred Net} generates trajectories and ultimately forecasts which (besides randomly generated numbers) only depend on the initial value of the parameters $\mathbf{A}_{{0}}^{{(1:K)}}, \ \mathbf{N}_{{0}}^{{(1:K)}},
 \bm{\Sigma}_{{0}}^{{(1:K)}}$ and the learning rates $\mathbf{H}^{{(1:K)}} = \big[ \eta^{{(1:K)}}_{{A}}, \eta^{{(1:K)}}_{{N}}, \eta^{{(1:K)}}_{{\Sigma}}, \varphi^{{(1:K)}}, \rho^{{(1:K)}} \big]$. In comparison to cells in conventional RNNs, the RegCell and PredCell have no weight or parameter to learn (via backpropagation through time). Despite this apparent simplicity, we observe in the case of FX rate time series, that the RegPred Net's regressed parameters and as a matter of consequence simulations and forecasts are all very sensitive to the values of the hyperparameters. The selection of the RegPred Net's hyperparameters is a thorny minimization problem of a loss function with many local minima for which global optimization is required. Note that the loss function $L_{avg} = L_{avg}^{\mathbb{E}} + L_{avg}^{\mathbb{V}}$to minimize is a) noisy, as it is calculated by Monte Carlo simulation, and b) costly to compute. Consequently, the optimization method chosen must be able to handle noise in the objective function $f$, be parsimonious in the number of evaluations of $f$, and ideally shall not require the evaluation of the derivative of $f$. All of these reasons make {Bayesian optimization} an adequate method to find optimal values of the hyperparameters. 

\subsection{Bayesian optimization} \label{sec:Bayesian optimization}

Bayesian optimization is a heuristic algorithm to solve a maximization problem:
\begin{equation} \label{eq:BayOpt_def}
\begin{aligned}
\mathbf x_{*} = \underset{\mathbf x \ \in \ \mathbb {R}^{d}}{argmax} \  f(\mathbf x) 
\end{aligned}
\end{equation}
where $f$ is an objective function taking its values in $\mathbb {R}$. Bayesian optimization is a sequential decision strategy for the efficient global optimization of black-box functions which does not require estimation of the function's derivative. Bayesian optimization used in this article to maximize the RegPred Net's negative loss $f(x) = -L_{avg} = -(L^{\mathbb E}_{avg} + L^{\mathbb V}_{avg})$ where $\mathbf x$ represents the network's hyperparameters $\mathbf x=\big[ \mathbf{A}_{{0}}^{{(1:K)}}, \ \mathbf{N}_{{0}}^{{(1:K)}},
 \bm{\Sigma}_{{0}}^{{(1:K)}},\mathbf{H}^{{(1:K)}} \big]$. 

Bayesian optimization sequentially improves its estimates $\mathbf x_n$ of the maximizer $\mathbf x_{*}$ of $f$. At each step $n$, the value of $f(\mathbf x_n)$ is calculated and collected in the set of observations $\mathcal{{D}}_{{1:n}}=\Big\{ \big(\mathbf x_{{i}}, f(\mathbf x_i) \big) \ \big| \ i=1,...,n \Big\}$. This data set is used to model the posterior distribution $p(f(\mathbf x') | \mathcal{{D}}_{{1:n}}, \mathbf x')$ of the unknown and random value $f(\mathbf x)$ for any arbitrary $\mathbf x'$.
In Gaussian Process Regression, it is assumed that the joint distribution of $f(\mathbf X)$ and $f(\mathbf x')$ is multivariate Gaussian with mean function zero and a covariance function or kernel $k:\mathbb{R} \times \mathbb{R} \rightarrow \mathbb{R}$:
\begin{equation} \label{GP_joint}
\begin{bmatrix} f(\mathbf{X}) \\ f(\mathbf{x}') \end{bmatrix} \sim \mathcal{N} \Bigg (0 \ , \ \begin{bmatrix} \bm{K}(\mathbf X \ , \ \mathbf{X}) & \bm{K}(\mathbf X \ , \ \mathbf{x}') \\ \bm{K}(\mathbf{x}', \ \mathbf X) & {k}( \mathbf{x}', \ \mathbf{x}') \\ \end{bmatrix} \Bigg) 
\end{equation}
where $\mathbf X = \big\{ \mathbf{x}_{{1}}, \ \cdots, \ \mathbf{x}_{{n}} \big\}$ and $\bm{K}(\mathbf X \ ,\ \mathbf X)$ represents the $n \times n$ covariance matrix:
\begin{equation}
\bm{K(\mathbf X \ ,\ \mathbf X)} = \begin{bmatrix} 
k(\mathbf{x}_{{1}}, \mathbf{x}_{{1}}) & \ldots & k(\mathbf{x}_{{1}}, \mathbf{x}_{{n}}) \\
\vdots & \ddots & \vdots \\
k(\mathbf{x}_{{n}}, \mathbf{x}_{{1}}) & \ldots & k(\mathbf{x}_{{n}}, \mathbf{x}_{{n}}) \\
\end{bmatrix}
\end{equation}
and similarly, $\bm{K}(\mathbf X \ , \ \mathbf x')$ is the $n \times 1$ covariance matrix computed for all possible combinations between vectors in $\mathbf X$ and $\mathbf x'$. $\bm{K}(\mathbf x' \ , \ \mathbf X) = \bm{K}(\mathbf X \ , \ \mathbf x')^{T}$, and ${k}(\mathbf x' \ , \ \mathbf x')=1$. A commonly used kernel is the Squared Exponential (SE), which is a function-space expression of Radial Basis Function (RBF) (\ref{RBF}):
\begin{equation} \label{SE}
cov(f(\mathbf{x}), f(\mathbf{x'})) = k(\mathbf{x}, \mathbf{x'}) = \text{exp} \Big(-\frac{1}{2l^{2}} ||\mathbf{x} - \mathbf{x'}||^{2} \Big)
\end{equation}
where $l$ is a parameter that denotes the kernel's width. A small $l$ makes the covariance smaller, and vice versa. Notice that the covariance between outputs $f(\mathbf{x})$ and $f(\mathbf{x'})$ is described as a function of the inputs $\mathbf{x}$ and $\mathbf{x'}$. It implies that the covariance between variables tends to $1$ if their inputs are similar and tends to $0$ if their inputs are different. Another commonly used covariance function is the Matern class \citep[Sec. 4.2.1]{GP}, defined as:
\begin{equation} \label{Matern_class}
k_{Matern}(||\mathbf x - \mathbf x'||) = \frac {2^{1-\nu}} {\Gamma(\nu)} \big( \frac {\sqrt{2\nu} ||\mathbf x - \mathbf x'||} {l} \big)^{\nu} K_{\nu} \big( \frac {\sqrt{2\nu} ||\mathbf x - \mathbf x'||} {l} \big)
\end{equation}
where $\nu$ is a positive parameter that controls the smoothness of the function and $l$ is a positive scale parameter. $\Gamma(\cdot)$ is a gamma function (\ref{Gamma}) and $K_{\nu}$ is the modified Bessel function \citep[Sec. 9.6]{Bassel}. When $\nu \rightarrow \infty$, Eq. (\ref{Matern_class}) is exactly the SE covariance function described in Eq. (\ref{SE}). The most commonly used values for the Matern class in Machine Learning are $\nu = 3/2$ and $\nu = 5/2$:
\begin{equation} \label{Matern_class_ml}
\begin{aligned}
k_{\nu=3/2}(||\mathbf x - \mathbf x'||) &= \Big(1 + \frac {\sqrt{3} ||\mathbf x - \mathbf x'||} {l} \Big) \ \text{exp} \Big(- \frac {\sqrt{3} ||\mathbf x - \mathbf x'||} {l} \Big) \\
k_{\nu=5/2}(||\mathbf x - \mathbf x'||) &= \Big(1 + \frac {\sqrt{5} ||\mathbf x - \mathbf x'||} {l} + \frac {\sqrt{5} ||\mathbf x - \mathbf x'||^{2}} {3l^{2}} \Big) \ \text{exp} \Big(- \frac {\sqrt{5} ||\mathbf x - \mathbf x'||} {l} \Big) 
\end{aligned}
\end{equation}
Other covariance functions can be found in \citep[Chap. 4]{GP}.

Under the Gaussian Process assumption, it can be proven \citep{GP} using Bayes' theorem that the posterior distribution of $f(\mathbf x')$ follows a normal distribution with mean $\mu_n$ and variance $\sigma_n^2$:
\begin{equation} \label{evaluate posterior x_n+1}
\begin{aligned}
&p\big(f(\mathbf  x') \ |  \ \mathcal{D}_{1:n}, \ \mathbf x' \big) = \mathcal N \big(\mu_{n}(\mathbf x') \ , \ \sigma_{n}^{2}(\mathbf x') \big) \\
& \mu_{n}(\mathbf x') = \bm K(\mathbf x', \ \mathbf{X})^{T}\bm K(\mathbf{X}, \ \mathbf{X})^{-1}f(\mathbf X) \\ 
& \sigma_{n}^{2}(\mathbf x') =  k(\mathbf x', \ \mathbf x') - \bm K (\mathbf x', \  \mathbf X)^{T} \bm K (\mathbf X, \ \mathbf X)^{-1}\bm K (\mathbf X, \ \mathbf x') \\ 
\end{aligned}
\end{equation}

After evaluating the posterior distribution of $f(\mathbf{x}')$, Bayesian optimization requires the use of an acquisition function $utility(\cdot)$ to guide the search of the maximizer $\mathbf x_{*}$. A high value of the acquisition function implies a potentially high value of the objective function. The maximizer of the acquisition function provides the next estimate $\mathbf x_{{n+1}}$:
\begin{equation} \label{x_next}
\begin{aligned}
\mathbf x_{{n+1}} = argmax_{\mathbf{x}} \ \ utility(\mathbf{x} \ | \ \mathcal{D}_{1:n})
\end{aligned}
\end{equation}
Many acquisition functions have been proposed in the past. One commonly used function is the Expected Improvement (EI) from \citep{Expected_Improvement}. It defines first an improvement function ${I}$:
\begin{equation} \label{I(x)}
\begin{aligned}
I(\mathbf{x}) = max \big\{ 0, \ f(\mathbf{x}) - f(\mathbf{x_{*}}) \big\} 
\end{aligned}
\end{equation}
where $f(\mathbf x_{*})$ denotes the best estimate the objective function so far. The expected improvement $\mathbb{E}(I)$ is defined by $\mathbb{E} \big( max\{ 0, \  f(\mathbf{x}) - f(\mathbf{x_{*}}) \} \ | \ \mathcal{D}_{1:n} \big)$. $\mathbb{E}(I)$ is then calculated by:
\begin{equation} \label{EI_int}
\begin{aligned}
\mathbb{E}(I) &= \int_{{I}=0}^{\text{I}=\infty} {I} \  \frac{1}{\sqrt{2\pi} \sigma(\mathbf{x})} \ exp \Bigg( -\frac{ \big ( \mu(\mathbf{x}) -f(\mathbf{x}_{*}) - {I} \big)^{2}}{2 \sigma^{2}(\mathbf{x})} \Bigg) \ d {I} \\
& = \sigma(\mathbf{x}) \Big[ \frac{ \mu(\mathbf{x}) -f(\mathbf{x}_{*}) }{\sigma(\mathbf{x})} \ \Phi \big( \frac{ \mu(\mathbf{x}) -f(\mathbf{x}_{*}) }{\sigma(\mathbf{x})} \big) + \phi \big( \frac{ \mu(\mathbf{x}) -f(\mathbf{x}_{*}) }{\sigma(\mathbf{x})} \big) \Big]
\end{aligned} 
\end{equation}
Eq. (\ref{EI_int}) can be analytically evaluated as: 
\begin{equation} \label{EI_no_explorative_parameter}
\begin{aligned}
&\mathbb{E}({I})=
    \begin{cases}
      \big( \mu(\mathbf{x}) -f(\mathbf{x}_{*}) \big) \Phi(Z) + \sigma(\mathbf{x}) \phi(Z), & \text{if}\ \sigma(\mathbf{x})>0 \\
      0, & \text{if}\ \sigma(\mathbf{x})=0
    \end{cases} \\
&Z = \frac{ \mu(\mathbf{x}) -f(\mathbf{x}_{*}) }{\sigma(\mathbf{x})} 
\end{aligned}
\end{equation}
where $\Phi$ and $\phi$ indicate the CDF and PDF of the standard normal distribution, respectively.

A more general acquisition function, which allows controlling the balance between the exploitation and exploration of the optimum of $f$ is: 
\begin{equation} \label{EI_xi}
\begin{aligned}
&\mathbb{E}({I})=
    \begin{cases}
      \big( \mu(\mathbf{x}) -f(\mathbf{x}_{*}) - \xi \big) \Phi(Z) + \sigma(\mathbf{x}) \phi(Z), & \text{if}\ \sigma(\mathbf{x})>0\\
      0, & \text{if}\ \sigma(\mathbf{x})=0
    \end{cases} \\
&Z = \frac{ \mu(\mathbf{x}) -f(\mathbf{x}_{*}) - \xi}{\sigma(\mathbf{x})} 
\end{aligned}
\end{equation}
where $\xi \geqslant 0$ is a parameter encouraging exploration in regions where the variance $\sigma(\mathbf x)$ is large. By maximizing $\mathbb{E}({I})$ with $\xi$, we find the next $\mathbf x_{{n+1}}$ that can lead to a higher value of $f$:
\begin{equation} \label{EI_argmax}
\begin{aligned}
\mathbf{x}_{n+1} &=
\begin{cases}
      argmax_{\mathbf{x}} \ \ \big( \mu(\mathbf{x}) -f(\mathbf{x}_{*}) - \xi \big) \Phi(Z) + \sigma(\mathbf{x}) \phi(Z), & \text{if}\ \sigma(\mathbf{x})>0 \\
      0, & \text{if}\ \sigma(\mathbf{x})=0
    \end{cases} \\
Z &= \frac{ \mu(\mathbf{x}) -f(\mathbf{x}_{*}) - \xi}{\sigma(\mathbf{x})} 
\end{aligned}
\end{equation}

\begin{figure}[pos=htbp, align=\centering]
\centering
\includegraphics[scale=0.32]{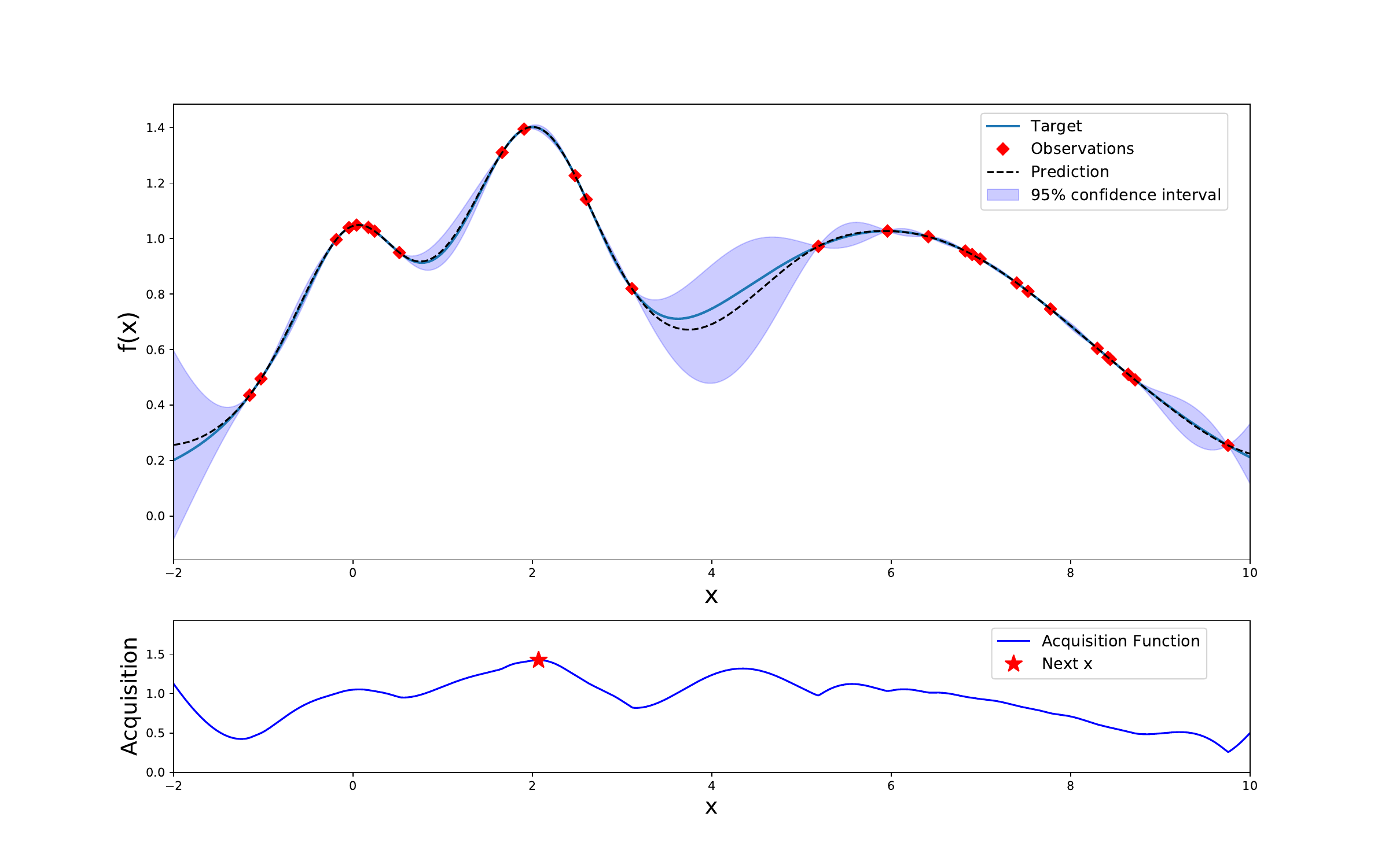}
\caption{Bayesian optimization of the function $f(x) = e^{-(x-2)^{2}} + e^{-{(x-6)^{2}}/{10}} + {1}/{x^{2}+1}$ in the search range $[-2, 10]$. The global maximum of $f$ is $x=2$ and is found after 27 iterations.
}\label{fig:BayOpt_example}
\end{figure}

Fig. \ref{fig:BayOpt_example} shows the Bayesian optimization of the function $f(x) = e^{-(x-2)^{2}} + e^{-{(x-6)^{2}}/{10}} + {1}/{x^{2}+1}$ that has a global maximum in $x=2$. The blue curve indicates the target (true function $f$), the red squares are the observations $(x_n, f(x_n))$, the dashed line represents the predicted mean of $f$ and the purple area is a $95\%$ confidence interval for the target. The acquisition function in the lower part of the figure suggests a new maximizer $x_{28}$ (red star) at iteration $n=27$ that is very close to the global optimum.

The Bayesian optimization algorithm transforms the optimization problem described in Eq. (\ref{eq:BayOpt_def}) to the optimization problem presented in Eq. (\ref{EI_argmax}), which is easier to solve. Many methods can be used to optimize Eq. (\ref{EI_argmax}), such as Quasi-Newton methods \citep{Quasi_Netwon}. Quasi-Newton methods like the Broyden–Fletcher–Goldfarb–Shanno algorithm (BFGS) \citep{BFGS} can iteratively solve unbounded optimization problems which the function is non-smooth. As can be understood from its name, Limited-memory BFGS with Bounds (L-BFGS-B) \citep{L-BFGS-B} is another commonly used variant of BFGS with limited memory and bounds. {Algorithm} \ref{alg:BayOpt} summarizes the procedure of Bayesian optimization to maximize the objective function $f$ with imposed bounds on $\mathbf x$. 

\bibliographystyle{plainnat}

\begin{algorithm}[h]
\caption{Bayesian optimization of $f(\mathbf x)$ with bounds on $\mathbf x$}
\begin{algorithmic}[1]
\renewcommand{\algorithmicrequire}{\textbf{Input:}}
\renewcommand{\algorithmicensure}{\textbf{Output:}}

\Require objective function $f$ to maximize, $N$ as the number of iterations, bounds for each element in $\mathbf{x}$ for L-BFGS-B, number of different initial guesses of L-BFGS-B $N_{s}$, $\xi$ as the exploration parameter of expected improvement

\Ensure $\mathbf{x}_{{*}}$, $f(\mathbf{x}_{{*}})$

\State \textbf{Initialization:} Initialize $\mathcal D$ as an empty list, the maximal function value so far as $f(\mathbf{x}_{*}) = -\infty$, $N_{s}$ initial input vectors within bounds for L-BFGS-B

\For{ $i = 1, \ldots, N$} 

\State Use L-BFGS-B with $N_{s}$ different initial inputs to optimize $\mathbb{E}(I)$ and assign the optimum to $\mathbf{x}_{{i}}$:

$\begin{aligned}
&\mathbb{E}(I) = 
\begin{cases}
      \big( \mu_{{i-1}}(\mathbf{x}) -f(\mathbf{x}_{*}) - \xi \big) \Phi(Z) + \sigma_{{i-1}}(\mathbf{x}) \phi(Z), & \text{if}\ \sigma_{{i-1}}(\mathbf{x})>0\\
      0, & \text{if}\ \sigma_{{i-1}}(\mathbf{x})=0
    \end{cases} \\
&Z = \frac{ \mu_{{i-1}}(\mathbf{x}) - f(\mathbf{x_{*}}) - \xi }{\sigma_{{i-1}}(\mathbf{x})}
\end{aligned}$

\State $\mathbf{x}_{{i}} = argmax_{\mathbf{x}} \  \mathbb{E}(I) $

\State Update the mean  $\mu_{{i}}(\mathbf{x}_{{i}})$ and covariance $\sigma_{{i}}^{2}(\mathbf{x}_{{i}})$ of the posterior distribution:

 $\begin{aligned}
&p(f(\mathbf{x}_{{i}}) \ | \ {D}_{{1:i-1}}, \mathbf{x}_{{i}}) = \mathcal{N}(\mu_{{i}}(\mathbf{x}_{{i}}), \  \sigma_{{i}}^{2}(\mathbf{x}_{{i}})) \\
&\mu_{{i}}(\mathbf{x}_{{i}}) = \bm K(\mathbf{x}_{{i}} \ , \ \mathbf{X}_{{1:i-1}})^{T}\bm K(\mathbf{X}_{{1:i-1}} \ , \ \mathbf{X}_{{1:i-1}})^{-1}f(\mathbf X_{{1:i-1}}) \\ 
&\sigma_{{i}}^{2}(\mathbf{x}_{{i}}) =  k(\mathbf{x}_{{i}} \ , \ \mathbf{x}_{{i}}) - \bm K (\mathbf{x}_{{i}} \ , \  \mathbf X_{{1:i-1}})^{T} \bm K (\mathbf X_{{1:i-1}} \ , \ \mathbf X_{{1:i-1}})^{-1}\bm K (\mathbf X_{{1:i-1}} \ , \ \mathbf{x}_{{i}})
\end{aligned}$

\State the kernel $k$ used is the Matern class function with $\nu=5/2$ and $l=1$:

$\begin{aligned}
&k_{{\nu=5/2, \ l=1}}(\mathbf x, \mathbf x') = \Big(1 + {\sqrt{5} ||\mathbf x - \mathbf x'||} + \frac {\sqrt{5} ||\mathbf x - \mathbf x'||^{2}} {3^{2}} \Big) \ \text{exp} \Big(-  {\sqrt{5} ||\mathbf x - \mathbf x'||} \Big) \\
&\bm K (\mathbf{X}_{{1:i-1}} \ , \ \mathbf{X}_{{1:i-1}}) = 
\begin{bmatrix} 
k(\mathbf{x}_{{1}}, \mathbf{x}_{{1}}) & \ldots & k(\mathbf{x}_{{1}}, \mathbf{x}_{{i-1}}) \\
\vdots & \ddots & \vdots \\
k(\mathbf{x}_{{i-1}}, \mathbf{x}_{{1}}) & \ldots & k(\mathbf{x}_{{i-1}}, \mathbf{x}_{{i-1}}) \\
\end{bmatrix} 
\end{aligned}$

\State Calculate $f(\mathbf{x}_{{i}})$ from objective function $f$

\If{ $f(\mathbf{x}_{{i}}) > f(\mathbf{x}_{{*}})$}

\State 
$\begin{aligned}
&\mathbf{x}_{{*}} = \mathbf{x}_{{i}} \\ 
&f(\mathbf{x}_{*}) = f(\mathbf{x}_{{i}})
\end{aligned}$

\EndIf

\State Add $\big(\mathbf{x}_{{i}}, {f}{(\mathbf{x}_{{i}})}\big)$ to observations data set $\mathcal{D}_{{1:i-1}}$

\State Update the covariance matrix of the Gaussian process model by calculating:

 $\bm K(\mathbf{X}_{{1:i}} \ , \ \mathbf{X}_{{1:i}}) = \begin{bmatrix} \bm{K}(\mathbf X_{{1:i-1}} , \mathbf{X}_{{1:i-1}}) & \bm{K}(\mathbf X_{{1:i-1}} \ , \ \mathbf{x}_{{i}}) \\ \bm{K}(\mathbf{x}_{{i}} \ , \ \mathbf X_{{1:i-1}}) & k( \mathbf{x}_{{i}} \ , \ \mathbf{x}_{{i}}) \\ \end{bmatrix}$

\EndFor

\State Return the maximizer $\mathbf{x}_{{*}}$ and maximum value $f(\mathbf{x}_{{*}})$

\end{algorithmic}\label{alg:BayOpt}
\end{algorithm}

\subsection{Layerwise training of RegPred Net with Bayesian optimization} \label{sec:layerwise_training}

Since the number of dimensions for the variable $\mathbf x$ that can be handled by the Bayesian optimization procedure is practically limited to roughly $20$ \citep[Sec. 1]{BayOpt_dimension} and $d(1)=3$, $d(2)=21$ but $d(3)=903$, we are constrained to a maximum of $K=2$ layers. Also, it is difficult to train simultaneously the hyperparameters of layers $k=1$ and $k=2$, as this represents a total of $24$ parameters. Thus, we adopt a layerwise training strategy and tune the hyperparameters of the RegPred Net for only one layer at a time. We first consider a single-layer network ($k=1$) and use Bayesian optimization to find the optimal hyperparameters in that layer. Then, the learned parameters of the first layer are fixed and we add a second layer to the network ($k=2$) and train only the newly added hyperparameters. We get the optimal parameters for the second layer and stop. The layerwise training procedure is summarized in {Algorithm} \ref{alg:layerwise_training}.

\begin{algorithm}[!htbp]
\caption{Layerwise training of RegPred Net with $K\leq 2$ layers by Bayesian optimization}
\begin{algorithmic}[1]
\renewcommand{\algorithmicrequire}{\textbf{Input:}}
\renewcommand{\algorithmicensure}{\textbf{Output:}}

\Require number of layers $K$, input time series $\mathbf{Y}_{0:T+N}$

\Ensure optimal hyperparameters $\mathbf{x}^{{(1\, : \, K)}} _{{*}} = \big\{ \mathbf{H}^{{(1\,: \,K)}}_{{*}}, \ \mathbf{Z}^{{(1\, : \,K)}}_{{0 \, *}} \big\}$ for the $K$ layers of RegPred Net

\For{$k = 1, \ldots, K$}

\State Add layer $k$ to RegPred Net with $\mathbf{x}^{(k)}$ as hyperparameters

\State Use Bayesian optimization in Algorithm \ref{alg:BayOpt}  with $f = -L_{avg} = -(L^{\mathbb{E}}_{avg} + L^{\mathbb{V}}_{avg})$ as objective function to find the k-th layer's optimal parameters $\mathbf{x}^{{(k)}}_{{*}}$, where the average losses $L^{\mathbb{E}}_{avg}$ and $L^{\mathbb{V}}_{avg}$ are computed by {Algorithm} \ref{alg:avg_loss}

\EndFor

\end{algorithmic} \label{alg:layerwise_training}
\end{algorithm}

\section{{Experimental validation}} \label{sec:intro_experiments}

In this section, we evaluate the performance of RegPred Net and compare it to other time series forecasting models. The data set used for FX rates is described in Sec. \ref{sec:dataset}. The models compared to RegPred Net include Deep Learning models ({LSTM}, {Auto-LSTM}) as well as traditional time series models ({ARMA}, {ARIMA}). We gather and analyze all performance results in Sec. \ref{experimental_results}. The {RegPred Net}, {LSTM} and {Auto-LSTM} were implemented in Python's Deep Learning Framework \textit{Tensorflow} \citep{tensorflow2015} and {ARMA} and {ARIMA} were implemented using \textit{statsmodels} \citep{seabold2010statsmodels}.

\subsection{{Data set}} \label{sec:dataset}

The data set used covers three major FX rates: {EUR/CNY}, {EUR/USD}, and {EUR/GBP}. The historical data for the corresponding time series span from {2000.01.04} to {2019.01.29}, amounting to 19 years of data and 4975 daily observations per time series. The data used in the experiments are daily closing values of the FX rates provided by Bloomberg.

\subsection{Experimental setting}

For each FX rate considered, the historical data are used to generate $n=95$ samples in window steps of $30$ days, as illustrated in Fig. \ref{fig:data_preprocessing}. Each sample contains a portion of the data over a window of $N_{train,valid}+N_{test}$ days, whereby the first $N_{train,valid}$ days are used for training and validating the model considered in the experiment and the last $N_{test}=100$ days are used for testing the performance of the model. The first $N_{train,valid}$ days of a sample are in turn sub-divided into $N_{train}=1830$ days for training and $N_{valid}=200$ days for validation. Then, for any given model and performance metric considered, the arithmetic mean of the metric is estimated on the basis of calculated values in the $n$ samples and serves as performance statistics for the experimental validation.
\tikzstyle{Y_block} = [draw, rectangle, fill=cyan!15, text width=15 em, 
                             text centered, minimum height=5mm,
                             font=\fontsize{8}{0}\selectfont, node distance=5em]
\tikzstyle{Y_L_block} = [draw, rectangle, fill=yellow!15, text width=8 em,  text centered, minimum height=5mm, font=\fontsize{8}{0}\selectfont, node distance=5em]
\tikzstyle{Y_P_block} = [draw, rectangle, fill=black!15, text width=4 em,  text centered, minimum height=5mm, font=\fontsize{8}{0}\selectfont, node distance=5em]
\tikzstyle{blank} = [ fill=white,  minimum height=1mm, font=\fontsize{8}{0}\selectfont, node distance=1em]

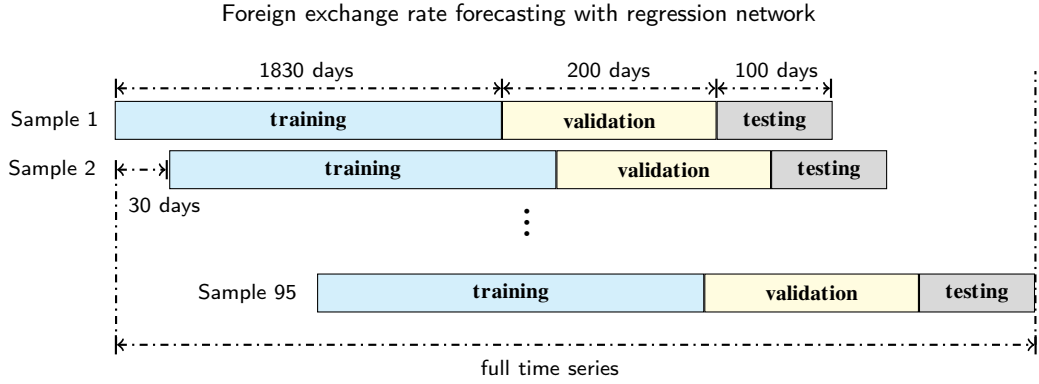
\begin{figure}[pos=htpb, width=12cm, align=\centering] \label{fig:data_preprocessing}
\centering
\captionsetup{font=footnotesize}
\begin{tikzpicture}

\node [black, Y_block] (instance_1) 
{
$\mathbf{training}$
  };

\node [black, blank, left=0.1cm of instance_1] (instance_1_left) 
{
Sample 1
  };

\node [black, Y_L_block, right=0cm of instance_1] (instance_1_2) 
{
$\mathbf{validation}$
  };

\node [black, Y_P_block, right=0cm of instance_1_2] (instance_1_3) 
{
$\mathbf{testing}$
  };

\node[black, text width=6em] at (1em,2em) {\footnotesize{1830 days}};
\draw[black, |<->|, thick, dash dot] (-7.85em, 1.3em) -- (7.9em, 1.3em);

\node[black, text width=6em] at (13.5em,2em) {\footnotesize{200 days}};
\draw[black, <->|, thick, dash dot] (7.9em, 1.3em) -- (16.6em, 1.3em);

\node[black, text width=6em] at (20.3em,2em) {\footnotesize{100 days}};
\draw[black, <->|, thick, dash dot] (16.6em, 1.3em) -- (21.25em, 1.3em);
 \node [black, blank] (blank_1) at (-6em,-2em) {};

 \node [black, Y_block, right=0cm of blank_1] (instance_2)
{
$\mathbf{training}$
  };

\node [black, blank, left=0.85cm of instance_2] (instance_2_left) 
{
Sample 2
  };

\node [black, Y_L_block, right=0cm of instance_2] (instance_2_2) 
{
$\mathbf{validation}$
  };

\node[black, blank, below left=0.2cm of instance_2_2] {\Large$\vdots$};

\node [black, Y_P_block, right=0cm of instance_2_2] (instance_2_3) 
{
$\mathbf{testing}$
  };

\node[black, text width=6em] at (-4.3em, -3.5em) {\footnotesize{30 days}};
\draw[black, |<->|, thick, dash dot] (-7.85em, -2em) -- (-5.7em, -2em);

 \node [black, blank] (blank_2) at (0em,-7em) {};
 \node [black, Y_block, right=0cm of blank_2] (instance_n)
{
$\mathbf{training}$
  };
\node [black, blank, left=0.15cm of instance_n] (instance_n_left) 
{
Sample 95
  };
\node [black, Y_L_block, right=0cm of instance_n] (instance_n_2) 
{
$\mathbf{validation}$
  };
\node [black, Y_P_block, right=0cm of instance_n_2] (instance_n_3) 
{
$\mathbf{testing}$
  };

\draw[black, thick, dash dot] (29.5em, -9.1em) -- (29.5em, 2em);
\draw[black, thick, dash dot] (-7.8em, -9.1em) -- (-7.8em, -2.5em);
\draw[black, |<->|, thick, dash dot] (-7.85em, -9.1em) -- (29.55em, -9.1em);
\node[black, text width=6em] at (10em,-10em) {\footnotesize{full time series}};
 \end{tikzpicture}
\caption{Sampling of the historical data used for computing performance statistics.}
\label{fig:data_preprocessing}
\end{figure}

\subsection{{Computing Infrastructure}} \label{sec:ci}
The computing infrastructure used in this work are one computer with Intel Core(TM) i7-6700K (4.00 GHz) CPU and Nvidia GeForce GTX 970 (6GB) GPU. 

\subsection{Setting for the models compared}
In this section, we explain how we set the experiments for different model comparisons.

\subsubsection{{RegPred Net}}\label{sec:experiment_RegPred}

We set the number of layers to ${K}=2$ and train RegPred Net layerwise according to {Algorithm} \ref{alg:layerwise_training}. The number of generated trajectories for each Monte Carlo simulation is $n_{W}=50$. For optimizing the acquisition function, {L-BFGS-B} is used with $5$ restart times and the bounds used are those detailed in Tab. \ref{tab:bounds_}. We set the number of iterations for Bayesian optimization to $D=200$. For the exploration parameter, we use $\xi = 0.01$ for {EUR/CNY} and {EUR/GBP} and $\xi = 0.05$ for {EUR/USD}. Tab. \ref{tab:bounds_} shows the bounds we used in L-BFGS-B algorithm for finding the optimal hyperparameters of RegPred Net for different samples of time series by Bayesian optimization. 

\begin{table}[width=.6\linewidth, cols=5, pos=htpb]
\centering
\caption{Bounds in the form of [min, max] used for finding the optimal hyperparameters of RegPred Net by {Bayesian optimization}. }
\begin{tabular*}{\tblwidth}{@{} LLLLL @{} }
\toprule
 & ${A}_{{0}}$, ${N}_{{0}}$ & {$\Sigma_{{0}}$} & $\eta_{A}, \eta_{N}, \eta_{\Sigma}$ & $\varphi, \rho$ \\
\midrule
Layer $1$ & [-0.3, 0.3] & [0.001, 0.01] & [0.001, 0.3] & [0.1, 1.0] \\
Layer $2$ & [-0.1, 0.1] & [-0.001, 0.001] & [0.001, 0.3] & [0.1, 1.0] \\
\bottomrule
\end{tabular*}  \label{tab:bounds_}
\end{table}


\subsubsection{LSTM and Auto-LSTM}
\label{sec:experiment_deeplearning}
Two Deep Learning models are considered: the LSTM and Auto-LSTM. For the LSTM, we test the architecture of \citep{Auto-LSTM} in both single-shot (predict in once) and autoregressive way (predict stepwisely), the single-shot way failed on predicting long-term multi-steps time series forecasting task. We performed optimization of the LSTM's hyperparameters by grid search, considering a number of layers ranging from $1$ to $5$, learning rates of $10^{-2}$, $10^{-3}$ and $10^{-4}$, LSTM cells with $32, 64$ and $128$ units. The LSTM with the best results is illustrated in part $(a)$ of Fig. \ref{fig:LSTM_Auto}, where each layer is described by layer type and index / layer size (units) / activation function. We connect the output of the last time steps of the LSTM $3$ with a dense layer to generate the prediction. In autoregressive mode the network only predicts 1 step at each time and predicts $N_{valid} = 200$ and $N_{test} = 100$ times.
We compared the sigmoid and relu activation functions for the last dense and chose the sigmoid for its superior performance. During training, an early stopping technique with patience equals to 50 was used.
\tikzstyle{block_reg} = [draw, 
                             rectangle, 
                             fill=cyan!15,
                             text width=15em, 
                             text centered, 
                             minimum height=1.5em,
                             font=\fontsize{8}{0}\selectfont, 
                             node distance=5em]
\tikzstyle{block_0} = [draw, 
                             rectangle, 
                             fill=yellow!15,
                             text width=18em, 
                             text centered, 
                             minimum height=1.5em,
                             font=\fontsize{8}{0}\selectfont, 
                             node distance=5em]
\tikzstyle{block_1} = [draw, 
                             rectangle, 
                             fill=yellow!15, 
                             text width=14em, 
                             text centered, 
                             minimum height=1.5em,
                             font=\fontsize{8}{0}\selectfont, 
                             node distance=5em]
\tikzstyle{block_2} = [draw, 
                             rectangle, 
                             fill=yellow!15, 
                             text width=12em, 
                             text centered, 
                             minimum height=1.5em,
                             font=\fontsize{8}{0}\selectfont, 
                             node distance=5em]
\tikzstyle{block_3} = [draw, 
                             rectangle, 
                             fill=yellow!15, 
                             text width=10em, 
                             text centered, 
                             minimum height=1.5em,
                             font=\fontsize{8}{0}\selectfont, 
                             node distance=5em]

\tikzstyle{blank} = [fill=white, text width=15em, text centered, minimum height=1mm, font=\fontsize{9}{0}\selectfont, node distance=6em]
\tikzstyle{blank2} = [fill=white, text width=1em, text centered, minimum height=1mm, font=\fontsize{9}{0}\selectfont, node distance=1em]

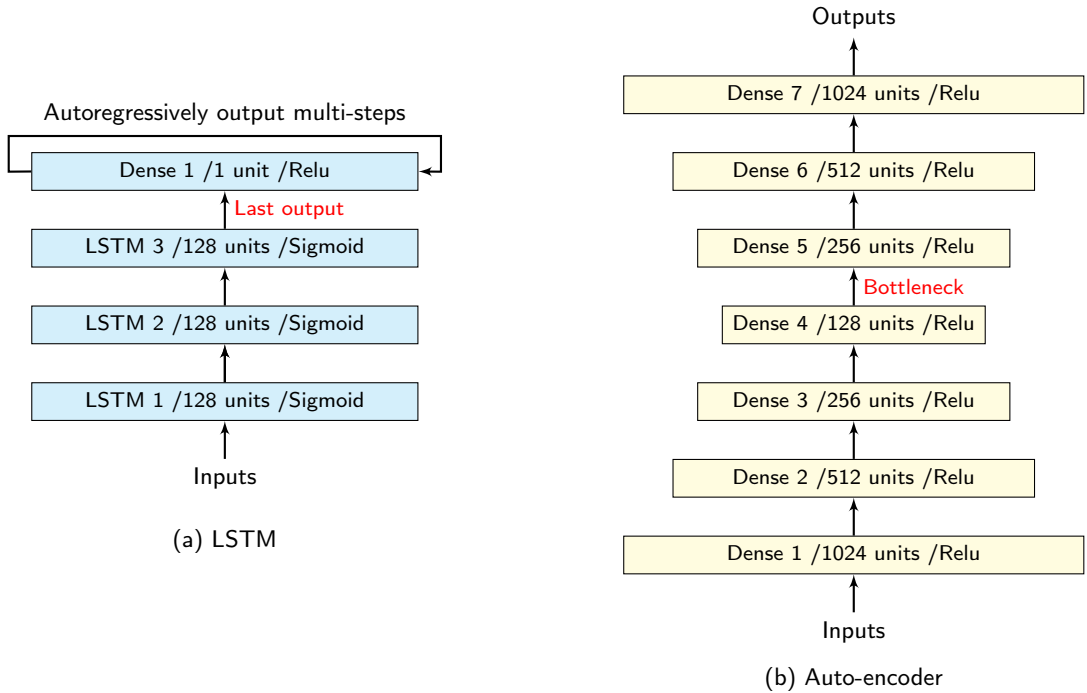
\begin{figure}[pos=htpb, width=15cm, align=\centering] 
\begin{tikzpicture} 
\node [black, block_reg] (layer_1) 
{
 LSTM 1 /128 units /Sigmoid
};

\node [black, block_reg, above=0.5cm of layer_1] (layer_2) 
{
LSTM 2 /128 units /Sigmoid
};

\node [black, block_reg, above=0.5cm of layer_2] (layer_3) 
{
 LSTM 3 /128 units /Sigmoid
};

\node [black, block_reg, above=0.5cm of layer_3] (layer_4) 
{
 Dense 1 /1 unit /Relu
};
\node [black, blank, below=0.5cm of layer_1] (input) {Inputs};
\node [black, blank, below=0.3cm of input] (a) {(a) LSTM};
\node [black, blank, above=0.25cm of layer_4] (output) {Autoregressively output multi-steps};

\draw [black, ->, thick, -latex'] (layer_1) -- (layer_2);
\draw [black, ->, thick, -latex'] (layer_1) --  (layer_2);
\draw [black, ->, thick, -latex'] (layer_2) -- (layer_3);
\draw [black, ->, thick, -latex'] (layer_3) -- node [right, midway, font=\fontsize{8}{0}\selectfont, text=red] {Last output} (layer_4);
\draw [black, ->, thick, -latex'] (input) --  (layer_1);
\draw [black, ->, thick, -latex'] (layer_4.west) -- (-2.85, 3.03) -- (-2.85, 3.5) -- (2.85, 3.5) -- (2.85, 3.03) -- (layer_4.east);

\node [black, block_3, right=4cm of layer_2] (bottleneck) 
{
Dense 4 /128 units /Relu
};

\node [black, block_2, below=0.5cm of bottleneck] (au_3) 
{
Dense 3 /256 units /Relu
};

\node [black, block_1, below=0.5cm of au_3] (au_2) 
{
Dense 2 /512 units /Relu
};

\node [black, block_0, below=0.5cm of au_2] (au_1) 
{
Dense 1 /1024 units /Relu
};

\node [black, block_2, above=0.5cm of bottleneck] (au_4) 
{
Dense 5 /256 units /Relu
};

\node [black, block_1, above=0.5cm of au_4] (au_5) 
{
Dense 6 /512 units /Relu
};

\node [black, block_0, above=0.5cm of au_5] (au_6) 
{
Dense 7 /1024 units /Relu
};
\node [black, blank, below=0.5cm of au_1] (input) {Inputs};
\node [black, blank, below=0.1cm of input] (b) {(b) Auto-encoder};
\node [black, blank, above=0.5cm of au_6] (output) {Outputs};

\draw [black, ->, thick, -latex'] (bottleneck) -- node [right, midway, font=\fontsize{8}{0}\selectfont, text=red] {Bottleneck} (au_4);
\draw [black, ->, thick, -latex'] (au_4) --  (au_5);
\draw [black, ->, thick, -latex'] (au_5) --  (au_6);
\draw [black, ->, thick, -latex'] (au_1) --  (au_2);
\draw [black, ->, thick, -latex'] (au_2) --  (au_3);
\draw [black, ->, thick, -latex'] (au_3) --  (bottleneck);
\draw [black, ->, thick, -latex'] (input) --  (au_1);
\draw [black, ->, thick, -latex'] (au_6) --  (output);
\end{tikzpicture}
\caption{Architecture of the LSTM and Autoencoder part of the Auto-LSTM used in the experimental validation.}
\label{fig:LSTM_Auto}
\end{figure}

To build an {Auto-LSTM}, we stacked the auto-encoder illustrated in part $(b)$ of Fig. \ref{fig:LSTM_Auto} on top of the {LSTM} shown in part $(a)$. The auto-encoder was pre-trained and used to extract features from the input time series. We then fed the time series into the auto-encoder and used the extracted features from the middle hidden layer (part $(b)$, bottleneck) as inputs for the LSTM.

\subsubsection{ARMA and ARIMA}
\label{sec:experiment_ARIMA}

Two of the most important statistical models for time series forecasting are considered: Autoregressive Moving Average ({ARMA}) \citep{ARMA_ori} and Autoregressive Integrated Moving Average ({ARIMA}) \citep{ARIMA_ori}. An {ARMA} model with orders $p$ and $q$ as hyperparameters is denoted {ARMA}(p, q) and is of the following form:  
\begin{equation} \label{ARMA}
X_{t} = c + \epsilon_{t} + \sum_{i=1}^{p} \varphi_{i}X_{t-i} + \sum_{i=1}^{q}\theta_{i}\epsilon_{t-i}
\end{equation}
where $X_{t}$ is the value of time series at time $t$, $c$ is a constant, $\epsilon_{t}$ is a noisy term whose values are assumed to be i.i.d. and normally distributed, $\varphi_{1}, \ldots, \varphi_{p}$ are $p$ parameters for the autoregressive part (AR) and $\theta_{1}, \ldots, \theta_{q}$ are $q$ parameters for the moving average part (MA) of the time series.
{ARIMA} is a generalized version of {ARMA}. {ARIMA} uses an additional hyperparameter $d$ that plays the role of number of differencing steps required to make the time series stationary. The differencing computes the differences between consecutive observations. This helps stabilize the time series and eliminate the trend. Therefore, an {ARIMA} is represented in the form {ARIMA}$(p, d, q)$. For the choice of $(p, d, q)$ order for the {ARIMA}, we referred to \citep{ARIMA_pdq_1} and also compared several values for $p$, $d$, and $q$. 

\subsection{Experimental results} \label{experimental_results}

The performance statistics are here-after presented separately for the three FX rates EUR/CNY, EUR/USD and EUR/GBP in Sec. \ref{sec:experimental_results_EUR_CNY},  \ \ref{sec:experimental_results_EUR_USD},  and \ref{sec:experimental_results_EUR_GBP}, respectively. The metrics we use to evaluate the forecasting performance of the models are the \textit{Pearson} correlation coefficient (\textit{Pearson}'s $R$), R-squared ($R^{2}$), Root mean square error ({RMSE}) and Mean directional accuracy (MDA). For RegPred Net, it took 5 minutes to train each single sample of each currency type in the data set using the 2 layers infrastructure described in Section \ref{sec:ci}. We use a batch size of the data set size of each currency to train LSTM and Auto-LSTM, it took in average 1000 iterations and 120 minutes to finish the training. For ARMA and ARIMA, we search the parameters $p, d, q$ from $0$ to $20$ and take the best of them for each sample, which costs around 5 minutes. 

\subsubsection{Experiment results of EUR/CNY} \label{sec:experimental_results_EUR_CNY}

According to the results reported in Tab. \ref{tab:CNY_results}, {RegPred Net} outperforms the other four models (LSTM, Auto-LSTM, ARMA and ARIMA) regardless of the performance metric considered. The RedPred Net's forecasts have a correlation (R) with the true value of the FX rate that is $2.2$ times higher than the second best correlated model (LSTM). RegPred Net has an error (RMSE) that is about $30\%$ lower than the error of the second most accurate model (ARIMA).  For MDA the gap between the methods is within $10\%$. For R-squared, all methods except RegPred are negative, which means that the long-term forecasts do not follow the trend of the actual values. Our dataset has several instances where exchange rates fall or rise sharply in the short term due to unexpected events (e.g., financial crises, wars). None of the models involved in the experiments incorporate mechanisms to deal with such situations and therefore can not predict such trends. When the forecasted FX rate has the same tendency as the actual exchange rate, R square is a number in the interval $(0, 1]$. In contrast, the R square will be a negative number with a large absolute value when the two trends are opposite. 
\begin{table}[cols=11, pos=htpb]
\centering
\caption{Comparison of forecasting performance obtained for {EUR/CNY}. The results are presented in form of mean $\pm$ std., the best results are indicated in bold font.}
\begin{tabular*}{\tblwidth}{@{} LLLLLL @{} }
\toprule
 & \multicolumn{1}{L}{\textit{Pearson's} R} & \multicolumn{1}{L}{\textit{R-squared}} & \multicolumn{1}{L}{RMSE} & \multicolumn{1}{L}{MDA} \\
\midrule
  RegPred Net & $\mathbf{0.344} \pm \mathbf{0.443} $ & $\mathbf{0.141} \pm \mathbf{0.862}$ & $\mathbf{0.225} \pm \mathbf{0.141}$ & $\mathbf{0.568} \pm \mathbf{0.046}$ \\
LSTM & ${0.156} \pm {0.638}$ & ${-4.888} \pm {6.264}$ & ${0.366} \pm {0.234}$ & ${0.517} \pm {0.080}$ \\
Auto-LSTM
&  ${0.141} \pm {0.691}$ 
& ${-5.120} \pm {6.330}$ 
& ${0.410} \pm {0.272}$ 
& ${0.488} \pm {0.084}$ \\
ARMA
& ${0.047}\pm {0.667}$
 & ${-2.359}\pm {2.571}$ 
 & ${0.323}\pm {0.230}$ 
 & ${0.502}\pm {0.048}$ \\
ARIMA 
& $0.036 \pm {0.668}$
 & $-2.279 \pm {2.740}$
  & ${0.318} \pm {0.225}$
   & $0.460 \pm {0.089}$ \\
\bottomrule
\end{tabular*}\label{tab:CNY_results}
\end{table}

\begin{figure}[pos=htpb, align=\centering]
\begin{minipage}{\linewidth}
\centering
\subfloat[RegPred, sample 2]{\includegraphics
[scale=0.36]{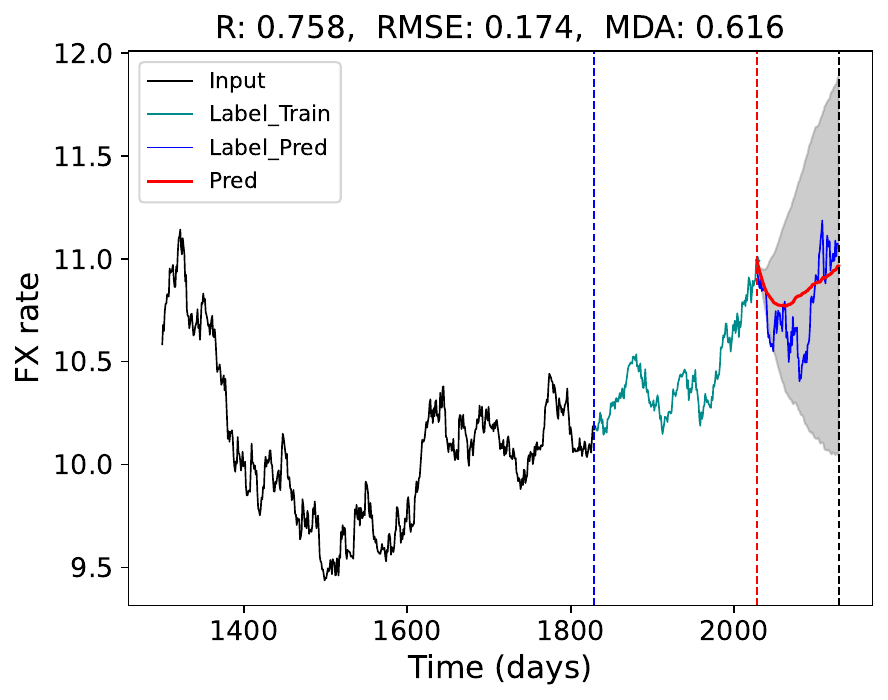}}
\subfloat[RegPred, sample 38]{\includegraphics
[scale=0.36]{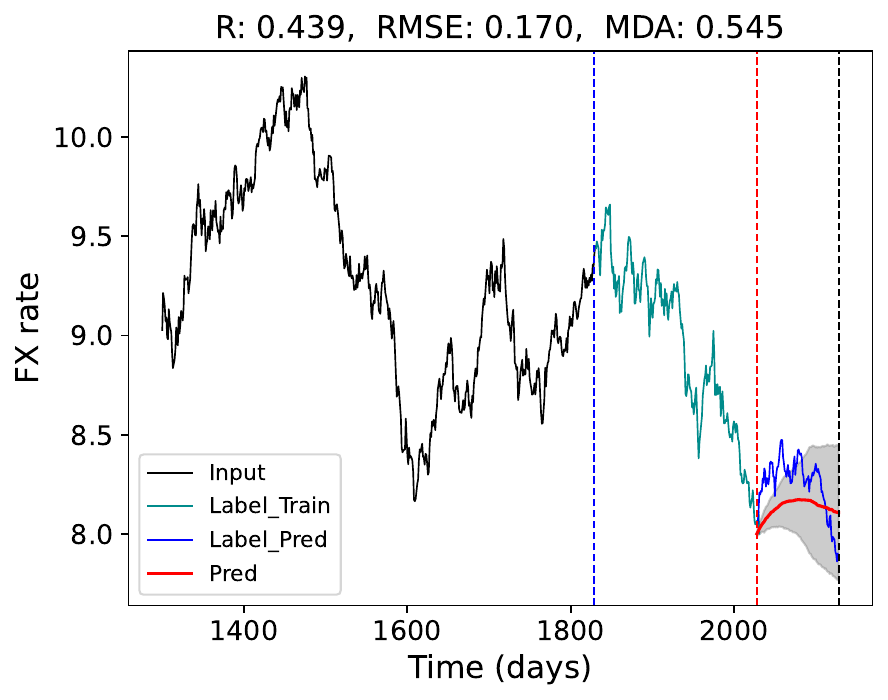}}
\subfloat[RegPred, sample 83]{\includegraphics
[scale=0.36]{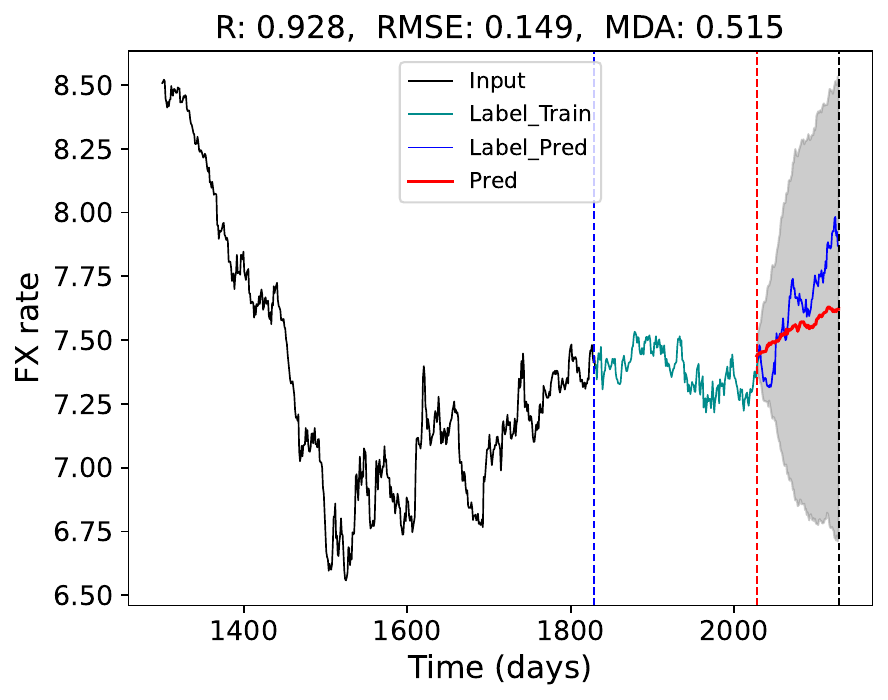}}
\end{minipage}%

\begin{minipage}{\linewidth}
\centering
\subfloat[LSTM (Autoregressive), sample 2]{\includegraphics[scale=0.36]{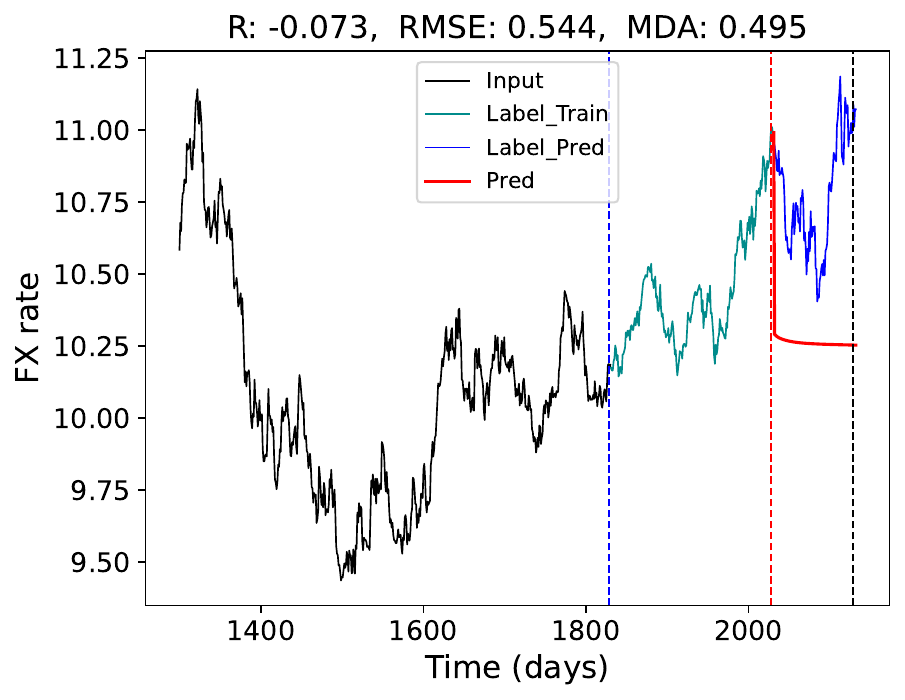}}
\subfloat[LSTM (Autoregressive), sample 38]{\includegraphics[scale=0.36]{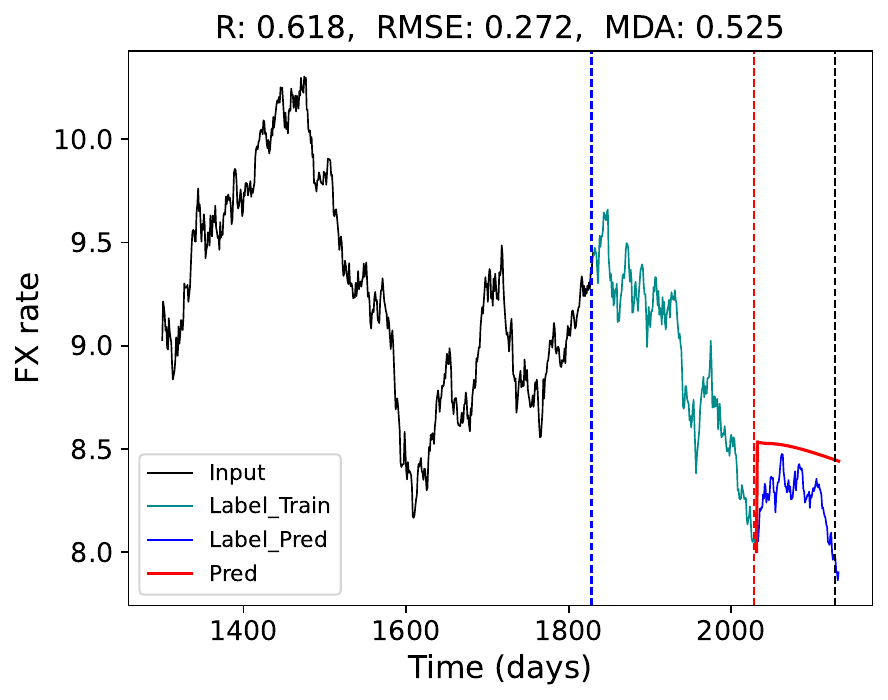}}
\subfloat[LSTM (Autoregressive), sample 83]{\includegraphics[scale=0.36]{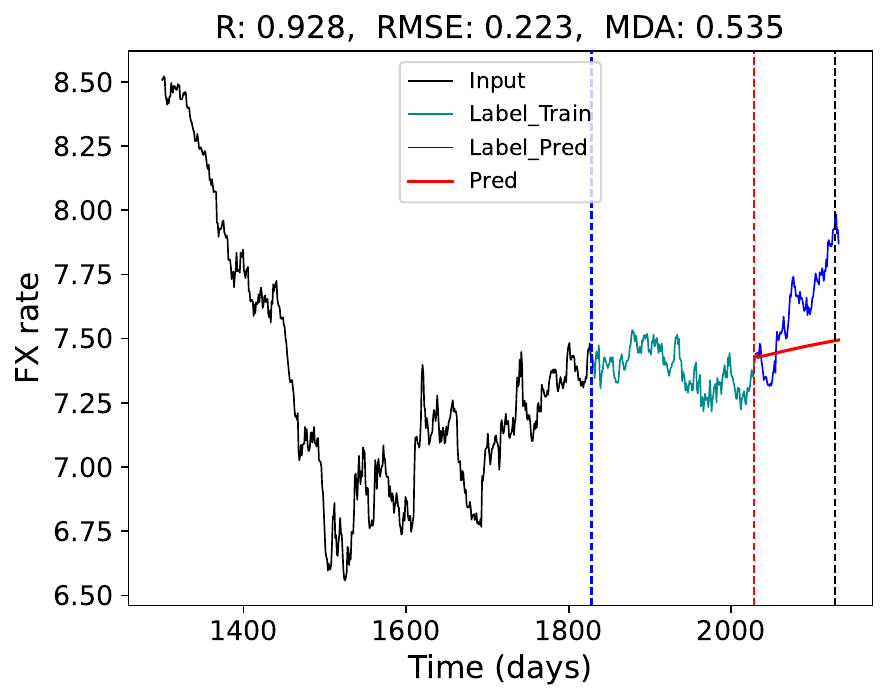}}
\end{minipage}

\begin{minipage}{\linewidth}
\centering
\subfloat[ARIMA, sample 2]{\includegraphics[scale=0.36]{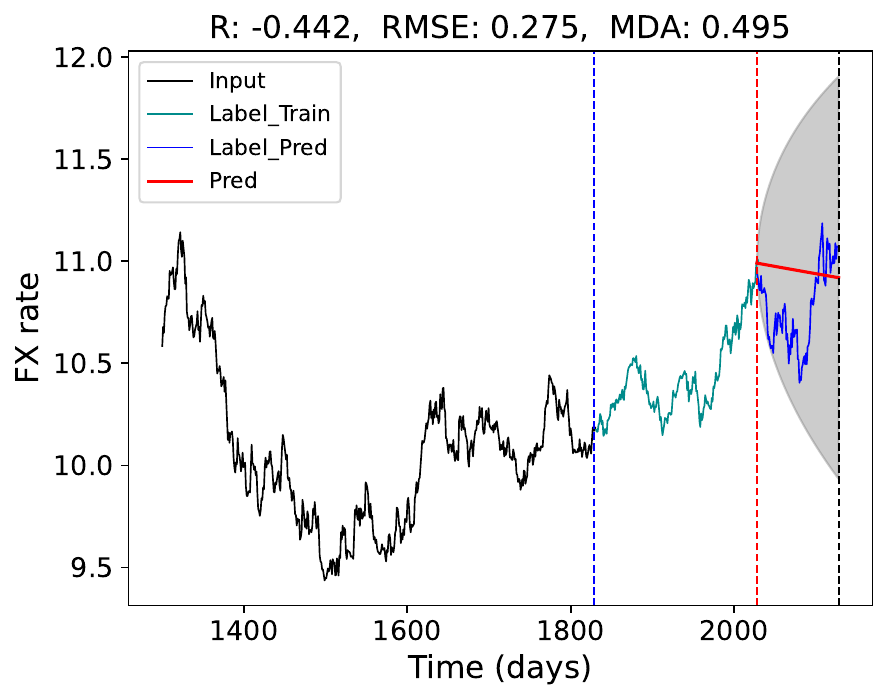}}
\subfloat[ARIMA, sample 38]{\includegraphics[scale=0.36]{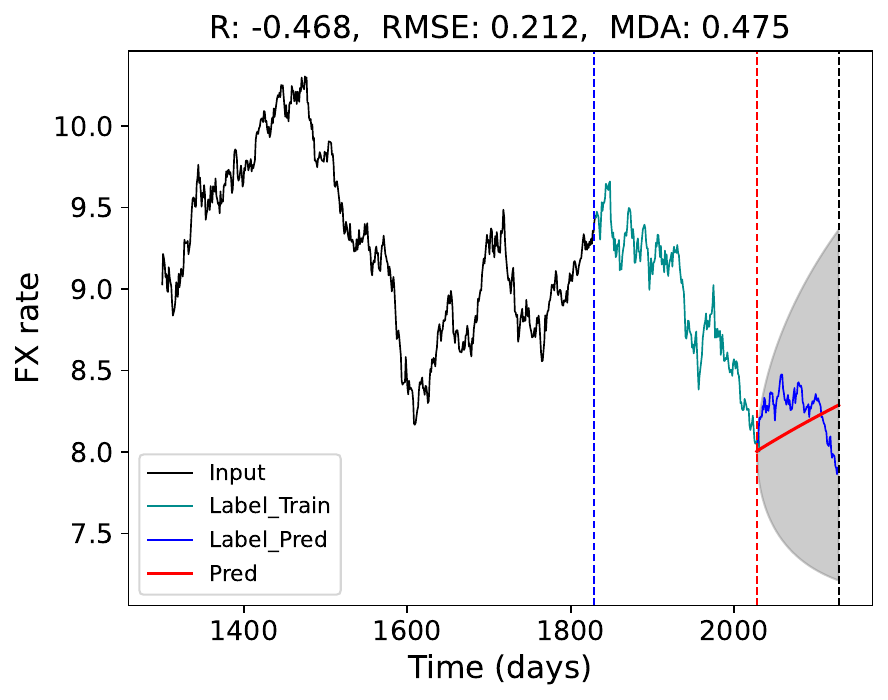}}
\subfloat[ARIMA, sample 83]{\includegraphics[scale=0.36]{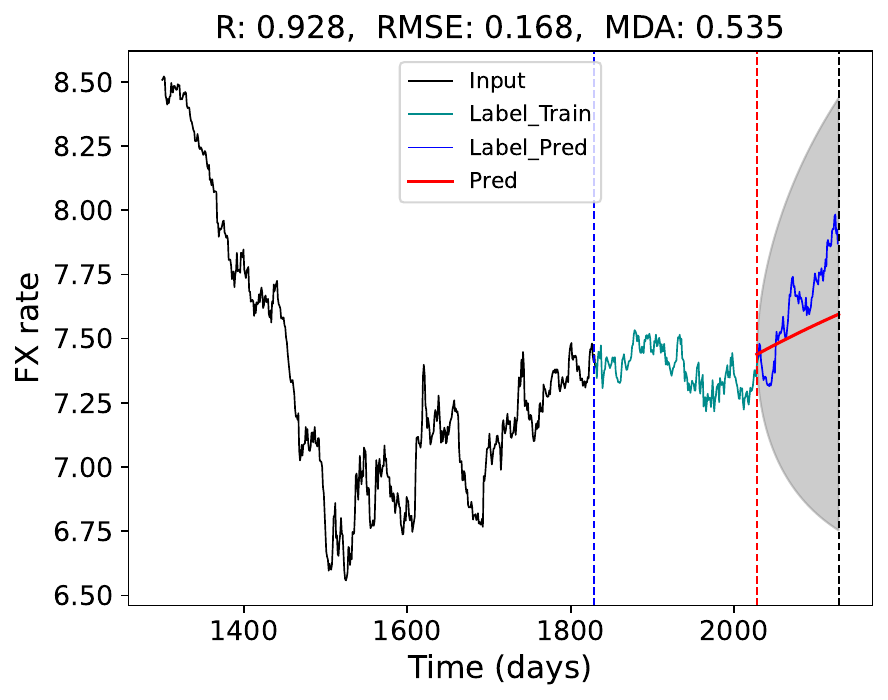}}
\end{minipage}
\caption{{Forecasts obtained for EUR/CNY with RegPred Net, LSTM (Single-shot) and ARIMA in $3$ situations.}}
\label{fig:CNY_pred}
\end{figure}

Fig. \ref{fig:CNY_pred} illustrates the forecasts obtained with RegPred Net, LSTM and ARIMA for 3 of the 95 data samples used. In the examples, we see the RegPred Net's better ability to predict the time-dependent shape and level of the FX rate's estimated trend (plotted in red) and the possibility offered by RegPred Net to interpret and visualize the FX rate's volatility via the $95\%$ confidence zone (colored in gray).

\subsubsection{Experiment results of EUR/USD} \label{sec:experimental_results_EUR_USD}

\begin{figure}[pos=htpb]
\begin{minipage}{\linewidth}
\centering
\subfloat[RegPred, sample 13]{\includegraphics
[scale=0.36]{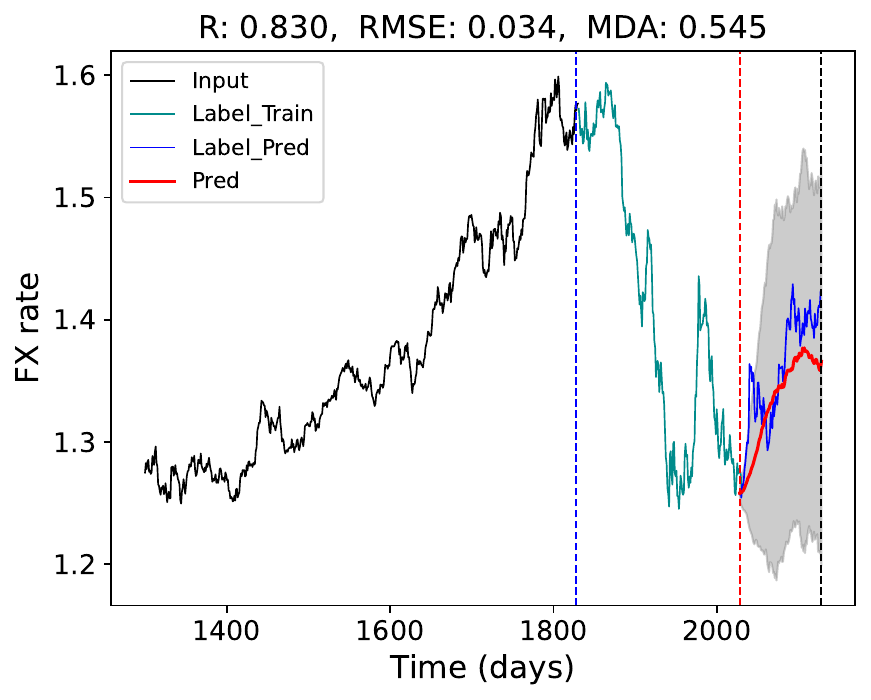}}
\subfloat[RegPred, sample 20]{\includegraphics
[scale=0.36]{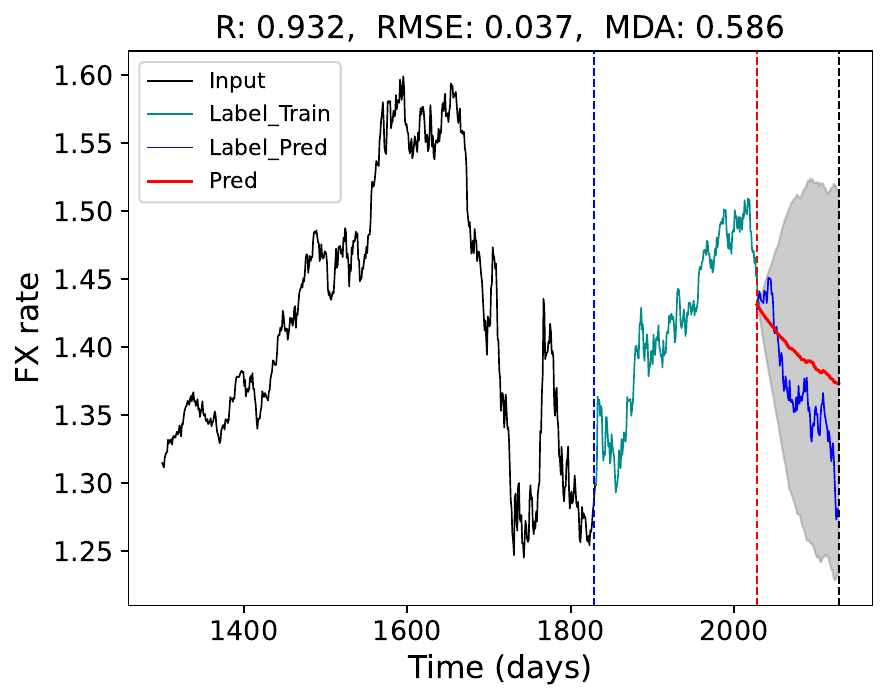}}
\subfloat[RegPred, sample 86]{\includegraphics
[scale=0.36]{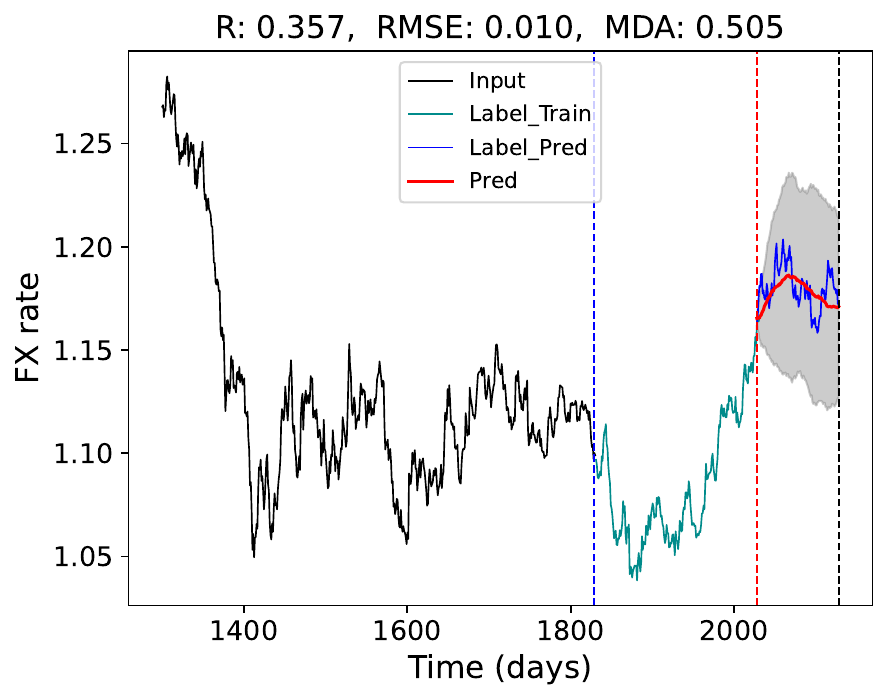}}
\end{minipage}%

\begin{minipage}{\linewidth}
\centering
\subfloat[LSTM sample 13]{\includegraphics[scale=0.36]{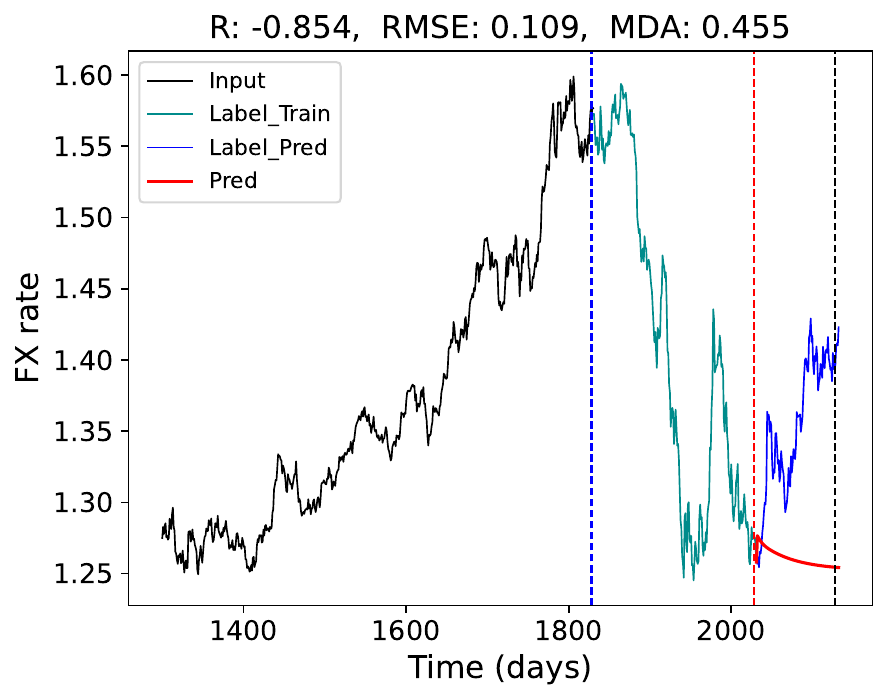}}
\subfloat[LSTM, sample 20]{\includegraphics[scale=0.36]{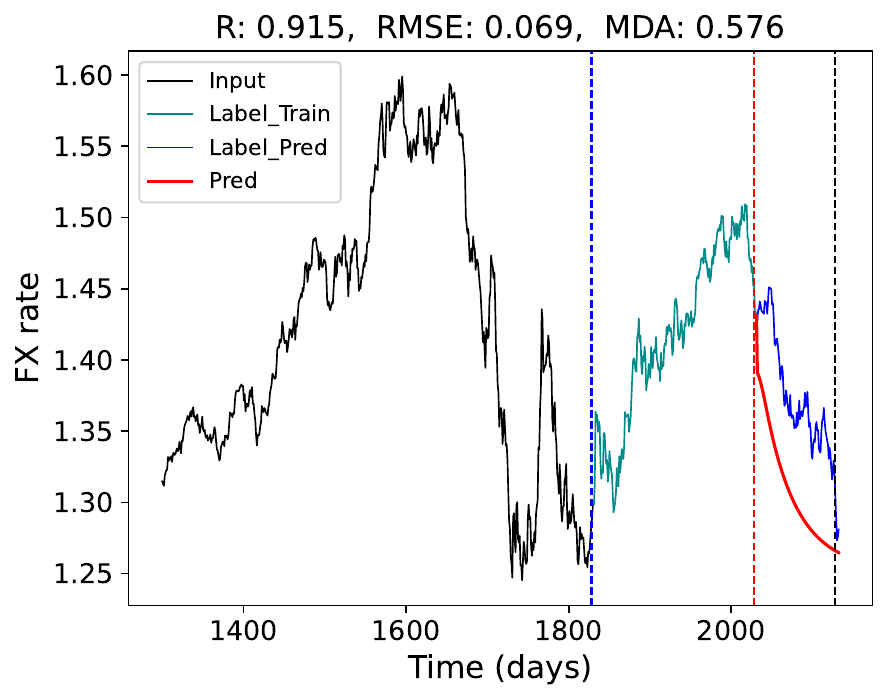}}
\subfloat[LSTM, sample 86]{\includegraphics[scale=0.36]{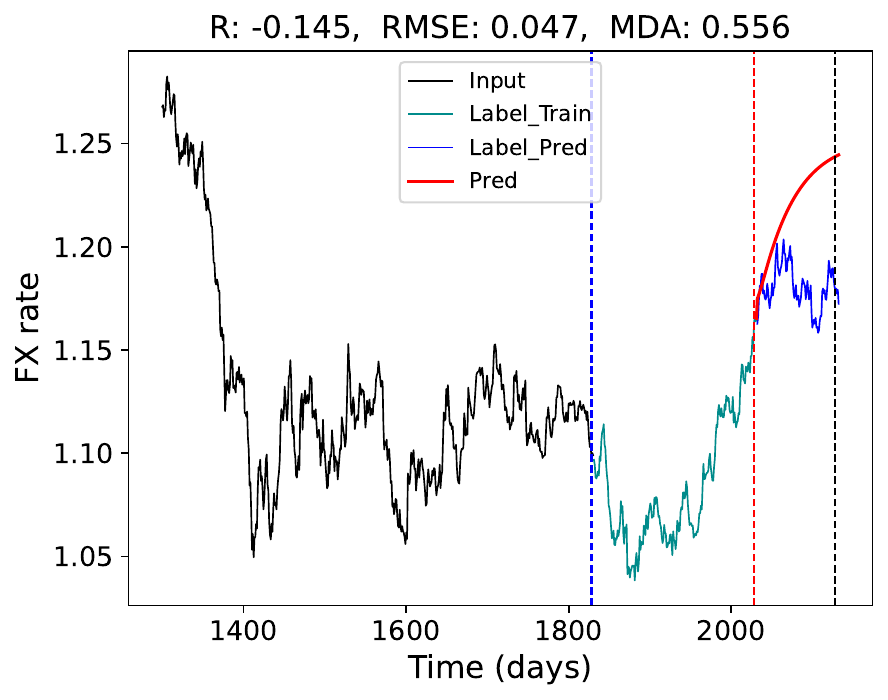}}
\end{minipage}

\begin{minipage}{\linewidth}
\centering
\subfloat[ARIMA, sample 13]{\includegraphics[scale=0.36]{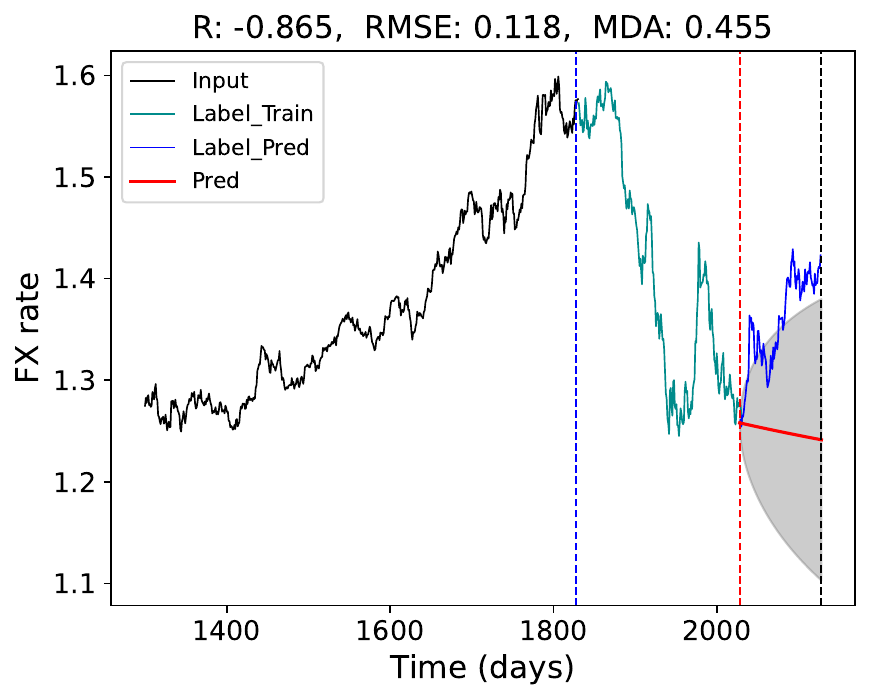}}
\subfloat[ARIMA, sample 20]{\includegraphics[scale=0.36]{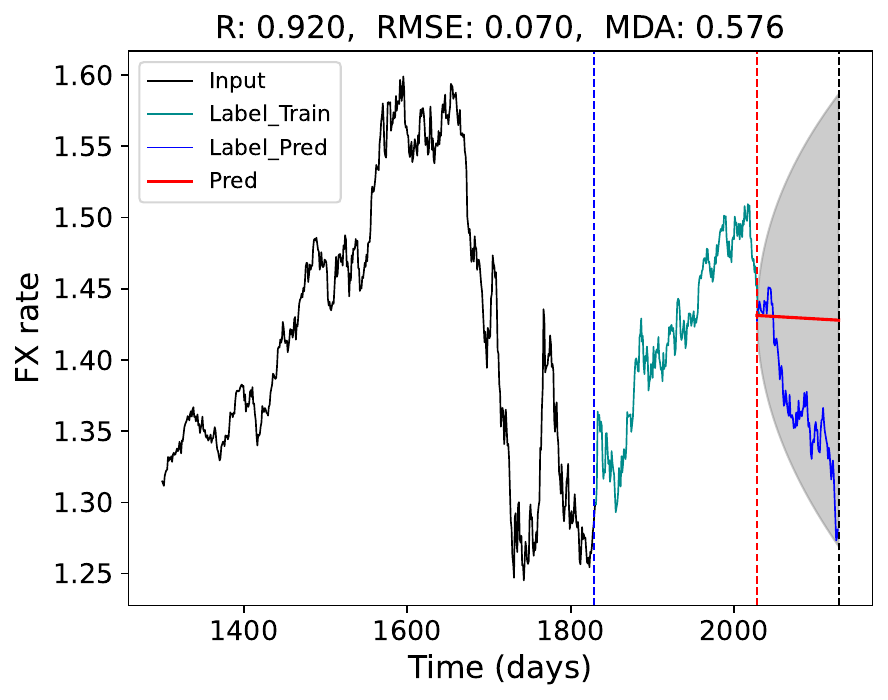}}
\subfloat[ARIMA, sample 86]{\includegraphics[scale=0.36]{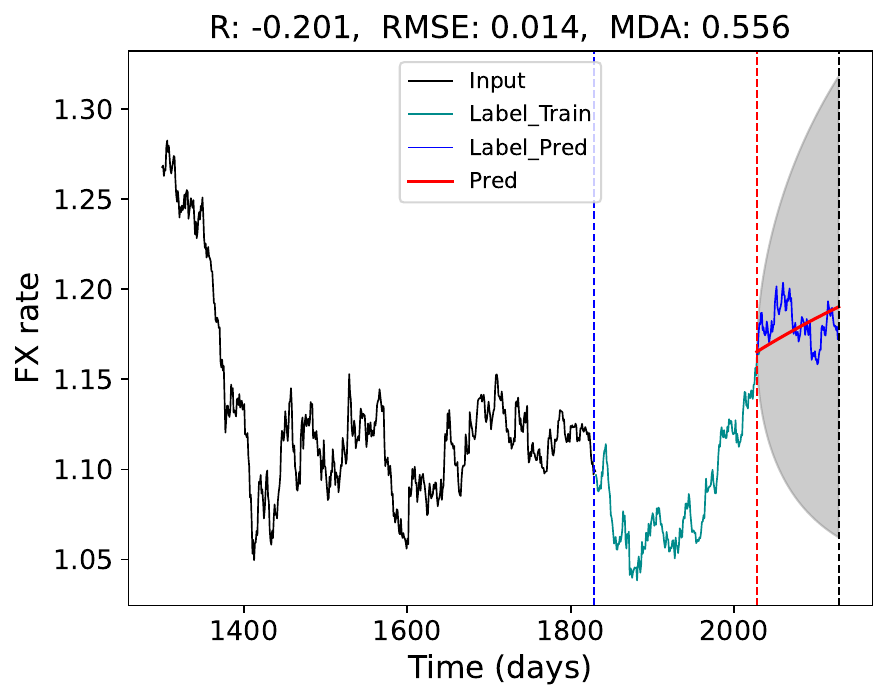}}
\end{minipage}
\caption{{Forecasts obtained for EUR/USD with RegPred Net, LSTM  and ARIMA in $3$ situations.}}
\label{fig:USD_pred}
\end{figure}

In our second experiment, we consider the EUR/USD: an interesting case of erratic time series characterised by the existence of major trend reversals, unstable and often high volatility levels and the frequent occurrence of large (positive or negative) jumps. According to the results reported in Tab. \ref{tab:USD_results}, {RegPred Net} also outperforms the other four models (LSTM, Auto-LSTM, ARMA and ARIMA) regardless of the performance metric considered. The RedPred Net's forecasts have a correlation $R$ with the true value of the FX rate that is $7$ times higher than the second best correlated model (LSTM). RegPred Net has an error (RMSE) that is about $25\%$ lower than the error of the second most accurate model (ARIMA). A few representative situations of the predictions of RegPred Net, LSTM and ARIMA can be found in Fig. \ref{fig:USD_pred}, as previously done for EUR/CNY.
\begin{table}[cols=11, pos=htpb]
\centering
\caption{Comparison of forecasting performance obtained for {EUR/USD}. The results are presented in form of mean $\pm$ std., the best results are indicated in bold font.} 
\begin{tabular*}{\tblwidth}{@{} LLLLLL @{} }
\toprule
 & \multicolumn{1}{L}{\textit{Pearson's} R} & \multicolumn{1}{L}{\textit{R-squared}} & \multicolumn{1}{L}{RMSE} & \multicolumn{1}{L}{MDA} \\
\midrule
RegPred Net &
{$\mathbf{0.342} \pm \mathbf{0.453}$} & 
  {$\mathbf{0.108} \pm \mathbf{0.964}$} &
    {$\mathbf{0.038}\pm \mathbf{0.021}$} &
     {$\mathbf{0.544}\pm \mathbf{0.048}$} \\
LSTM & 
${0.043} \pm {0.656}$ & 
  ${-21.983} \pm {30.188}$ & 
    ${0.093} \pm {0.050}$ &
  ${0.501} \pm {0.052}$ \\
Auto-LSTM &  
${-0.013} \pm {0.062}$ &
 ${-27.212} \pm {30.544}$ & 
 ${0.116} \pm {0.083}$ & 
  ${0.344} \pm {0.051}$ \\
ARMA & 
$0.051\pm {0.684}$ & 
 $-2.695\pm {3.110}$ & 
 $0.053\pm {0.033}$ & 
 $0.498\pm {0.055}$ \\
ARIMA & 
$-0.020\pm {0.700}$ &
 $-2.042\pm {1.744}$ & 
${0.051}\pm {0.035}$ & 
${0.476}\pm {0.095}$ \\\bottomrule
\end{tabular*}\label{tab:USD_results}
\end{table}

\subsubsection{Experiment results of {EUR/GBP}} \label{sec:experimental_results_EUR_GBP}

\begin{figure}[pos=htpb]
\begin{minipage}{\linewidth}
\centering
\subfloat[RegPred, sample 13]{\includegraphics
[scale=0.36]{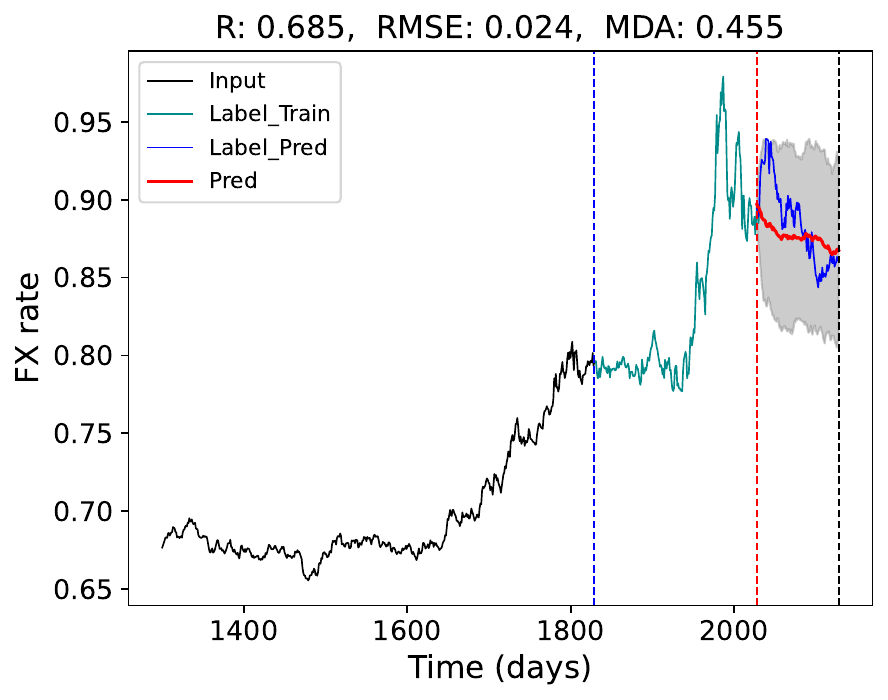}}
\subfloat[RegPred, sample 48]{\includegraphics
[scale=0.36]{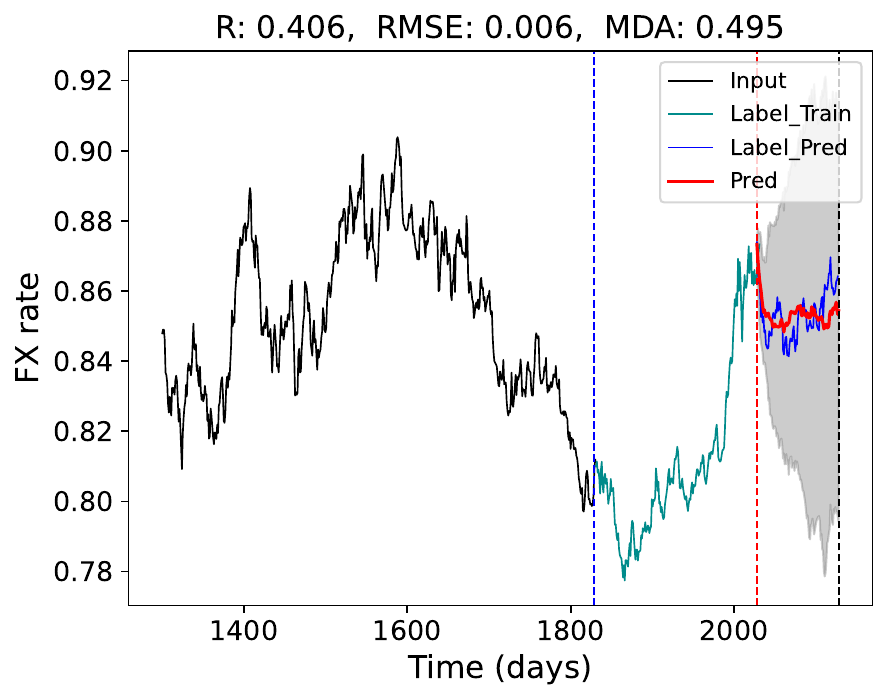}}
\subfloat[RegPred, sample 87]{\includegraphics
[scale=0.36]{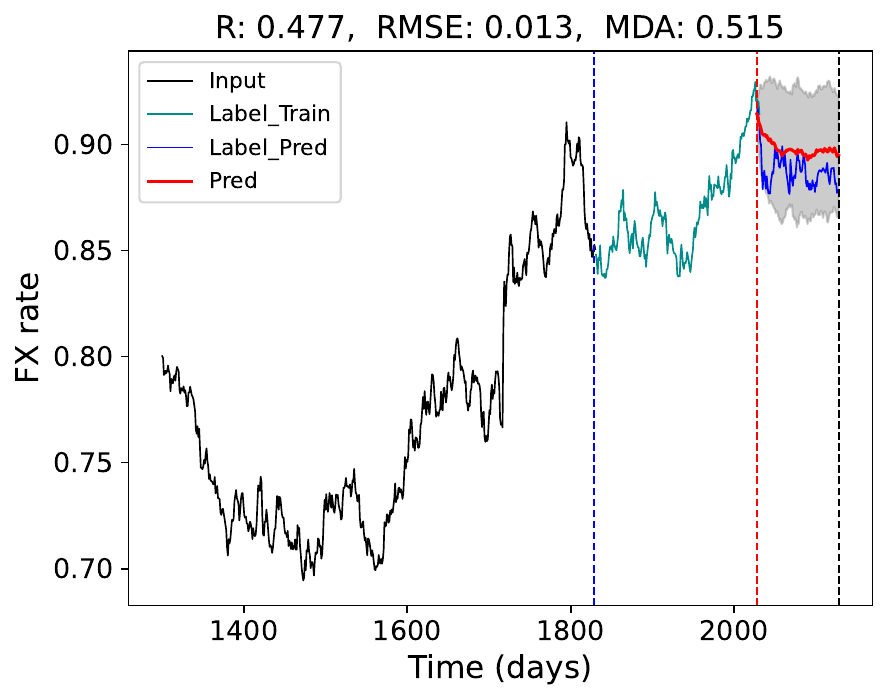}}
\end{minipage}%

\begin{minipage}{\linewidth}
\centering
\subfloat[LSTM, sample 13]{\includegraphics[scale=0.36]{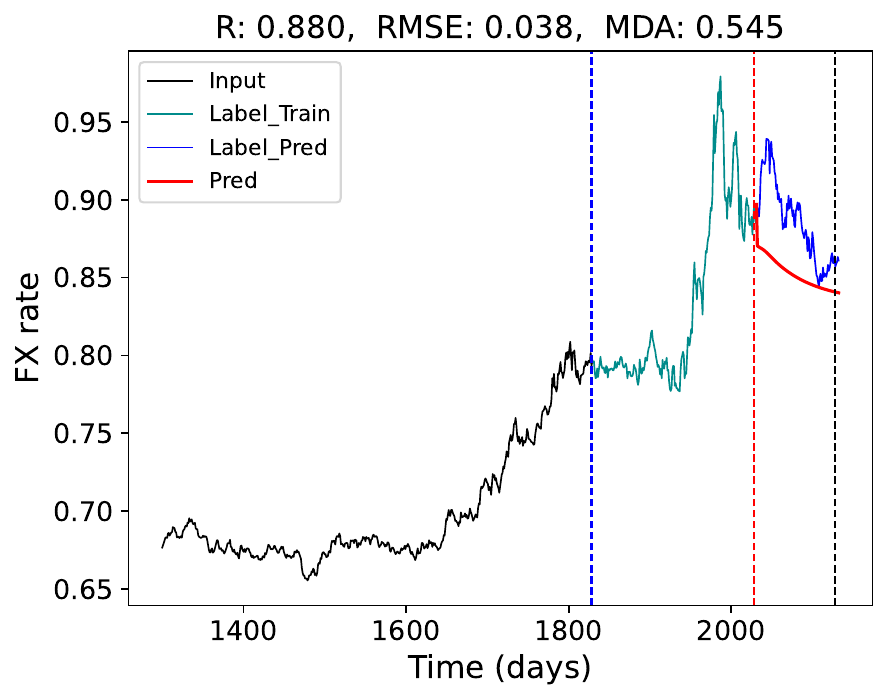}}
\subfloat[LSTM, sample 48]{\includegraphics[scale=0.36]{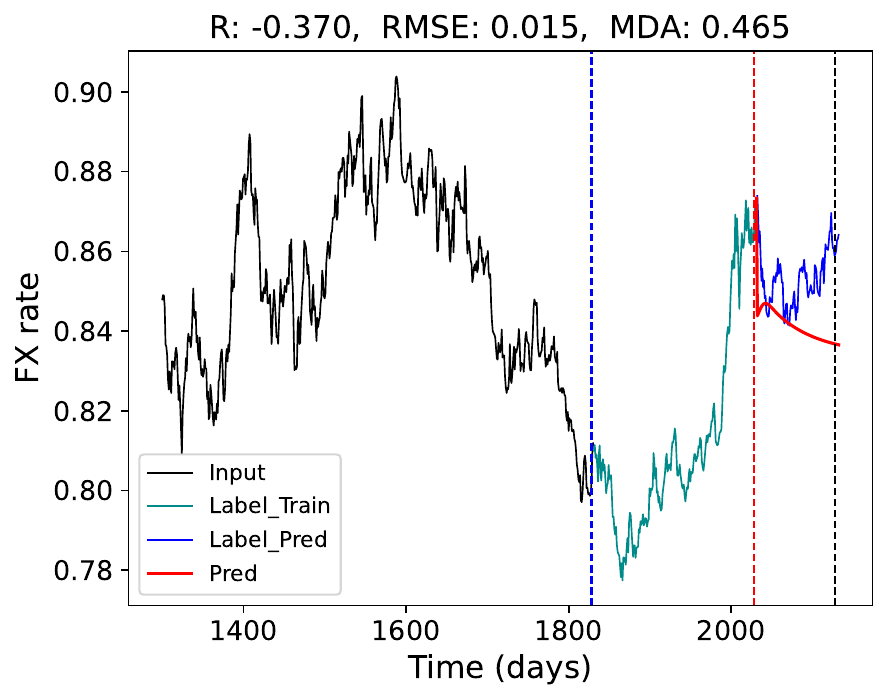}}
\subfloat[LSTM, sample 87]{\includegraphics[scale=0.36]{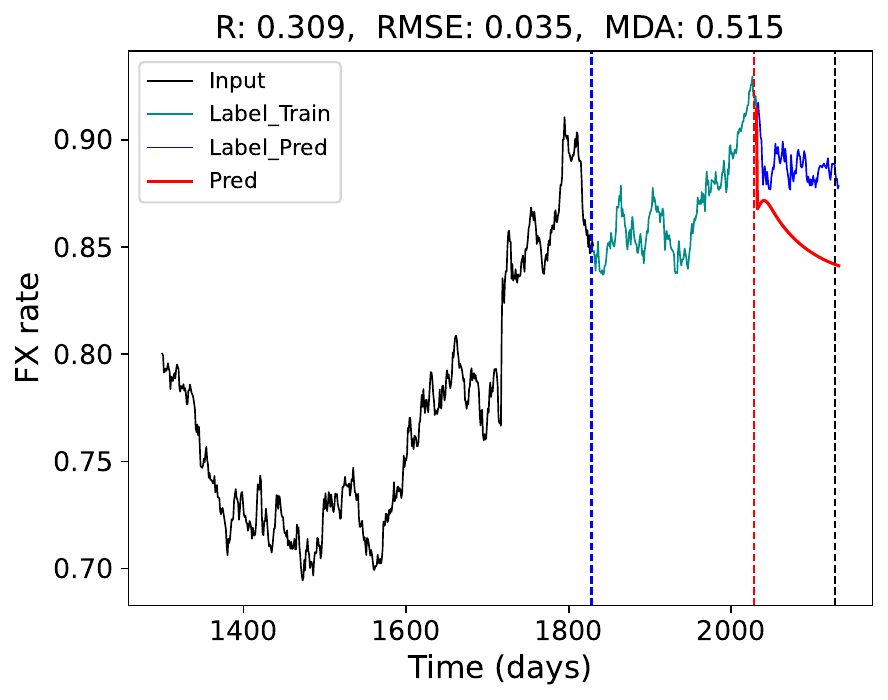}}
\end{minipage}

\begin{minipage}{\linewidth}
\centering
\subfloat[ARIMA, sample 13]{\includegraphics[scale=0.36]{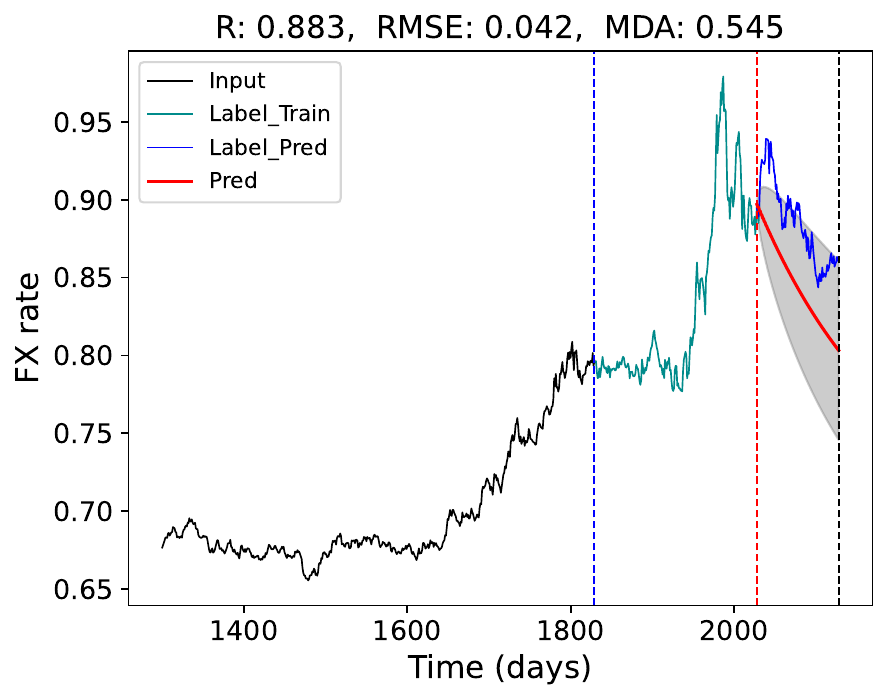}}
\subfloat[ARIMA, sample 48]{\includegraphics[scale=0.36]{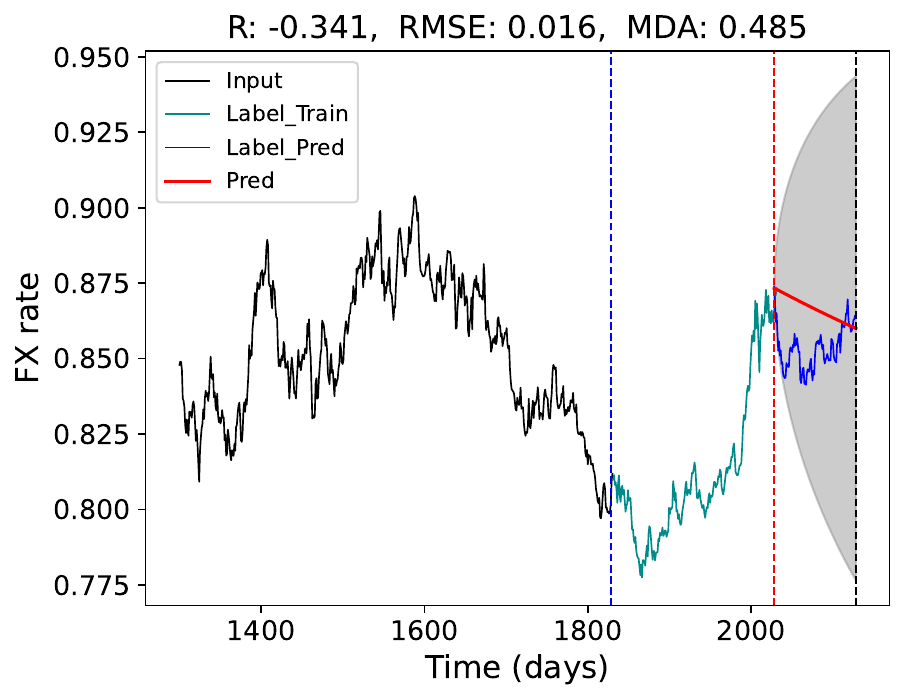}}
\subfloat[ARIMA, sample 87]{\includegraphics[scale=0.36]{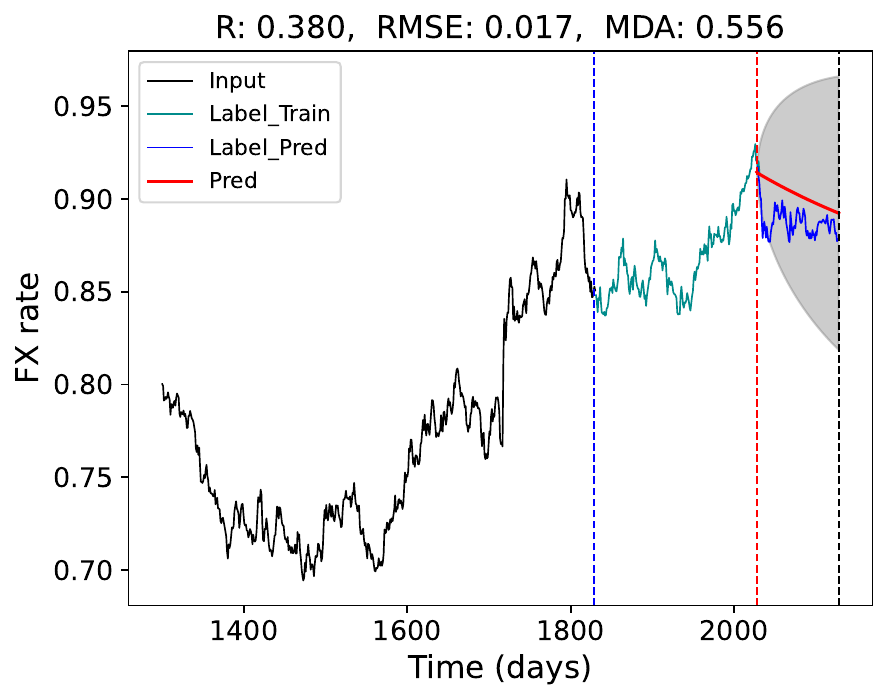}}
\end{minipage}
\caption{{Forecasts obtained for EUR/GBP with RegPred Net, LSTM  and ARIMA in $3$ situations.}}
\label{fig:GBP_pred}
\end{figure}

The experimental results for {EUR/GBP} are shown in Tab. \ref{tab:GBP_results}. From the table we observe results consistent with those of EUR/CNY and EUR/USD, with a correlation to the true value of the target variable increased by a factor $4$ and a reduction of error of $30\%$. Again, some examples of predictions are shown in Fig. \ref{fig:GBP_pred}.

\begin{table}[cols=11, pos=htpb]
\centering
\caption{Comparison of forecasting performance obtained for {EUR/GBP}. The results are presented in form of mean $\pm$ std., the best results are indicated in bold font.}
\begin{tabular*}{\tblwidth}{@{} LLLLLL @{} }
\toprule
 & \multicolumn{1}{L}{\textit{Pearson's} R} & \multicolumn{1}{L}{\textit{R-squared}} & \multicolumn{1}{L}{RMSE} & \multicolumn{1}{L}{MDA} \\
\midrule
RegPred Net &
{$\mathbf{0.434}\pm \mathbf{0.384}$} &
  {$\mathbf{0.193}\pm \mathbf{0.871}$}  &
    {$\mathbf{0.019}\pm \mathbf{0.011}$}  & 
     {$\mathbf{0.537}\pm \mathbf{0.052}$} \\
LSTM & 
 ${0.168} \pm {0.680}$ & 
  ${-9.569} \pm {14.620}$ & 
    ${0.035} \pm {0.020}$ & 
  ${0.507} \pm {0.063}$ \\
Auto-LSTM &  
${-0.001} \pm {0.440}$ & 
 ${-10.150} \pm {16.233}$ & 
 ${0.050} \pm {0.034}$ & 
  ${0.407} \pm {0.071}$ \\
ARMA & $0.049\pm0.644$ & 
$-2.051\pm2.575$ & 
$0.027\pm0.017$ & 
$0.495\pm0.054$ \\
ARIMA &
 $-0.111 \pm0.630$ & 
 $-2.106 \pm2.436$ & 
  ${0.025} \pm0.016$ & 
  ${0.447} \pm0.084$ \\
\bottomrule
\end{tabular*} \label{tab:GBP_results}
\end{table}

\section{{Conclusion}} 
\label{sec:conclusion}

In this article, we proposed a novel regression network baptised {RegPred Net} to forecast daily FX rates in the long term ($100$ days or more) in an explainable way, by exploiting the regressed and time-dependent parameters of a generalized mean-reverting or \textit{Ornstein-Uhlenbeck} (OU) process. A layerwise procedure based on Bayesian optimization was
designed to efficiently train the network in this hard domain of application of Machine Learning.

Despite the strong non-stationarity of FX rates (absence of clear trends and unstable volatility levels), RegPred Net allows to robustly derive via Monte Carlo simulation some accurate and interpretable long term forecasts. In the experiments conducted with $3$ of the most traded currencies worldwide (US dollar, Euro and Chinese Yuan) over a history of $19$ years, a RegPred Net with 2 layers significantly outperformed other Deep Learning-based models (LSTM, Auto-LSTM) and traditional time series forecasting models (ARMA, ARIMA), reducing the forecasting error (RMSE) by $25$-$30\%$ and increasing the statistical correlation (R) between forecast and actual value of FX rate by a factor of $2$ to $7$. The RegPred Net's R-squared coefficient is positive while the others are negative and its MDA is in average $10\%$ higher than the others. 

This Deep Learning and generative model of FX rates can in principle be used for simulating general stochastic and non-stationary environment variables following Brownian motion or mean-reverting processes, such as is often considered to be the case in Finance with bond or stock prices and FX rates. Such a model can thus be employed for solving a range of risk analysis and sequential decision making problems.
\section*{Declaration of competing interest}
The authors declare that they have no known competing financial interests or personal relationships that could have influenced the work reported in this paper.
\section*{Acknowledgements}
This work was supported by the University of Munich (LMU) and the Daimler AG. 

\appendix
\section{Mathematical Background} \label{Appendix A}
\subsection{Matrix / Vector Differentiation}
\begin{tabular}{cp{0.6\textwidth}}
  $\mathbf{A}$ & \quad Matrix \\
  $\mathbf{a}$ & \quad Vector (column-vector) \\
  $a$ & \quad Scalar \\
\end{tabular}

\begin{equation} \label{matrix_derivative_1}
\quad \frac{ \partial \mathbf{x}^{T} \mathbf{a} } { \partial \mathbf{x} } = \frac{ \partial \mathbf{a}^{T} \mathbf{x} } { \partial \mathbf{x} } = \mathbf{a}
\end{equation}
\begin{equation} \label{matrix_derivative_2}
\quad \frac{ \partial \mathbf{a}^{T} \mathbf{X} \mathbf{a} } { \partial \mathbf{a} } = (\mathbf{X} + \mathbf{X}^{T})\mathbf{a}
\end{equation}
\begin{equation} \label{matrix_derivative_3}
\quad \frac{ \partial \mathbf{a}^{T} \mathbf{X} \mathbf{b} } { \partial \mathbf{X} } = \mathbf{a}\mathbf{b}^{T}
\end{equation}
\begin{equation} \label{matrix_derivative_4}
\quad \frac{ \partial \mathbf{a}^{T} \mathbf{X}^{T} \mathbf{b} } { \partial \mathbf{X} } = \mathbf{b}\mathbf{a}^{T}
\end{equation}
\begin{equation} \label{matrix_derivative_5}
\quad \frac{ \partial \mathbf{a}^{T} \mathbf{X}^{T} \mathbf{X} \mathbf{b} } { \partial \mathbf{X} } = \mathbf{X} (\mathbf{a} \mathbf{b}^{T} + \mathbf{b} \mathbf{a}^{T})
\end{equation}
\begin{equation} \label{matrix_derivative_6}
\quad \frac{ \partial \mathbf{a}^{T} \mathbf{X} \mathbf{a} } { \partial \mathbf{X} } = \frac{ \partial \mathbf{a}^{T} \mathbf{X}^{T} \mathbf{a} } { \partial \mathbf{X} } =  \mathbf{a} \mathbf{a}^{T} 
\end{equation}

\subsection{Gaussian Distirbution} 
\label{normal}

The Gaussian or normal distribution is given by the following probability density function:
\begin{equation}
f(x \ | \ \mu, \sigma^{2}) = \frac{1}{\sqrt{2 \pi \sigma^{2}}} e^{-\frac{(x-\mu)^{2}}{2\sigma^{2}}}
\end{equation}
where $\mu$ is the mean or expectation of the distribution, $\sigma$ is the standard deviation and $\sigma^{2}$ the variance.

\subsection{Multivariate Gaussian Distirbution} 
\label{mg}

The multivariate Gaussian (normal) distribution generalizes the univariate normal distribution to higher dimensions. A vector-valued random variable $x \in \mathbb R^{n}$ is considered as multivariate normal distribution of mean $\mu \in \mathbb R^{n}$ and covariance matrix $\Sigma \in \mathbb S^{n}_{++}$ if its probability density distribution follows
\begin{equation} \label{eq:mg}
p(x; \ \mu, \Sigma) = \frac {1} {{(2\pi)}^{n/2} |\Sigma|^{1/2}} \ exp\Bigg(- \frac{1} {2}(x-\mu)^{T} \Sigma^{-1}(x-\mu)\Bigg)
\end{equation}
Eq. (\ref{eq:mg}) can be written as $x \sim \mathcal N (\mu, \Sigma)$. $\mathbb S^{n}_{++}$ refers to the space of symmetric positive definite $n \times n$ matrices. 

\subsection{Radial Basis Function} 
\label{RBF}

The Radial Basis Function (RBF) kernel, commonly used in kernelized learning algorithms, e.g. SVMs is defined as:
\begin{equation}
k(\mathbf{x}, \mathbf{x'}) = \text{exp} \big(-\frac{1}{2\theta^{2}} ||\mathbf{x} - \mathbf{x'}||^{2} \big)
\end{equation}
where $||\mathbf{x} - \mathbf{x'}||^{2}$ is the squared Euclidean distance between two feature vectors $\mathbf x$ and $\mathbf x'$. $\theta$ is a free parameter which indicates the width of the kernel. Since the output range of RBF is in $[0\ ,\ 1]$, which is inversely proportional to the distance between vectors, RBF is often used as a similarity measure. 

\subsection{Gamma function} 
\label{Gamma}

The Gamma function is the generalization of the factorial function to complex numbers and is defined as:
\begin{equation} 
\Gamma(z) = \int_{0}^{\infty} x^{z-1}e^{-x} dx
\end{equation}
where $z$ is a complex number with positive real part ($Re({z})>0$). The function has the follow property:
\begin{equation}
\Gamma(z+1) = z \Gamma(z)
\end{equation}

\section{Mathematical Derivations} \label{Appendix B}

\subsection{Simplification of the Covariance Matrix} 
\label{cov_mat}

Each element $cov ( \epsilon_{{i, t}}, \epsilon_{{j, t}} )$ of the $d\times d$ covariance matrix $\bm{K}_{{\bm{\epsilon} \bm{\epsilon}, t}}$ is equal to:
\begin{equation} \label{eq:appx_cov_ep}
\begin{split}
cov ( \epsilon_{{i, t}}, \epsilon_{{j, t}} ) &= \mathbb{E} \Big[\big(\epsilon_{{i, t}} - \mathbb{E}(\epsilon_{{i, t}})\big)\big(\epsilon_{{j, t}} - \mathbb{E}(\epsilon_{{j, t}})\big) \Big] \\
&= \mathbb{E}\big(\epsilon_{{i, t}}\cdot \epsilon_{{j, t}}\big) \ \quad\quad\quad\quad\quad\,\,\, \,\text{since $\mathbb{E}(\epsilon_{{i, t}}) = 0$ (Eq. (\ref{ols_OU}))} \\
&= \mathbb{E} \bigg( \sum_{{k=1}}^{{d}}\sum_{{k'=1}}^{{d}} \sigma_{{i, k}} \cdot \sigma_{{j, k'}} \cdot \Delta W_{{k,t}} \cdot \Delta W_{{k',t}} \bigg) \\
&= \mathbb{E} \bigg( \sum_{{k=1}}^{{d}} \sigma_{{i, k}} \cdot \sigma_{{j, k}}\Delta W_{{k,t}}^{{2}} \bigg) 
\quad \begin{aligned} &\text{since $\mathbb{E}(\Delta W_{{k,t}} \cdot \Delta W_{{k', t}}) = 0$ when $k \neq k'$,} \\ &\text{because $\Delta W_{{i,t}}$ are independent and $\mathbb{E}(\Delta W_{{i,t}}) = 0$ } \end{aligned} \\
&= \sum_{{k=1}}^{{d}} \sigma_{{i, k}} \cdot \sigma_{{j, k}} \cdot \mathbb{E}\Big(\Delta W_{{k,t}}^{{2}}\Big) 
\\
&= \sum_{{k=1}}^{{d}} \sigma_{{i, k}} \cdot \sigma_{{j, k}}\quad\quad\quad\quad\quad\quad \text{since $\Delta W_{{k, \ t}} \sim \mathscr{N}(0, \ 1)$, (Sec. \ref{sec:GOU})}\\
\end{split}
\end{equation}

\subsection{Computation of the Gradients of $\mathbf{A}, \mathbf{N}, \mathbf{\Sigma}$} \label{appx_gradients}
The loss $L_{{t}} \big( \mathbf{A}_{{t-1}}; \mathbf{N}_{{t-1}} \big)$ in Eq. (\ref{eq:L_AN}) can be further computed as:
\begin{equation} \label{appx_L_AN}
\begin{aligned}
L_{{t}} \big( \mathbf{A}_{{t-1}}; \mathbf{N}_{{t-1}} \big)  = &\Big[ \Delta \mathbf{Y}_{{t}} - \big( \mathbf{A}_{{t-1}} \cdot \mathbf{Y}_{{t-1}} + \mathbf{N}_{{t-1}} \big) \Big]^{T} \Big[ \Delta \mathbf{Y}_{{t}} - \big( \mathbf{A}_{{t-1}} \cdot \mathbf{Y}_{{t-1}} + \mathbf{N}_{{t-1}} \big) \Big] \\
 =& \Delta \mathbf{Y}_{{t}}^{T} \Delta \mathbf{Y}_{{t}} - \Delta \mathbf{Y}_{{t}}^{{T}}\mathbf{A}_{{t-1}}\mathbf{Y}_{{t-1}} - \Delta \mathbf{Y}_{{t}}^{{T}}\mathbf{N}_{{t-1}} -  \mathbf{Y}_{{t-1}}^{{T}}\mathbf{A}_{{t-1}}^{{T}}\Delta \mathbf{Y}_{{t}} \\ 
& + \mathbf{Y}_{{t-1}}^{{T}}\mathbf{A}_{{t-1}}^{{T}}\mathbf{A}_{{t-1}}\mathbf{Y}_{{t-1}} + \mathbf{Y}_{{t-1}}^{{T}}\mathbf{A}_{{t-1}}^{{T}}\mathbf{N}_{{t-1}} - \mathbf{N}_{{t-1}}^{{T}}\Delta \mathbf{Y}_{{t}} + \mathbf{N}_{{t-1}}^{{T}}\mathbf{A}_{{t-1}}\mathbf{Y}_{{t-1}} + \mathbf{N}_{{t-1}}^{{T}}\mathbf{N}_{{t-1}}\\
\end{aligned}
\end{equation}
According to the matrix differentiation rules in Eq. (\ref{matrix_derivative_3}), (\ref{matrix_derivative_4}), (\ref{matrix_derivative_5}) and the definition of $\bm{\epsilon}_{t}$ in Eq. (\ref{update_epsilon}), the partial derivative of $L_{{t}}(\mathbf{A}_{{t-1}}; \mathbf{N}_{{t-1}})$ with respect to $\mathbf{A}_{{t-1}}$ is calculated as:
\begin{equation} \label{appx_partial_A}
\begin{split}
\frac{ \partial L_{{t}}(\mathbf{A}_{{t-1}}; \mathbf{N}_{{t-1}}) } { \partial \mathbf{A}_{{t-1}} } &= - 2 \cdot \Delta \mathbf{Y}_{{t}} \mathbf{Y}_{{t-1}}^{{T}} + 2 \cdot \mathbf{A}_{{t-1}}\mathbf{Y}_{{t-1}}\mathbf{Y}_{{t-1}}^{{T}} + 2 \cdot \mathbf{N}_{{t-1}}\mathbf{Y}_{{t-1}}^{{T}} \\
 &= - 2 \cdot \bm{\epsilon}_{{t}} \mathbf{Y}_{{t-1}}^{{T}}
\end{split}
\end{equation}
Similarly, the partial derivative of $L_{{t}}(\mathbf{A}_{{t-1}}; \mathbf{N}_{{t-1}})$ with respect to $\mathbf{N}_{{t-1}}$ is:
\begin{equation} \label{appx_partial_N}
\begin{split}
\frac{ \partial L_{{t}}(\mathbf{A}_{{t-1}}; \mathbf{N}_{{t-1}}) } { \partial \mathbf{N}_{{t-1}} } &= - 2 \cdot \Delta \mathbf{Y}_{{t}} + 2 \cdot \mathbf{A}_{{t-1}}\mathbf{Y}_{{t-1}} + 2 \cdot \mathbf{N}_{{t-1}} \\
& = -2 \cdot \bm{\epsilon}_{{t}} 
\end{split}
\end{equation}

\makeatletter
\newcommand{\vast}{\bBigg@{4}}
To calculate the derivative of the loss $L_{{t}}(\bm{\Sigma}_{{t-1}})$ (Eq. (\ref{eq:L_sigma})) with respect to $\bm{\Sigma}_{{t-1}}$, first calculate the derivative of $L_{{t}}(\bm{\Sigma}_{{t-1}})$ with respect to a single coefficient $\sigma_{{ij, t-1}}$ of the matrix $\bm{\Sigma}_{{t-1}}$: 
\begin{equation} \label{appx_derivative_online_Sigma}
\begin{aligned}
\frac{ \partial L_{{t}}(\bm{\Sigma}_{{t-1}}) }{ \partial \sigma_{{ij, t-1}} } &= \frac{ \partial \big|\big|\enskip \bm{\Sigma}_{{t-1}} \bm{\Sigma}^{{T}}_{{t-1}} - \hat{cov}(\bm{\epsilon}_{{t}})\enskip \big|\big|^{2}_{2} }{ \partial {\sigma}_{{ij, t-1}} } \\
&= \frac{ \partial \sum\limits_{{l,m}} \Big({\big( \bm{\Sigma}_{{t-1}} \bm{\Sigma}_{{t-1}}^{{T}} \big)}_{{l,m}} - \hat{cov}{(\bm{\epsilon}_{{t}})}_{{l,m}} \Big)^{2} } { \partial {\sigma}_{{ij, t-1}} } \\
 &= \sum\limits_{{l,m}} \frac{ \partial \Big( \big(\bm{\Sigma}_{{t-1}} \bm{\Sigma}_{{t-1}}^{{T}} \big)_{{l,m}} - \hat{cov}{(\bm{\epsilon}_{{t}})}_{{l,m}} \Big)^{2}}{ \partial {\sigma}_{{ij, t-1}} } \\
 &= \sum_{{l,m}} 2 \cdot \Big( \big(\bm{\Sigma}_{{t-1}} \bm{\Sigma}_{{t-1}}^{{T}}  \big)_{{l,m}} - \hat{cov}{(\bm{\epsilon}_{{t}})}_{{l,m}} \Big) \cdot \underset{(1)} { \vast( \frac{  \partial \Big( \big( \bm{\Sigma}_{{t-1}} \bm{\Sigma}_{{t-1}}^{{T}} \big)_{{l,m}} - \hat{cov}{(\bm{\epsilon}_{{t}})}_{{l,m}} \Big )} { \partial {\sigma}_{{ij, t-1}} } \vast) }\\
\end{aligned}
\end{equation}
where the subscript ${{l,m}}$ represents the position at line $l$ and column $m$. $\big( \bm{\Sigma}_{{t-1}}\bm{\Sigma}_{{t-1}}^{{T}} \big)_{{l,m}}$ and $cov(\bm{\epsilon}_{{t}})_{{l,m}}$ are:
\begin{equation} \label{appx_SigmaSigmaT_lm}
\big( \bm{\Sigma}_{{t-1}}\bm{\Sigma}_{{t-1}}^{{T}} \big)_{{l,m}} = \sum \limits_{{k=1}}^{{d}} \sigma_{{lk, t-1}} \cdot \sigma_{{mk, t-1}}
\end{equation}
\begin{equation} \label{appx_cov_epsilon_lm}
cov(\bm{\epsilon}_{{t}})_{{l,m}} = cov(\epsilon_{{l,t}}, \epsilon_{{m,t}})
\end{equation}
respectively.\\

Substituting Eq. (\ref{appx_SigmaSigmaT_lm}) into term $(1)$ of Eq. (\ref{appx_derivative_online_Sigma}), we get:
\begin{equation}
\begin{aligned}
\frac{ \partial \Big( \big(\bm{\Sigma}_{{t-1}} \bm{\Sigma}_{{t-1}}^{{T}}  \big)_{{l,m}}  - { \hat{cov}(\bm{\epsilon}_{{t}})_{{l,m}} }  \Big)} { \partial {\sigma}_{{ij, t-1}} } & = 
\frac{ \partial \sum\limits_{{k=1}}^{{d}} \enskip \sigma_{{lk, t-1}} \cdot \sigma_{{mk, t-1}}} { \partial \sigma_{{ij, t-1}} } \\
 &= \delta_{{(i,l)} } \cdot \sigma_{{mj, t-1} } + \delta_{{(i,m)} } \cdot \sigma_{{lj, t-1} }  \quad\quad \text{ $\delta_{{(a,b)} } = 1$ when $a=b$, else $\delta_{{(a,b)} } = 0$ }
\end{aligned}
\end{equation}
Thus, Eq. (\ref{appx_derivative_online_Sigma}) simplifies as:
\begin{equation} \label{appx_derivative_expansion_Sigma}
\begin{aligned}
\frac{ \partial L_{{t} } \big(\bm{\Sigma}_{{t-1} } \big) }{ \partial {\sigma}_{{ij, t-1} } } &= \sum_{{l,m} } 2 \cdot \Big( \big( \bm{\Sigma}_{{t-1} } \bm{\Sigma}_{{t-1} }^{{T} } \big)_{{l,m} } - \hat{cov}{(\bm{\epsilon}_{{t} } )}_{{l,m} } \Big)  \Big( \delta_{{(i,l)} } \cdot \sigma_{{mj, t-1} } + \delta_{{(i,m)} } \cdot \sigma_{{lj, t-1} } \Big) \\
 &= \underset{(1)}{ \sum\limits_{{m} } 2 \cdot \Big( \big(\bm{\Sigma}_{{t-1} } \bm{\Sigma}_{{t-1} }^{{T} }  \big)_{{i,m} } - \hat{cov}{(\bm{\epsilon}_{{t} })}_{{i,m} } \Big) \cdot \sigma_{{mj, t-1} } } + \underset{(2)} { \sum \limits_{{l} } 2 \cdot \Big( \big(\bm{\Sigma}_{{t-1} } \bm{\Sigma}_{{t-1} }^{{T} } \big)_{{l,i} } - \hat{cov}{(\bm{\epsilon}_{{t} })}_{{l,i} } \Big) \cdot \sigma_{{lj, t-1} } }
\end{aligned}
\end{equation}
Term (1) in Eq. (\ref{appx_derivative_expansion_Sigma}) is $2 \cdot \big(\bm{\Sigma}_{{t-1} } \bm{\Sigma}_{{t-1} }^{{T} } - \hat{cov}(\bm{\epsilon}_{{t} })\big)_{{\text{line}\, i}} \cdot {({\sigma}_{{t-1}} )}_{{\text{column} \, j} }$ and term (2) is $2 \cdot \big(\bm{\Sigma}_{{t-1} } \bm{\Sigma}_{{t-1} }^{{T} } - \hat{cov}(\bm{\epsilon}_{{t} })\big)^{{T} }_{{\text{line}\, i} } \cdot {({\sigma}_{{t-1} })}_{{\text{column} \, j} }$, where $\big(\bm{\Sigma}_{{t-1} }\bm{\Sigma}_{{t-1} }^{{T} }  - \hat{cov}(\bm{\epsilon}_{{t} })\big)^{{T} } = \bm{\Sigma}_{{t-1} }\bm{\Sigma}_{{t-1} }^{{T} }  - \hat{cov}(\bm{\epsilon}_{{t} })$. Therefore,
\begin{equation} \label{appx_derivative_sigma_ij}
\begin{split}
\frac{ \partial L_{{t} }(\bm{\Sigma}_{{t-1} }) }{ \partial {\sigma}_{{ij, t-1} } } 
&= 4 \cdot \Big(\big( \bm{\Sigma}_{{t-1} } \bm{\Sigma}_{{t-1} }^{{T} }  - \hat{ cov}(\bm{\epsilon}_{{t} }) \big) \cdot \bm{\Sigma}_{{t-1} } \Big)_{{i,j} } \end{split}
\end{equation}
and the derivative of $L_{{t} }(\bm{\Sigma}_{{t-1} })$ with respect to the matrix $\bm{\Sigma}_{{t-1} }$ is finally
\begin{equation}
\frac{ \partial L_{{t} }(\bm{\Sigma}_{{t-1} }) }{ \partial \bm{\Sigma}_{{t-1} } } =
4 \cdot \big(\bm{\Sigma}_{{t-1} } \bm{\Sigma}_{{t-1} }^{{T} } - \hat{ cov}(\bm{\epsilon}_{{t} })\big) \cdot \bm{\Sigma}_{{t-1} }
\end{equation}

\printcredits

\bibliographystyle{cas-model2-names}

\bibliography{cas-refs}

\end{document}